\documentclass{article}[12pt]
\usepackage{ifthen}
\usepackage{mdwlist}
\usepackage{amsmath,amssymb,amsfonts,amsthm,bbm}
\usepackage{bm}
\usepackage{soul}
\usepackage[colorlinks=true,linkcolor=black,citecolor=black,urlcolor=blue]{hyperref}
\usepackage{enumitem}
\usepackage{graphicx}
\usepackage{xspace}
\usepackage{verbatim}
\usepackage{algorithm}
\usepackage{algorithmic}
\usepackage[margin=1in]{geometry}
\usepackage{color}
\usepackage{thm-restate}
\usepackage{latexsym}
\usepackage{epsfig}
\usepackage{subcaption}
\usepackage{tabularx}
\usepackage{mathrsfs}
\usepackage{xcolor}
\usepackage{tcolorbox}
\usepackage{appendix}
\usepackage{titlesec}
\usepackage[colorinlistoftodos,textsize=scriptsize]{todonotes}
\usepackage[normalem]{ulem}
\usepackage{tikz}
\usetikzlibrary{bayesnet}
\usepackage{natbib}
\usepackage{booktabs}
\usepackage{dsfont}
\definecolor{bgcolor}{RGB}{254, 252, 232}
\definecolor{OliveGreen}{rgb}{0,0.6,0}
\def\withcolors{1}
\def\withnotes{1}

\usepackage[noabbrev,capitalise]{cleveref}
\creflabelformat{equation}{#2\textup{#1}#3}
\numberwithin{equation}{section}
\numberwithin{figure}{section}
\numberwithin{table}{section}

\usepackage{setspace}
\newtheorem{theorem}{Theorem}

\newtheorem{condition}[theorem]{Condition}
\newtheorem{example}[theorem]{Example}

\numberwithin{theorem}{section} 
\numberwithin{nontheorem}{section} 
\numberwithin{proposition}{section} 
\numberwithin{observation}{section} 
\numberwithin{remark}{section} 
\numberwithin{fact}{section} 
\numberwithin{lemma}{section} 
\numberwithin{claim}{section} 
\numberwithin{corollary}{section} 
\numberwithin{case}{section} 
\numberwithin{dfn}{section} 
\numberwithin{definition}{section} 
\numberwithin{question}{section} 
\numberwithin{openquestion}{section} 
\numberwithin{res}{section} 
\numberwithin{condition}{section}
\numberwithin{example}{section} 
\numberwithin{conjecture}{section}

\ifnum\withcolors=1

\else

\fi

\ifnum\withnotes=1

\else
  \newcommand{\gnote}[1]{}
  \newcommand{\gfootnote}[1]{}

\fi
\newcommand{\ignore}[1]{\leavevmode\unskip} 

\newcommand{\Esymb}{\mathbb{E}}

\DeclareMathOperator*{\E}{\Esymb}

\newcommand\bdot\bullet

\newcommand{\R}{\mathbb R}

\renewcommand{\leq}{\leqslant}

\renewcommand{\geq}{\geqslant}

\let\epsilon=\varepsilon
\newcommand\MYcurrentlabel{xxx}
\newcommand{\MYstore}[2]{%
	\global\expandafter \def \csname MYMEMORY #1 \endcsname{#2}%
}
\newcommand{\MYload}[1]{%
	\csname MYMEMORY #1 \endcsname%
}
\newcommand{\MYnewlabel}[1]{%
	\renewcommand\MYcurrentlabel{#1}%
	\MYoldlabel{#1}%
}
\newcommand{\MYdummylabel}[1]{}
\newcommand{\torestate}[1]{%
	\let\MYoldlabel\label%
	\let\label\MYnewlabel%
	#1%
	\MYstore{\MYcurrentlabel}{#1}%
	\let\label\MYoldlabel%
}
\newcommand{\restatetheorem}[1]{%
	\let\MYoldlabel\label
	\let\label\MYdummylabel
	\begin{theorem*}[Restatement of \cref{#1}]
		\MYload{#1}
	\end{theorem*}
	\let\label\MYoldlabel
}
\newcommand{\restatelemma}[1]{%
	\let\MYoldlabel\label
	\let\label\MYdummylabel
	\begin{lemma*}[Restatement of \cref{#1}]
		\MYload{#1}
	\end{lemma*}
	\let\label\MYoldlabel
}
\newcommand{\restateprop}[1]{%
	\let\MYoldlabel\label
	\let\label\MYdummylabel
	\begin{proposition*}[Restatement of \cref{#1}]
		\MYload{#1}
	\end{proposition*}
	\let\label\MYoldlabel
}
\newcommand{\restatefact}[1]{%
	\let\MYoldlabel\label
	\let\label\MYdummylabel
	\begin{fact*}[Restatement of \cref{#1}]
		\MYload{#1}
	\end{fact*}
	\let\label\MYoldlabel
}
\newcommand{\restate}[1]{%
	\let\MYoldlabel\label
	\let\label\MYdummylabel
	\MYload{#1}
	\let\label\MYoldlabel
}

\allowdisplaybreaks
\newcommand{\indep}{\perp\!\!\!\perp}

\makeatletter
\newcommand{\distas}[1]{\mathbin{\overset{#1}{\kern\z@\sim}}}%
\newsavebox{\mybox}\newsavebox{\mysim}
\newcommand{\distras}[1]{%
	\savebox{\mybox}{\hbox{\kern3pt$\scriptstyle#1$\kern3pt}}%
	\savebox{\mysim}{\hbox{$\sim$}}%
	\mathbin{\overset{#1}{\kern\z@\resizebox{\wd\mybox}{\ht\mysim}{$\sim$}}}%
}
\makeatother

\def\*#1{\mathbf{#1}}

\def\HiLi{\leavevmode\rlap{\hbox to \hsize{\color{yellow!50}\leaders\hrule height 1.2 \baselineskip depth .5ex\hfill}}}

\definecolor{dred}{rgb}{0.75, 0.0, 0.2}
\definecolor{dgreen}{rgb}{0.33, 0.42, 0.18}
\definecolor{dblue}{rgb}{0.0, 0.2, 0.4}
\definecolor{crimson}{rgb}{0.86, 0.08, 0.24}

\title{Bayesian model criticism using uniform parametrization checks}
\author{Christian T. Covington\thanks{Department of Biostatistics, Harvard T.H. Chan School of Public Health, Boston, MA, \texttt{ccovington@g.harvard.edu}}
        \and 
        Jeffrey W. Miller\thanks{Department of Biostatistics, Harvard T.H. Chan School of Public Health, Boston, MA, \texttt{jwmiller@hsph.harvard.edu}}}
\date{}

\graphicspath{ {results/} }

\begin{document}

\maketitle

\begin{abstract}

Models are often misspecified in practice, making model criticism a key part of Bayesian analysis.
It is important to detect not only when a model is wrong, but which aspects are wrong, and to do so in a computationally convenient and statistically rigorous way.   
We introduce a novel method for model criticism based on the fact that if the parameters are drawn from the prior, 
and the dataset is generated according to the assumed likelihood, then a sample from the posterior will be distributed 
according to the prior.  Thus, departures from the assumed likelihood or prior can be detected by testing whether a 
posterior sample could plausibly have been generated by the prior.  
Building upon this idea, we propose to reparametrize all random elements of the likelihood and prior in terms of independent uniform random variables, or u-values.
This makes it possible to aggregate across arbitrary subsets of the u-values for data points and parameters to test for model departures using 
classical hypothesis tests for dependence or non-uniformity.
We demonstrate empirically how this method of 
uniform parametrization checks (UPCs) facilitates model criticism in several examples, and we develop supporting theoretical results.

\end{abstract}
\section{Introduction}

Bayesian statistics proceeds by defining a model---consisting of a prior and likelihood---and drawing posterior inferences based on the assumption that this model is correct.
However, if the model is not correct then the resulting inferences may be misleading.
In practice, it can be difficult to know whether a model is sufficiently accurate to provide reliable inferences and, if not, which aspects of the model need improvement.
The task of detecting a model's inadequacies is called ``model criticism'' \citep{Box80,gelman2013bayesian,blei2014build}.

While many methods have been proposed for Bayesian model criticism (see \cref{section:previous-work}), posterior predictive checks (PPCs) are currently the most commonly used technique.
PPCs compare the observed value of a test statistic---or more generally, a test quantity that may depend on the data and parameters---to its distribution under the posterior predictive \citep{guttman1967use,rubin1984bayesianly,meng1994posterior,gelman2013bayesian}.
However, it is well known that the ``p-values'' produced by PPCs are not valid, in the sense that they are not uniformly distributed under the null hypothesis that the model is correct, even asymptotically \citep{bayarri1999quantifying,bayarri2000p,robins2000asymptotic}.
Valid PPC p-values can be obtained by using a partial posterior or conditional predictive \citep{bayarri1999quantifying,bayarri2000p}, however, these may be hard to implement in many models.
Alternatively, \citet{moran2019population} and \citet{li2022calibrated} propose using data splitting to obtain valid PPC p-values, however, this (i) entails a loss of information since the split-data posterior does not use the full dataset, and (ii) typically involves several posterior inference runs over various splits.
Furthermore, an even more fundamental difficulty of using PPCs is that they require one to design test quantities to detect the various types of misspecification of concern in the model at hand.
There are infinitely many possible test quantities, and choosing which ones to consider requires (i) confidence about the kinds of misspecification that may be present and (ii) statistical insight into what makes a good PPC test quantity, including subtle considerations of sufficiency and ancillarity \citep{gelman2013bayesian,mimno2015posterior,bolsinova2016posterior}.
This puts a major burden on the analyst, hindering the adoption of PPCs in practice. 

In this paper, we introduce a novel method for model criticism that overcomes these limitations.
The method is inspired by the observation that if the parameters $\theta$ are drawn from the prior, a dataset $Y\mid\theta$ is drawn according to the assumed likelihood, and $\widetilde{\theta}\mid Y$ is drawn from the posterior, then the marginal distribution of $\widetilde{\theta}$ (integrating out $\theta$ and $Y$) is equal to the prior.
Thus, if a posterior sample could not plausibly have been generated by the prior, then this indicates misspecification of some part of the model likelihood or prior \citep[Section 6.4]{gelman2013bayesian}.
Building upon this basic idea, we propose to reparametrize all random elements in the model---including data and parameters---in terms of independent and identically distributed uniform random variables (``u-values'') on the unit interval.
Then, under the null that the model is correct, a posterior sample of the u-values is exactly distributed as i.i.d.\ uniform.
This enables one to easily perform hypothesis tests probing for misspecification of various parts of the model simply by testing for departures from independence and uniformity of the u-values in various ways.
It is important to emphasize that \textit{multiple} posterior samples given the same dataset will \textit{not} be independent, because the dataset introduces dependence across samples.  Thus, to use multiple posterior samples---for example, from a Markov chain Monte Carlo (MCMC) run---we use a p-value aggregation technique for dependent p-values.

The proposed method, referred to as ``uniform parametrization checks'' (UPCs), shares a number of attractive features with PPCs.
Like PPCs, UPCs help reveal not only whether a model is wrong, but which parts are wrong, and how to improve it.
Also like PPCs, UPCs are computationally tractable and easy to implement, requiring only one posterior inference run (for example, using MCMC) that yields samples from the standard posterior given the observed data.
And like PPCs, UPCs are applicable to a very wide range of models.

Additionally, UPCs have several advantages compared to PPCs.
First, UPCs yield uniform p-values under the null that the model is correct, since---up to posterior inference approximations---the posterior u-values are exactly i.i.d.\ uniform when the model is correct.
Thus, Type I error rate is correctly controlled---regardless of the dataset size---and one can even perform iterative model building in a principled way via alpha spending.
Second, there are natural default choices of UPC tests that apply to any model, and it is usually intuitively clear how to design new customized tests.
Third, in contrast to PPCs, which only provide model criticism based on a collection of selected statistics, UPCs provide a more comprehensive understanding of where a model is going wrong, since we know the exact joint distribution of all u-values under the null.

The paper is organized as follows.
In \cref{section:methodology}, we introduce the UPC methodology.
\cref{section:previous-work} briefly reviews previous work, and in \cref{sec:theory}, we establish theoretical properties of UPCs.
In \cref{section:examples}, we demonstrate the UPC method in several examples involving real and simulated data.
Finally, \cref{section:discussion} concludes with a brief discussion and directions for future work.

\section{Methodology}
\label{section:methodology}

Suppose $\Pi$ is a prior distribution on the parameter $\theta$ and $P_\theta$ is a hypothesized distribution of the dataset $Y$ given $\theta$. Letting $\theta\sim\Pi$ and $Y\sim P_\theta$ given $\theta$, this defines a joint distribution on parameters and data, $(\theta,Y)$, which we refer to as the \textit{hypothesized model}.
Assume we can write $(\theta,Y) = g(U)$ where $g$ is a known function and $U\sim\mathrm{Uniform}_D(0,1)$, that is, $U = (U_1,\ldots,U_D)$ and $U_1,\ldots,U_D$ i.i.d.\ $\sim \mathrm{Uniform}(0,1)$. 
Most Bayesian models used in practice can be written in this way, including, for example, complex hierarchical models with continuous and discrete variables and identifiability constraints.
We refer to $U_1,\ldots,U_D$ as the \textit{u-values}.
For simplicity, we abuse terminology slightly by referring to $P_\theta$ as the likelihood.

To perform model criticism, we view the hypothesized model as the null hypothesis.
To test for departures from this hypothesis, we sample from the posterior of $U$, that is, from the conditional distribution $U|Y$ that arises from the joint distribution of $(U,Y)$ defined by the assumptions that $(\theta,Y) = g(U)$ and $U\sim \mathrm{Uniform}_D(0,1)$.
Sampling from $U|Y$ can either be done by sampling $\theta|Y$ and then $U|\theta,Y$, or by directly targeting $U|Y$; either option can often be implemented by taking an algorithm for sampling from $\theta|Y$ and making minor modifications (see \cref{section:computing-uvalues}).
Now, the key observation is that if $Y$ is sampled from the hypothesized model and $U$ is sampled from $U|Y$, then $U\sim\mathrm{Uniform}_D(0,1)$ marginally, integrating out $Y$.
In other words, if the model is correct, then a single posterior sample of $U = (U_1,\ldots,U_D)$ is uniformly distributed, that is, $U_1,\ldots,U_D$ i.i.d.\ $\sim \mathrm{Uniform}(0,1)$.
There is no approximation here -- if the model is correct, then a posterior draw of $U$ is exactly uniform, in complete generality.
Consequently, if we detect that $U_1,\ldots,U_D$ are not i.i.d.\ $\mathrm{Uniform}(0,1)$ under the posterior, then this implies misspecification of some aspect of the model, that is, there is mismatch between the true distribution and some aspect of the prior $\Pi$ or likelihood $P_\theta$.
This enables one to perform model criticism in a simple yet powerful way by testing for departures from uniformity or independence in various respects; see \cref{section:choice-of-tests}.

For posterior inference, we typically draw multiple posterior samples, say, $U^{(1)},\ldots,U^{(T)}\in (0,1)^D$, where each $U^{(t)} = (U_1^{(t)},\ldots,U_D^{(t)})\in (0,1)^D$ is drawn from $U|Y$.
A subtle but crucial point is that, while $U_1^{(t)},\ldots,U_D^{(t)}$ are i.i.d.\ $\mathrm{Uniform}(0,1)$ for any given $t$, it does not hold that $U_d^{(t)}\sim \mathrm{Uniform}(0,1)$ i.i.d.\ for all $t$ and $d$.  This is because the dataset $Y$ creates dependence among the samples $U^{(1)},\ldots,U^{(T)}$.
If an independent dataset $Y^{(t)}$ were used to generate each $U^{(t)}$, then $U^{(1)},\ldots,U^{(T)}$ would indeed be independent, but this is not the case since we only observe one dataset $Y$.
In \cref{section:combining-samples}, we describe how to combine across samples.

\begin{example}{AR(1) model.}
\label{example:AR1}
Consider an autoregression model where 
$Y_1 = \sigma \varepsilon_1$ and $Y_i = \phi Y_{i-1} + \sigma \varepsilon_i$ for $i = 2,\ldots,n$, where $\varepsilon_1,\ldots,\varepsilon_n$ i.i.d.\ $\sim\mathcal{N}(0,1)$, with prior $\phi\sim \mathrm{Uniform}(-0.5,0.5)$  and $\sigma\sim \mathrm{Exponential}(1)$ independently.  Then we can write $(\phi,\sigma,Y_1,\ldots,Y_n) = g(U_1,\ldots,U_{2+n})$, where $U_1,\ldots,U_{2+n}$ i.i.d.\ $\sim\mathrm{Uniform}(0,1)$, by setting
$\phi = F_\phi^{-1}(U_1)$, $\sigma = F_{\sigma}^{-1}(U_2)$, $\varepsilon_i = \Phi^{-1}(U_{d_i})$, $Y_1 = \sigma \varepsilon_1$, and $Y_i = \phi Y_{i-1} + \sigma \varepsilon_i$ for $i = 2,\ldots,n$,
where $F_\phi$, $F_\sigma$, and $\Phi$ denote the cumulative distribution functions (CDFs) of the $\mathrm{Uniform}(-0.5,0.5)$, $\mathrm{Exponential}(1)$, and $\mathcal{N}(0,1)$ distributions, respectively, and $d_i = 2+i$.
\end{example}

\subsection{Choice of tests}
\label{section:choice-of-tests}

In terms of model criticism, one advantage of using a uniform parametrization is that it greatly simplifies the construction of tests.
In particular, since any subset of u-values is i.i.d.\ uniform under the null hypothesis that the model is correct, (i) the u-values can be grouped in various ways and (ii) the same tests can be applied to any model.
Some natural choices of test are as follows. 
In each case, the test is performed on the u-values $U_1,\ldots,U_D$ from a single posterior sample; see \cref{section:combining-samples} for combining across samples.
\begin{enumerate}
    \item \textit{Testing for extreme values.} Outliers or poor choices of prior can often be detected by looking at individual u-values.  For instance, in the AR(1) example, if $U_2$ is very close to 1 then this indicates that the inferred value of $\sigma$ is much larger than expected under the $\mathrm{Exponential}(1)$ prior. To test for extreme values, we use $2 \min\{U_d, 1-U_d\}$ as a p-value; note that this is $\mathrm{Uniform}(0,1)$ under the null.
    
    \item \textit{Testing for non-uniformity.} Misspecification of model distributions can often be detected by testing for non-uniformity of a relevant subset of u-values. In the AR(1) example, if the empirical distribution of $U_{d_1},\ldots,U_{d_n}$ is significantly non-uniform (where $d_i=2+i$), then this suggests possible misspecification of either the normal distribution or the first-order linear assumption encoded in the equation $Y_i = \phi Y_{i-1} + \sigma\varepsilon_i$. To test for non-uniformity, we use the Anderson--Darling test with $\mathrm{Uniform}(0,1)$ as the null hypothesis \citep{anderson1954test}, but any goodness-of-fit test could be used.

    \item \textit{Testing for internal dependence.}  Structural misspecification can sometimes be detected by testing for dependence between relevant subsets of u-values. For instance, in the AR(1) example, dependence between $U_{d_i}$ and $U_{d_i+2}$ (for $i = 1,\ldots,n-2$) suggests that higher-order dependence is present, rather than just the first-order dependence assumed in the AR(1) model. 
    To test for internal dependence between sets of u-values, we use Hoeffding's test~\citep{Hoeff48}.

    \item \textit{Testing for external dependence.}  Missing structural features can sometimes be detected by testing for dependence with  external variables such as covariates.
    In the AR(1) example, dependence between $U_{d_i}$ and the index $i$ (for $i=1,\ldots,n$) suggests an unmodeled dependence on time, such as heteroskedasticity or a trend.
    More generally, if $x_i$ is a covariate, then dependence between $U_{d_i}$ and $x_i$ suggests the presence of unmodeled dependence on the covariate.
    To test for dependence (i) between two continuous variables, we use Hoeffding's test~\citep{Hoeff48}; (ii) between continuous and binary variables, we use 
    the Mann--Whitney U test~\citep{MW47}; and (iii) between continuous and non-binary discrete variables, we use 
    the Kruskal--Wallis H test~\citep{KW52}.
\end{enumerate}

See \cref{section:examples} for detailed illustrations, and see \cref{sec:AR1} in particular for the AR(1) example.
Typically, one will want to perform a number of tests, so it is necessary to adjust for multiple testing, for instance, using the Bonferroni, Holm, or Benjamini--Hochberg procedures; see \cref{sec:multiple-testing}.

\subsection{Choice of uniform parametrization}
\label{section:parametrization}

An attractive feature of UPCs is the intuitive way in which a test can connect with the type of misspecification indicated, however, the choice of parametrization plays an important role in this connection.
There are many possible ways to uniformly parametrize a given model, since there is not a unique choice of function $g$ such that $(\theta,Y) = g(U)$.
To see why, note that there are many distribution-preserving operations on $U$, such as mapping $U_d$ to $1 - U_d$ as a simple example.
Usually, though, the model specification will suggest a natural way of defining $g$ by following the generative process that defines the model; see \cref{example:AR1} and more examples in \cref{section:examples}.
It is less clear how to choose $g$ when there are constraints on $\theta$ or $Y$.
For handling constraints, we generally recommend defining a u-value for each interpretable univariate quantity and then mapping these through a function that enforces the constraint, rather than defining u-values for only a subset of variables and then setting the remaining variables to satisfy the constraint.
For example, in regression models, it is common to have constraints of the form $\sum_{j=1}^J \alpha_j = 0$, for which we suggest defining $\tilde{\alpha}_j = F_j^{-1}(U_j)$ and $\alpha_j = \tilde{\alpha}_j - \frac{1}{J}\sum_{j=1}^J \tilde{\alpha}_j$, for appropriately chosen CDFs $F_j$.
While this leads to non-identifiability of these u-values, it improves their interpretability.

\subsection{Combining multiple posterior samples}
\label{section:combining-samples}

So far, we have described UPC tests based on a single posterior sample of $U = (U_1,\ldots,U_D)\in (0,1)^D$ given the dataset $Y$. 
Since there is randomness in any one posterior sample, it is preferable to aggregate across many posterior samples $U^{(1)},\ldots,U^{(T)}\in (0,1)^D$ drawn from the conditional distribution of $U|Y$.
However, care must be taken to combine these in a valid way, because $U^{(1)},\ldots,U^{(T)}$ are not marginally independent, integrating out $Y$.
Thus, one cannot simply pool all of the u-values together to perform tests.

To understand why, consider the posterior of the $\phi$ parameter in the AR(1) model in \cref{example:AR1}, given a random dataset $Y$ generated according to the hypothesized model for $(\theta,Y)$.
When the number of data points $n$ is large, the posterior of $\phi$ will tend to be concentrated.
Likewise, the posterior of the corresponding u-value $U_1$ will also be concentrated.  Hence, the posterior samples $U_1^{(1)},\ldots,U_1^{(T)}$ will tend to be clustered together, and thus, for any given dataset $Y$, 
their empirical distribution will clearly not be close to uniform.
This is not a contradiction: Each $U_1^{(t)}$ is marginally $\mathrm{Uniform}(0,1)$ when integrating out $Y$, but the dependence among $U_1^{(1)},\ldots,U_1^{(T)}$ induced by $Y$ 
makes them tend to take similar values. 
See \cref{fig:newcomb-upc,fig:bernoulli_upc} for illustrations of this effect in examples.

We use the following approach to combine posterior samples in a valid way for any given test.
For each $t = 1,\ldots,T$, we perform the test on the posterior sample $U^{(t)} = (U_1^{(t)},\ldots,U_D^{(t)})$ to obtain a p-value $p^{(t)}$. 
Under the null that the model is correct, we know $U_1^{(t)},\ldots,U_D^{(t)}$ i.i.d.\ $\sim \mathrm{Uniform}(0,1)$, 
so a valid test will produce a uniformly distributed p-value $p^{(t)} \sim \mathrm{Uniform}(0,1)$.
Then, we combine the p-values $p^{(1)},\ldots,p^{(T)}$ using the Cauchy combination method for dependent p-values \citep{liu2020cauchy}. Specifically, we compute 
\begin{align}
    \label{eq:cauchy_combination}
     p^* = 1 - F_\mathrm{Cauchy}\Big(\frac{1}{T}\sum_{t=1}^T \tan\big((0.5 - p^{(t)})\pi\big)\Big),
\end{align}
where $F_\mathrm{Cauchy}$ is the CDF of the Cauchy distribution.
We then compare the aggregated p-value $p^*$ to a pre-specified level $\alpha$ to decide whether to reject the null that the model is correct.
Although $p^*$ is not uniformly distributed on all of $(0,1)$ under the null, empirically we find that $\mathbb{P}(p^* \leq \alpha)\approx \alpha$ for $\alpha\in(0, 0.05)$. Thus, for practically relevant values of $\alpha$, the Type I error is near the target level.
Several methods for aggregating dependent p-values have been proposed---for instance, the method of \citet{GWR24} yields similar results on the examples we consider---but overall we find that the Cauchy combination method tends to exhibit the greatest power while still controlling Type I error rate.

\subsection{Performing multiple tests}
\label{sec:multiple-testing}

To perform multiple different UPC tests, we apply the p-value aggregation technique in \cref{section:combining-samples} for each test,
yielding aggregated p-values $p_1^*,\ldots,p_M^*$.
Standard multiple testing adjustment procedures can then be applied to $p_1^*,\ldots,p_M^*$.
For instance, one can control family-wise error rate (FWER) with the Bonferroni or Holm procedures \citep{holm1979simple}. 
One can control false discovery rate (FDR) with the procedure of \citet{benjamini1995controlling}
in the case of independent tests, which holds when the tests are computed from disjoint sets of u-values;
more generally, the procedure of \citet{benjamini2001control} can be used in the case of dependent tests.

Furthermore, UPCs can be used to iteratively criticize and improve the model \citep{Box80,blei2014build}
in a principled way that controls Type I error.
Specifically, one can use \emph{alpha spending}, in which the rejection thresholds $\alpha_1,\ldots,\alpha_M$ for a sequence of tests 
are pre-selected such that they satisfy $\alpha = \sum_{m=1}^M \alpha_m$, where $\alpha$ is the overall FWER that one wishes to permit.
We illustrate with a logistic regression model in \cref{sec:logistic}.
Similarly, \emph{alpha investing} can be used to control FDR for a sequence of tests \citep{foster2008alpha}.

\subsection{Interpretation of tests}
\label{sec:interpretation-of-tests}

When there is an explicit generative description of the model, defining $(\theta,Y) = g(U)$ accordingly provides a natural correspondence between the entries of $U$ and the entries of $\theta$ and $Y$, which aids in the interpretation of the u-values and tests. 
As described in \cref{section:choice-of-tests}, a rejection of the null under a test for extreme values, non-uniformity, internal dependence, or external dependence suggests 
a possible issue with the corresponding part of the model from which the tested u-values arise.
However, the interpretation is not always straightforward, since misspecification of one part of a model can lead to departures from i.i.d.\ uniformity of u-values in other parts of the model.

For instance, in the AR(1) example, suppose the true distribution of $\varepsilon_1,\ldots,\varepsilon_n$ is Cauchy rather than $\mathcal{N}(0,1)$.
Then a posterior sample of the u-values corresponding to $\varepsilon_1,\ldots,\varepsilon_n$ will exhibit non-uniformity, and outliers with u-values close to 0 or 1 will likely be observed.
However, $\sigma$ will also tend to be wildly overestimated, causing the $\sigma$ u-value to be very close to 1.
Thus, there is not always a direct link between departure from i.i.d.\ uniformity of u-values and misspecification of the corresponding part of the model. 
Nonetheless, empirically we find that the strongest departures from i.i.d.\ uniformity tend to correspond to the aspects of the model that are misspecified.

\subsection{Computation of u-values}
\label{section:computing-uvalues}

In this section, we assume the following condition on the definition of the function $g$.
\begin{condition}
\label{condition:decomposition}
$\theta = g_{\mathrm{p}}(U_{1:K})$ and $Y = g_{\mathrm{d}}(U_{K+1:D}; \theta)$ for some $K$ and some functions $g_\mathrm{p}$ and $g_\mathrm{d}$. 
\end{condition}
When this holds, we refer to $U_{1:K}$ as the \textit{parameter u-values} and $U_{K+1:D}$ as the \textit{data u-values}.
Sampling the parameter u-values from $U_{1:K}|Y$ can either be done directly with standard Bayesian techniques or by post-processing from samples of $\theta$; see \cref{sec:computation-of-parameter-uvalues} for details.  Sampling the data u-values from $U_{K+1:D}\mid U_{1:K},Y$ requires post-processing; see \cref{sec:computation-of-data-u-values}.
In the AR(1) model (\cref{example:AR1}), $U_1$ and $U_2$ are the parameter u-values and $U_3,\ldots,U_{2+n}$ are the data u-values.

\subsubsection{Computation of parameter u-values}
\label{sec:computation-of-parameter-uvalues}
One method of sampling $U_{1:K}|Y$ is simply to reparametrize -- that is, to view $U_{1:K} \sim \mathrm{Uniform}_K(0,1)$ as the prior and $Y | U_{1:K}$ as defining the likelihood.  Letting $\mathcal{L}(\theta; Y)$ denote the likelihood function of $P_\theta$ for data $Y$, the posterior is then $\pi(u_{1:K}\mid Y) \propto \mathcal{L}(g_{\mathrm{p}}(u_{1:K}); Y)$ since the prior is uniform.
Thus, when using algorithms based directly on the target density, such as Hamiltonian Monte Carlo as in Stan and PyMC \citep{carpenter2017stan,abril2023pymc} or Langevin algorithms \citep{roberts1996exponential}, it is trivial to modify an MCMC sampler targeting $\theta|Y$ to instead target $U_{1:K}|Y$.
For instance, in the AR(1) example, the posterior of the parameter u-values is $\pi(u_1,u_2 \mid Y) \propto \prod_{i=1}^n \mathcal{N}(Y_i\mid F_\phi^{-1}(u_1) Y_{i-1},\, F_\sigma^{-1}(u_2)^2)$, if we define $Y_0 = 0$.

Sometimes this reparametrization approach might be undesirable, for instance, if the model is not conducive to sampling based on evaluation of the posterior density or if one wishes to use an existing MCMC algorithm for sampling $\theta|Y$.
In such cases, one can first sample $\theta|Y$ and then sample $U_{1:K}|\theta,Y$ in a post-processing step.
Recall that $U_{1:K}\sim\mathrm{Uniform}_K(0,1)$ and, in this section, we assume $\theta = g_\mathrm{p}(U_{1:K})$.
The simplest situation is when $g_\mathrm{p}$ is an invertible function, in which case we can deterministically transform each posterior sample of $\theta$ into a sample of the parameter u-values via $U_{1:K} = g_\mathrm{p}^{-1}(\theta)$.
This is the case in the AR(1) model in \cref{example:AR1}, for which $U_1 = F_\phi(\phi)$ and $U_2 = F_\sigma(\sigma)$.

Often, however, $\theta = g_\mathrm{p}(U_{1:K})$ is not invertible, such as when there are discrete latent variables or identifiability constraints.
Then one can stochastically generate the parameter u-values from the conditional distribution $U_{1:K}|\theta,Y$.
This is straightforward when the non-invertibility is solely due to discreteness of one or more entries of $\theta$, since then $U_{1:K}|\theta,Y$ is uniformly distributed subject to the constraint that $\theta = g_\mathrm{p}(U_{1:K})$.
A common situation is that $\theta = (\theta_1,\ldots,\theta_K)$ and there are functions $g_1,\ldots,g_K$ such that $\theta_1 = g_1(U_1)$ and $\theta_k = g_k(\theta_1,\ldots,\theta_{k-1},U_k)$ for $k=2,\ldots,K$.
Then a sample of $U_{1:K}|\theta,Y$ can be obtained by sampling $U_k \mid \theta_{1:k}$ sequentially for $k=1,\ldots,K$.  For instance, if $\theta_k$ is discrete then we draw $U_k \sim \mathrm{Uniform}(\{u\in(0,1) : \theta_k = g_k(\theta_1,\ldots,\theta_{k-1},u)\})$;
this is equivalent to the randomized probability integral transform described by \citet{CGH09}.
We use this technique for the component assignment variables in the mixture model in \cref{sec:logistic}. 
More generally, when $g_\mathrm{p}$ is non-invertible due to more than just discreteness of latent variables, care must be taken to determine valid conditional distributions for $U_{1:K}|\theta,Y$; see  \citet{chang1997conditioning} for a general framework for conditioning.

\subsubsection{Computation of data u-values}
\label{sec:computation-of-data-u-values}

Computing the data u-values is very similar to computing the parameter u-values.
First, if $Y = g_\mathrm{d}(U_{K+1:D}; \theta)$ is invertible (as a function from $U_{K+1:D}$ to $Y$) for all $\theta$, then we can simply transform the data into the data u-values via $U_{K+1:D} = g_\mathrm{d}^{-1}(Y; \theta)$ for any given posterior sample of $\theta$.
Again, this is the case in the AR(1) model, where we have $\theta = (\phi,\sigma)$ and $U_{d_i} = \Phi\big((Y_i - \phi Y_{i-1})/\sigma\big)$ for $i=1,\ldots,n$, where $Y_0 = 0$ and $d_i = 2+i$.

On the other hand, if $g_\mathrm{d}$ is not invertible, then just like in the case of the parameter u-values, we sample from $U_{K+1:D}\mid \theta,Y,U_{1:K}$.
Similar to before, a common situation is that for some ordering of the univariate entries of the data, say, $Y = (Y_1,\ldots,Y_n)$, there are functions $g_{K+1},\ldots,g_{K+n}$ such that $Y_i = g_{K+i}(Y_1,\ldots,Y_{i-1},U_{K+i};\, \theta)$ for $i=1,\ldots,n$, where $K+n = D$.
Then we can sample from the joint distribution of the data u-values $U_{K+1:D}\mid\theta,Y,U_{1:K}$
by drawing $U_{K+i}\mid \theta,Y_{1:i}$ sequentially.
For instance, if $Y_i$ is a discrete random variable, then we draw
$U_{K+i}\sim \mathrm{Uniform}(\{u\in(0,1) : Y_i = g_{K+i}(Y_1,\ldots,Y_{i-1},u;\, \theta)\})$.

\begin{example}[Bernoulli model]
As a simple example, consider an i.i.d.\ Bernoulli model where $\theta$ is a discrete random variable such that $\mathbb{P}(\theta = 1/4) = \mathbb{P}(\theta = 3/4) = 0.5$, and $Y_1,\ldots,Y_n\mid\theta$ i.i.d.\ $\sim \mathrm{Bernoulli}(\theta)$.
We can write this model as $U_1,\ldots,U_{n+1}$ i.i.d.\ $\sim\mathrm{Uniform}(0,1)$, 
$\theta = g_1(U_1) = (1/4) + (1/2)\mathds{1}(U_1\geq 0.5)$, 
and $Y_i = g_{i+1}(U_{i+1}; \theta) = \mathds{1}(U_{i+1}\geq 1-\theta)$ for $i=1,\ldots,n$.
Then, given data $Y = (Y_1,\ldots,Y_n)$ and a posterior sample of $\theta$, we can sample the parameter u-value $U_1|\theta,Y$ by drawing $U_1 \sim \mathrm{Uniform}(\{u\in(0,1) : g_1(u) = \theta\})$, that is, $U_1\sim\mathrm{Uniform}(0,0.5)$ if $\theta = 1/4$ and $U_1\sim\mathrm{Uniform}(0.5,1)$ if $\theta = 3/4$.
Likewise, we can sample the data u-values $U_{2:n+1}|\theta,Y,U_1$  by drawing $U_{i+1} \sim \mathrm{Uniform}(\{u\in(0,1) : g_{i+1}(u; \theta) = Y_i\})$ independently for $i = 1,\ldots,n$, that is, $U_{i+1}\sim \mathrm{Uniform}(0,\,1-\theta)$ if $Y_i=0$ and $U_{i+1}\sim \mathrm{Uniform}(1-\theta,\,1)$ if $Y_i = 1$.
\end{example}

\section{Previous work}
\label{section:previous-work}

The observation that each posterior draw is marginally distributed according the prior is a simple consequence of the definition of the joint distribution of the parameters and the data.
This fact has previously been used to assess the validity of posterior sampling algorithms using simulation-based calibration (SBC) \citep{Geweke2004,CGR06,talts2018validating,MMK+23}.
Specifically, these authors propose to generate simulated datasets from the hypothesized model, perform posterior inference for each simulated dataset,
and compare the resulting approximate posteriors to the prior.
Since, in this case, the datasets are known to be drawn from the model, any discrepancy between the posteriors and the prior is attributable to 
errors in the posterior approximation algorithm.
This is fundamentally different from our proposed method, since (i) they are not performing model criticism, 
(ii) they simulate datasets from the hypothesized model, and 
(iii) they run the posterior inference algorithm many times, once for each dataset.
\citet{talts2018validating} refer to the marginal distribution of $\theta$, integrating out the data distribution, as the ``data-averaged posterior''.

There is an extensive literature on Bayesian model criticism, also referred to as model checking.
The dominant approach is based on posterior predictive checks (PPCs), first introduced by \citet{guttman1967use}.
The modern formulation of PPCs was developed by \citet{rubin1984bayesianly}, generalized by \citet{meng1994posterior} to include test quantities that are functions of the data and parameters, and further developed by \citet{gelman1996posterior} on more complex models.
Issues with the lack of uniformity of PPC p-values were demonstrated by \citet{bayarri1999quantifying,bayarri2000p} and \citet{robins2000asymptotic}, who considered techniques for obtaining asymptotically valid PPC p-values using partial posterior or conditional predictive approaches.
Valid PPC p-values can also be obtained by splitting the data, as shown by \citet{moran2019population} and \citet{li2022calibrated}.
While we require p-values to be uniform under the null by definition, there is not universal agreement on this definition; see \citet{Gelman23} for other perspectives.
The PPC approach has several limitations that are resolved by UPCs, as discussed in the introduction.

To our knowledge, the nearest precedent to our proposed method appears in Section 6.4 of \citet{gelman2013bayesian} on ``Graphical Posterior Predictive Checks'', 
in which parameters of a certain hierarchical model were given $\mathrm{Beta}(2,2)$ priors, and \citet{gelman2013bayesian}
visually compare a posterior sample of these parameters to the $\mathrm{Beta}(2,2)$ prior, noting that the prior clearly does not match the posterior samples.
This example was the initial seed of the idea which eventually led to our proposed methodology.
\citet{johnson2007bayesian} and \citet{yuan2012goodness} develop a related method based on the use of pivotal discrepancy measures (PDMs), which are test quantities $T(Y,\theta)$ whose distribution is invariant to the value of $\theta$ when $Y$ is distributed according to the hypothesized model with parameter $\theta$; also see \citet{Gosselin11} and \citet{Zhang14}, who consider sampled posterior p-values (SPPs) based on a particular class of PDMs.
PDMs are used for model criticism by comparing this invariant distribution to the distribution of $T(Y^0,\tilde{\theta})$, where $Y^0$ is the observed data and $\tilde{\theta}$ is drawn from the posterior given $Y^0$.
Like UPCs, PDMs only require posterior sampling given the observed data and the null distribution is known exactly, but similar to PPCs, the burden is on the analyst to design PDMs that simultaneously have the required invariance property and are useful for detecting misspecification in a given model.
For continuous data, each data u-value can be viewed as a PDM of the form $T(y,\theta) = \mathbb{P}(Y_i \leq y_i \mid \theta)$. 
However, not all UPCs are PDMs, and not all PDMs are UPCs.

\citet{johnson2004bayesian} proposes another related idea, using what we refer to as the data u-values to construct a goodness-of-fit test based on a chi-squared test statistic; however, \citet{johnson2004bayesian} does not consider parameter u-values or any other types of test, and in general, Johnson's test statistic needs to be calibrated by sampling from the posterior predictive and then from the resulting posteriors given these simulated datasets. Furthermore, the associated theory is restricted to the asymptotic setting and requires regularity conditions.
Many other model checking methods have been proposed as well, for instance, based on 
simulation-based approaches \citep{dey1998simulation,hjort2006post},
cross-validation \citep{gelfand1992model,marshall2003approximate}, and assessing within-model conflict \citep{ohagan2003hsss,dahl2007robust},
but these tend to be either computationally intensive, model-specific, or do not provide well-calibrated tests.

Model criticism methods make it possible to refine a model by identifying and correcting its inadequacies.
\citet{Box80} proposed to iteratively perform model criticism and improvement, in a process dubbed ``Box's loop'' by \citet{blei2014build}.  
\citet{Box80} employed what later became known as prior predictive checks \citep{meng1994posterior}, but the same iterative refinement process can be implemented with other model checks such as PPCs \citep{belin1995analysis} or with our proposed UPC method, as we demonstrate in \cref{sec:logistic}.
Since UPCs control Type I error rate, they provide a theoretically well-founded method for implementing Box's loop while controlling the overall error rate (\cref{sec:multiple-testing}).
Interestingly, \citet{Box80} and \citet{gelfand1992model} argue in favor of using predictive distributions for model criticism rather than the posterior, since the posterior alone cannot reveal any lack of fit. While this may be true for the usual posterior on parameters, the posterior on u-values provides additional context since (i) the u-values are on a model-based scale that captures information about goodness-of-fit, and (ii) the data u-values also provide more information than the posterior alone.

\section{Theory}
\label{sec:theory}

Let $\Theta\subseteq\R^L$ and $\mathcal{Y}\subseteq\R^n$ be measurable subsets; we use the Borel sigma-algebra on all topological spaces, unless otherwise specified. 
Let $g:(0,1)^D \to \Theta\times\mathcal{Y}$ be a measurable function, and 
define $(\bm{\theta},Y) = g(U)$
where $U\sim\mathrm{Uniform}_D(0,1)$, that is, $U = (U_1,\ldots,U_D)$
and $U_1,\ldots,U_D$ i.i.d.\ $\sim\mathrm{Uniform}(0,1)$.
Let $\Pi$ denote the resulting distribution of $\bm{\theta}$ and let $P_\theta$ denote the conditional distribution of $Y\mid \bm{\theta}=\theta$.
In this section, we use boldface $\bm{\theta}$ to denote the random vector and $\theta$ for particular values.  All random elements are assumed to be defined on a common probability space $(\Omega,\mathcal{A},\mathbb{P})$.

The interpretation is that $\Pi$ and $(P_\theta : \theta\in\Theta)$ represent the analyst's prior and likelihood, and $(\bm{\theta},Y)$ is jointly distributed according to this hypothesized model for the parameter $\theta$ and dataset $Y$.
Write $\Pi_{\theta|y}$ to denote the  posterior resulting from dataset $y$, that is, the conditional distribution of $\bm{\theta}\mid Y = y$, 
and let $\Pi_{u|\theta,y}$ denote the conditional distribution of $U\mid \bm{\theta}=\theta, Y = y$.
The conditional distributions for $Y|\bm{\theta}$, $\bm{\theta}|Y$, and $U|\bm{\theta},Y$ are guaranteed to exist almost everywhere due to the existence of regular conditional distributions in standard Borel spaces \citep{durrett2019probability}.
We assume, further, that these conditional distributions exist for all values of $\theta\in\Theta$ and $y\in\mathcal{Y}$, and that the maps $\theta \mapsto P_\theta(A)$, $y\mapsto \Pi_{\theta|y}(B)$, and $(\theta,y)\mapsto \Pi_{u|\theta,y}(E)$ are measurable for all measurable $A\subseteq \mathcal{Y}$, $B\subseteq\Theta$, and $E\subset (0,1)^D$; these are very mild conditions in practice.

Now, we envision that the true distribution, which we refer to as \emph{nature's model}, arises from some probability measures $\Pi^0$ and $(P_{\theta}^0 : \theta\in\Theta)$ on $\Theta$ and $\mathcal{Y}$, respectively.
Let $\bm{\theta}^0\sim\Pi^0$ and $Y^0 \sim P_{\theta}^0$ given $\bm{\theta}^0 = \theta$, 
so that $(\bm{\theta}^0, Y^0)$ is jointly distributed according to nature's model; here, it is assumed that $\theta\mapsto P_{\theta}^0(A)$ is measurable for all measurable $A\subseteq \mathcal{Y}$.
In practice, the observed dataset is generated from nature's model as $Y^0$ and one uses the hypothesized model to perform posterior inference using $\Pi_{\theta|y}$, setting $y$ equal to $Y_0$. 
To represent this, we introduce a new random vector $\widetilde{\bm{\theta}}$ with conditional distribution $\Pi_{\theta|y}$ given $Y^0 = y$.
Likewise, we introduce $\widetilde{U}$ with conditional distribution $\Pi_{u|\theta,y}$ given $\widetilde{\bm{\theta}}=\theta$ and $Y^0 = y$, where $\Pi_{u|\theta,y}$ is the conditional distribution under the hypothesized model.
Then, we have that $(\widetilde{U},\widetilde{\bm{\theta}},Y^0)$ represents the true joint distribution of the observed dataset $Y^0$ and posterior draws of parameters $\widetilde{\bm{\theta}}$ and u-values $\widetilde{U}$ under the hypothesized model given the observed dataset.

\subsection{Uniformity and independence of the u-values}
This section contains the properties justifying the UPC method.
Our first result, \cref{theorem:self-consistency}, provides the primary basis for the UPC method.
All proofs are provided in \cref{subsection:appendix:theory}.

\begin{theorem}
\label{theorem:self-consistency}
    Suppose $Y^0 \stackrel{d}{=} Y$. Then $\widetilde{\bm{\theta}} \stackrel{d}{=} \bm{\theta}$ and $\widetilde{U} \stackrel{d}{=} U$, that is, $\widetilde{\bm{\theta}}\sim\Pi$
and $\widetilde{U}\sim\mathrm{Uniform}_D(0,1)$.
\end{theorem}

Here, $\stackrel{d}{=}$ denotes equality in distribution. 
In other words, \cref{theorem:self-consistency} says that if the marginal distribution of the dataset is the same under nature's model and the hypothesized model,
then posterior draws of the parameter and u-values are distributed according to their respective priors,
integrating out the dataset.
In particular, if the model is correct then a draw from the posterior of the u-values is i.i.d.\ uniform.
Thus, if we can reject the hypothesis that the u-values are i.i.d.\ uniform, then this implies the model is incorrect.
Our next result shows that when the function $g:(0,1)^D\to\Theta\times\mathcal{Y}$ is bijective, the converse also holds.

\begin{theorem}
    \label{theorem:partial-converse}
    If $\widetilde{U} \stackrel{d}{=} U$ and $g$ is a bijection, then $Y^0 \stackrel{d}{=} Y$.
\end{theorem}

Therefore, when $g$ is bijective, $\widetilde{U}\stackrel{d}{=} U$ is a necessary and sufficient condition for the hypothesized model to be correct, at least in the sense that it matches nature's model in terms of the marginal distribution of the dataset. Thus, when performing UPCs in such cases, if our tests fail to reject the null hypothesis that $\widetilde{U}\sim\mathrm{Uniform}_D(0,1)$ then---although we can never ``accept'' the null---this provides an indication that the model appears to be reasonable.
The next result justifies our approach to testing for dependence with external variables $X$ such as covariates.
Although covariates are usually treated as fixed non-random quantities, here we treat them as random in order to have a formal notion of independence.

\begin{theorem}
\label{theorem:external}
If $X$ is a random element such that $X\indep Y^0$ and $(\widetilde{U},\widetilde{\bm{\theta}})\indep X\mid Y^0$, then $(\widetilde{U},\widetilde{\bm{\theta}} )\indep X$.
\end{theorem}

The interpretation of \cref{theorem:external} is that if (i) nature's model is independent of $X$, that is, $X\indep Y^0$, and (ii) the hypothesized model does not use $X$ when computing the posterior, that is, $(\widetilde{U},\widetilde{\bm{\theta}})\indep X\mid Y^0$, then the u-values $\widetilde{U}$ and parameter $\widetilde{\bm{\theta}}$ are independent of $X$.
Roughly speaking, under the null hypothesis that $X$ is irrelevant, the u-values are independent of $X$.
Consequently, if we observe dependence between $X$ and the u-values, then this suggests that $X$ or some other related variable may need to be added to the model.

\subsection{Identifiability of the u-values}

The u-values $U$ are not necessarily identifiable, even if the parameter vector $\theta$ is identifiable.
For instance, if there are discrete variables in the model, then it is clear that the u-values will not be uniquely determined.
Another common situation is when \cref{condition:decomposition} holds and the vector of parameter u-values $U_{1:K}$ is in a higher dimensional space than the parameter vector $\theta\in\mathbb{R}^L$ (that is, $K > L$), in which case the u-values usually will not be uniquely determined; in practice this may occur when defining $\theta$ values that satisfy constraints.
Likewise, the data u-values are not uniquely determined under similar circumstances.

However, if $\theta$ is identifiable and the parameter u-values $U_{1:K}$ are uniquely determined by $\theta$ (which is typically the case when all the entries of $\theta$ are continuous and there are no constraints on them), then $U_{1:K}$ is identifiable.
If, further, the data u-values are uniquely determined by $\theta$ and the dataset $Y$, then they are identifiable as well for any given dataset.
The following theorem formally states these results.

\begin{theorem}
    \label{theorem:identifiability}
    Assume $\theta$ is identifiable, that is, if $\theta\neq\theta'$ then $P_\theta\neq P_{\theta'}$.
    (i) If $\bm{\theta} = g_{\mathrm{p}}(U_{1:K})$ for some one-to-one function $g_\mathrm{p}$ and some $K$, then $U_{1:K}$ is identifiable, in the sense that these u-values are uniquely determined by $P_\theta$.
    (ii) If $(\bm{\theta},Y) = g(U)$ for some one-to-one function $g$, then all the u-values $U = U_{1:D}$ are uniquely determined by $P_\theta$ and $Y$.
\end{theorem}

\section{Examples}
\label{section:examples}

We demonstrate the UPC methodology on examples involving 
a univariate normal model (\cref{sec:newcomb}), Bernoulli trials (\cref{sec:bernoulli}), logistic regression (\cref{sec:logistic}), and an autoregression model (\cref{sec:AR1}).

\subsection{Normal model for Newcomb's speed of light data}
\label{sec:newcomb}

\begin{figure}
\centering

\includegraphics[width=0.3 \textwidth]{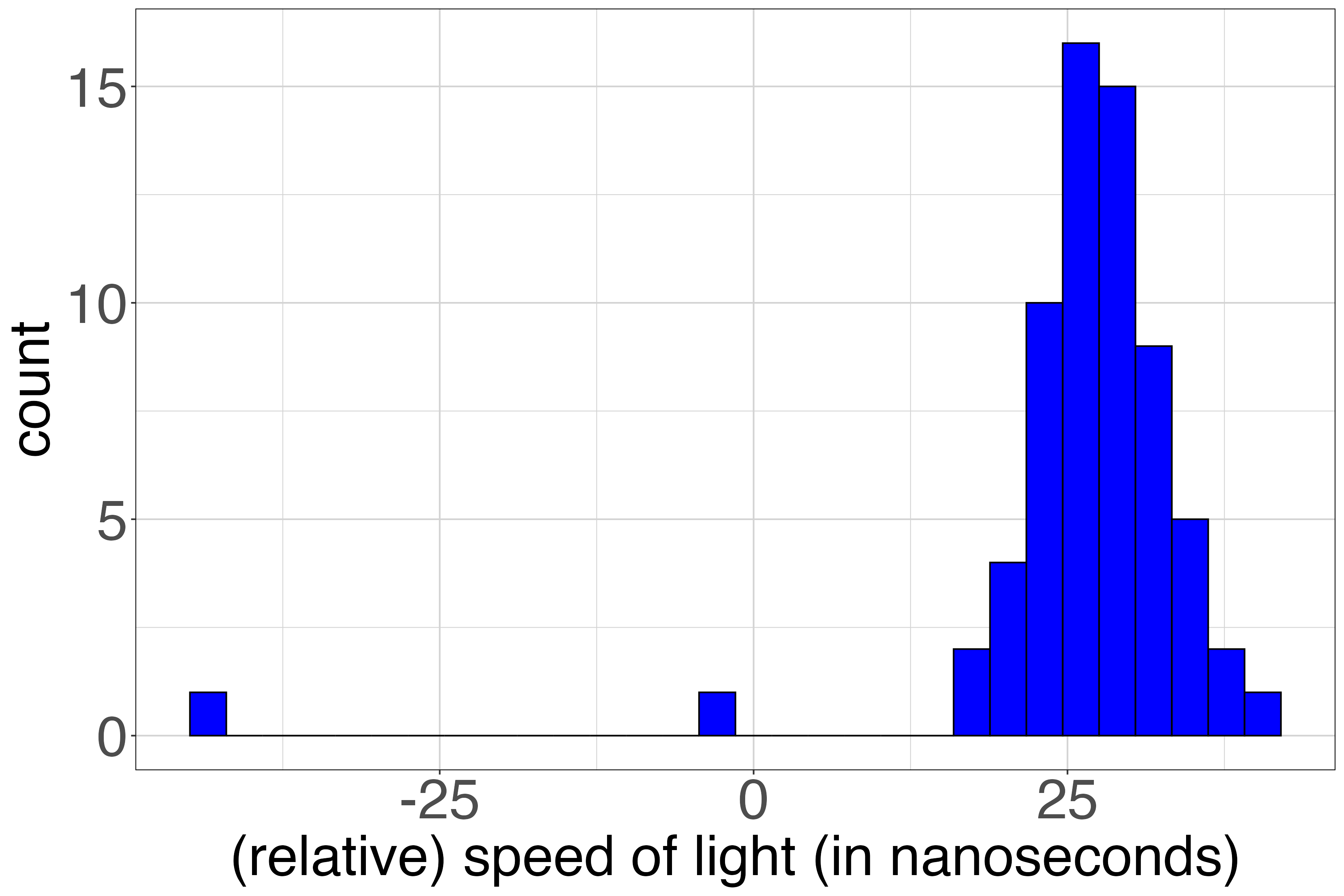}
\hfill 
\includegraphics[width=0.3 \textwidth]{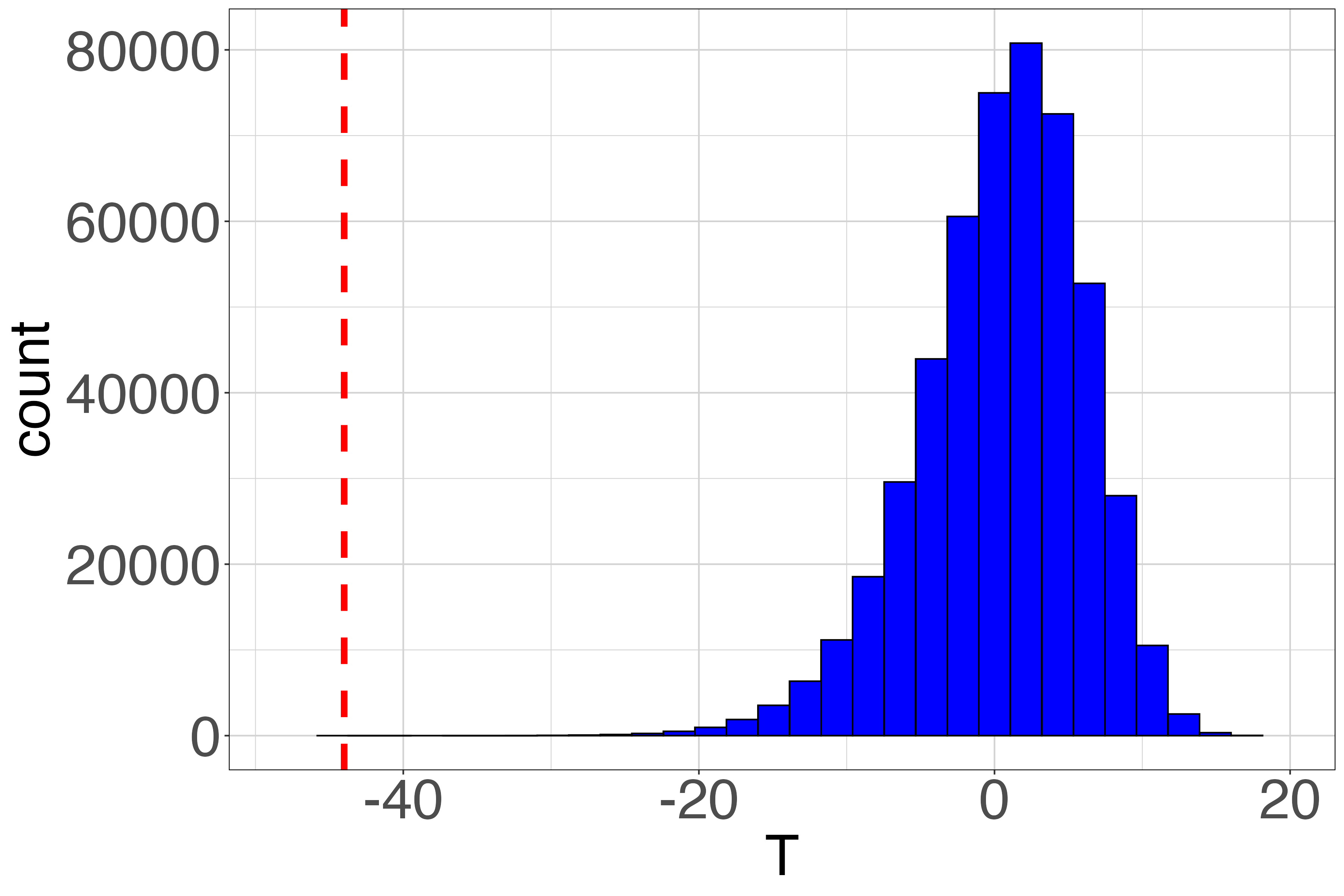}
\hfill
\includegraphics[width=0.3 \textwidth]{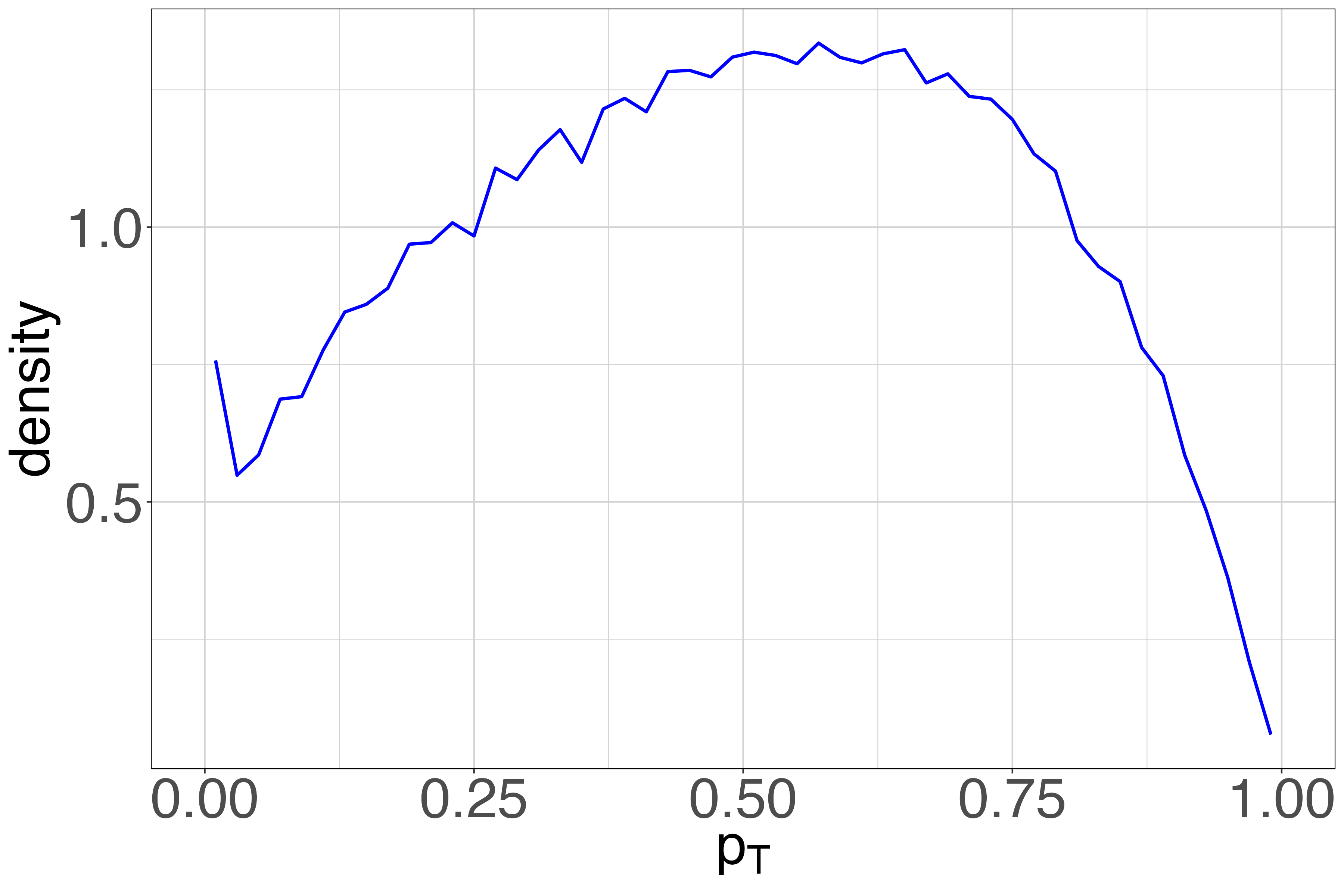}

\caption{Newcomb data and results using PPCs. (a) Histogram of Newcomb's speed of light data, relative to $24{,}800$ nanoseconds (ns). (b) Histogram of the distribution of $T = \min\{Y_1,\ldots,Y_n\}$ under the posterior predictive, along with the observed value of $T$ on the Newcomb data (vertical red dashed line).  (c) Density of the PPC p-values when sampling from the hypothesized model.}
\label{fig:newcomb-ppc}
\end{figure}

We begin by considering Simon Newcomb's $66$ measurements of the speed of light from 1882, a standard example for model criticism methods \citep{gelman2013bayesian}.
As seen in \cref{fig:newcomb-ppc}(a), Newcomb's data look plausibly normal except for two outlying values in the left tail of the distribution. 
Following \citet[Section 6.3]{gelman2013bayesian}, we consider modeling the data as $Y_1,\ldots,Y_n$ i.i.d.\ $\sim \mathcal{N}(\mu,\sigma^2)$, where $n = 66$. We use a weakly informative Normal-InverseGamma prior: $\mu\mid\sigma^2\sim\mathcal{N}(\mu_0, \sigma^2/\kappa_0)$ and $\sigma^2\sim\mathrm{InvGamma}(\alpha_0,\beta_0)$, with $\mu_0 
= 0$, $\kappa_0 = 1/10$, $\alpha_0 = 2$, and $\beta_0 = 300$.  
We draw $500{,}000$ samples from the posterior distribution, which is also a Normal-InverseGamma distribution by conjugacy.

\citet{gelman2013bayesian} suggest using $T = \min\{Y_1,\ldots,Y_n\}$ as a posterior predictive check (PPC) test statistic.
As shown in \cref{fig:newcomb-ppc}(b), the observed value of $T$ is not representative of the posterior predictive distribution, 
successfully detecting that there is an issue with the model.
However, as we can see from \cref{fig:newcomb-ppc}(c), the distribution of PPC p-values is not uniform under the null that the model is correct,
so this is not a well-calibrated test.
A deeper issue with this PPC is that this choice of $T$ was apparently made by looking at the data.
Without looking at the data, it would be difficult to know whether use the maximum instead of the minimum, or perhaps the interquartile range, or some other statistic.

To implement the UPC approach, we write $\sigma^2 = F_{\sigma^2}^{-1}(U_2)$, $\mu = \mu_0 + \sigma \kappa_0^{-1/2} \Phi^{-1}(U_1)$, and $Y_i = \mu + \sigma \Phi^{-1}(U_{d_i})$ where $F_{\sigma^2}$ is the CDF of $\sigma^2$ (that is, the $\mathrm{InvGamma}(\alpha_0,\beta_0)$ CDF), $\Phi$ is the standard normal CDF, and $d_i = 2+i$. For each posterior sample of $(\mu,\sigma^2)\mid Y_{1:n}$, we compute the u-values by inverting these functions, that is, $\widetilde{U}_1 = \Phi((\mu - \mu_0) / (\sigma\kappa_0^{-1/2}))$, $\widetilde{U}_2 = F_{\sigma^2}(\sigma^2)$, and $\widetilde{U}_{d_i} = \Phi((Y_i - \mu)/\sigma)$.

As described in \cref{section:choice-of-tests}, we test for extreme values of $\mu$ and $\sigma$
by computing p-values $p_\mu = 2\min\{\widetilde{U}_1, 1-\widetilde{U}_1\}$ and $p_\sigma = 2\min\{\widetilde{U}_2, 1-\widetilde{U}_2\}$,
and we test for non-uniformity of the data u-values $\widetilde{U}_{d_1},\ldots,\widetilde{U}_{d_n}$ by performing an Anderson--Darling test to obtain a p-value $p_\mathrm{data,unif}$.
Aggregating across posterior samples using the Cauchy combination method (\cref{section:combining-samples}), we obtain 
$p_\mu^* = 0.45$, 
$p_\sigma^* = 0.83$, and
$p_\mathrm{data,unif}^* = 1.60 \times 10^{-4}$.
This suggests no issues with the priors, but indicates that the normal model may be misspecified.

\begin{figure}
\centering

\includegraphics[width=0.3 \textwidth]{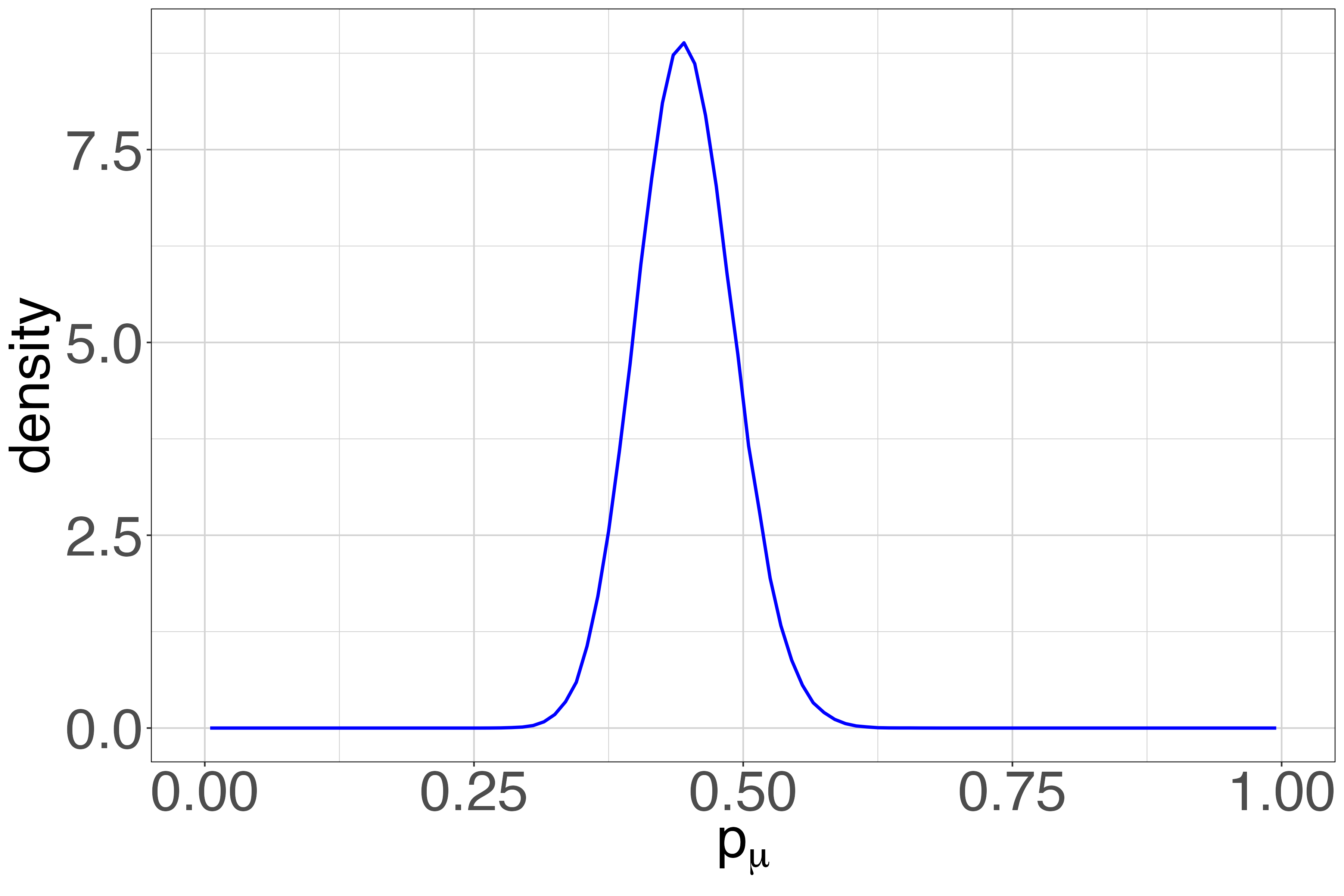}
\hfill 
\includegraphics[width=0.3 \textwidth]{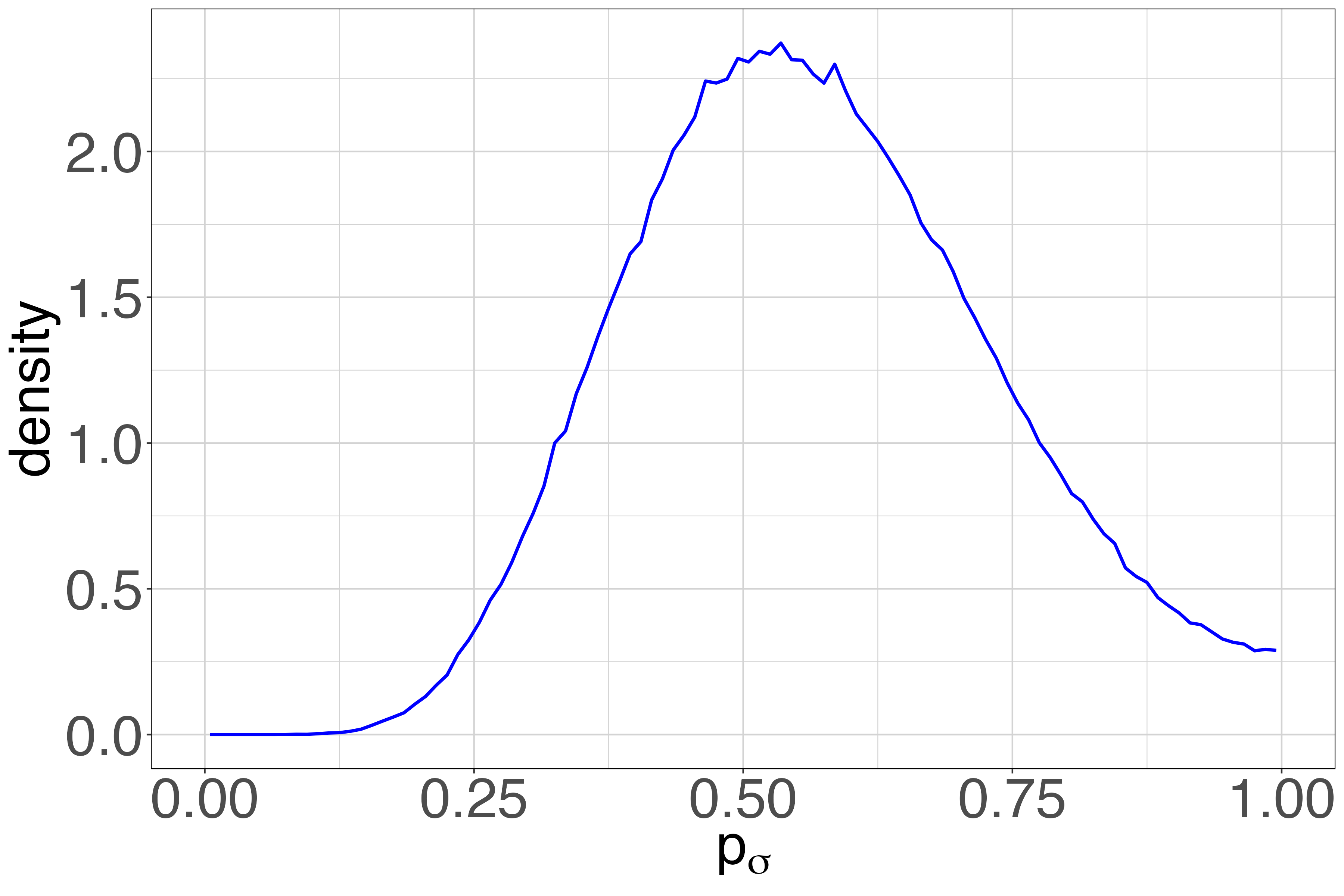}
\hfill
\includegraphics[width=0.3 \textwidth]{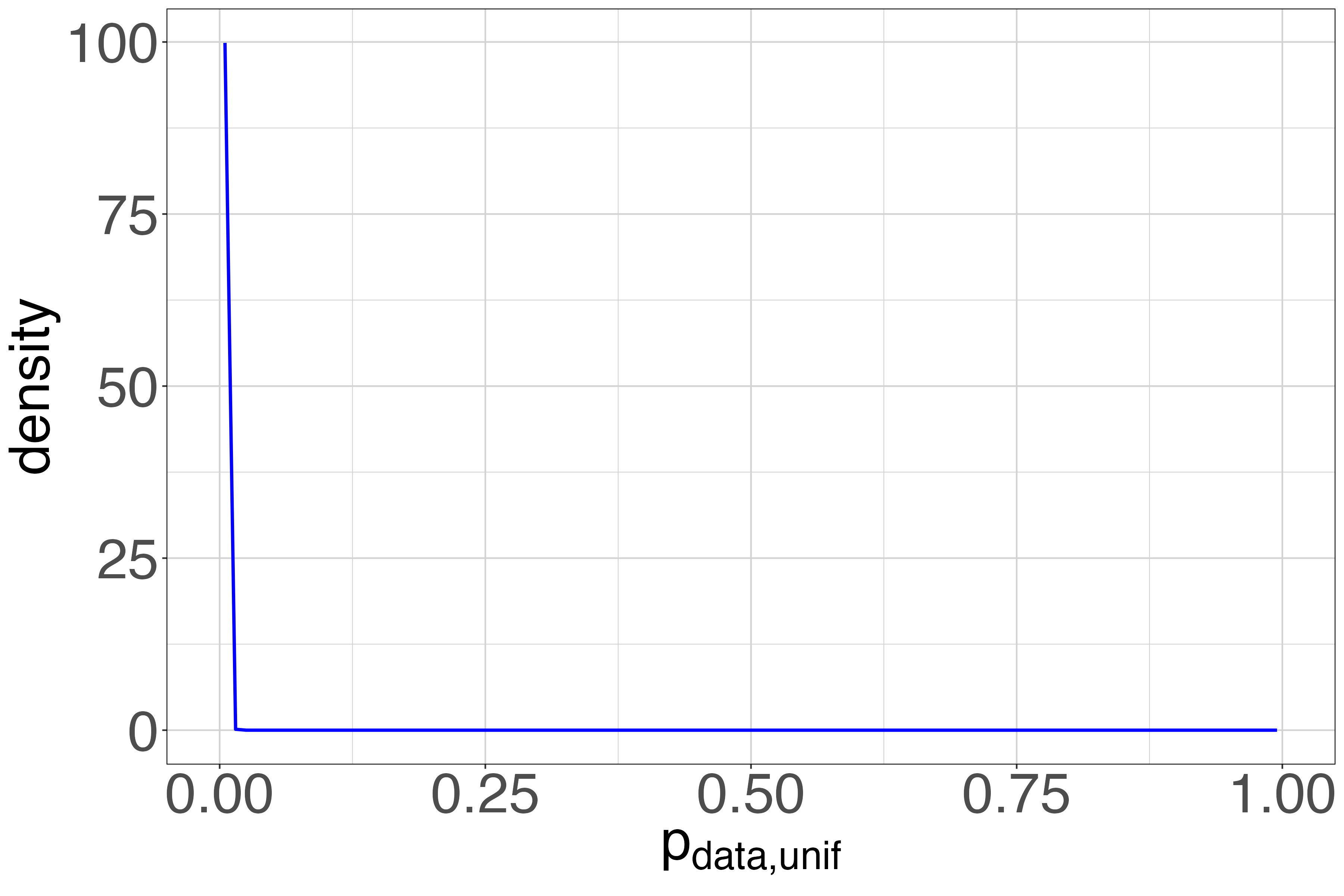}

\includegraphics[width=0.3 \textwidth]{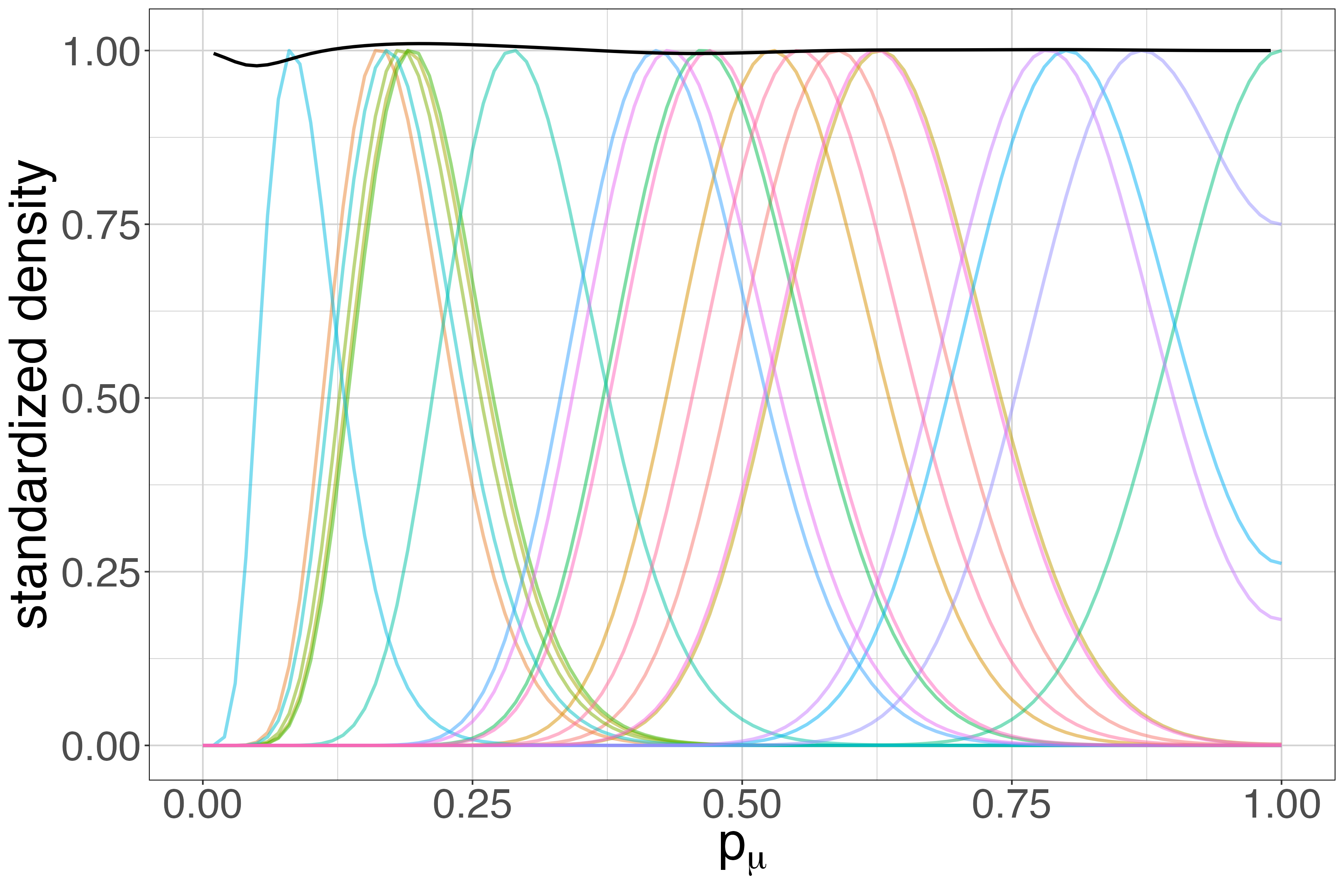}
\hfill 
\includegraphics[width=0.3 \textwidth]{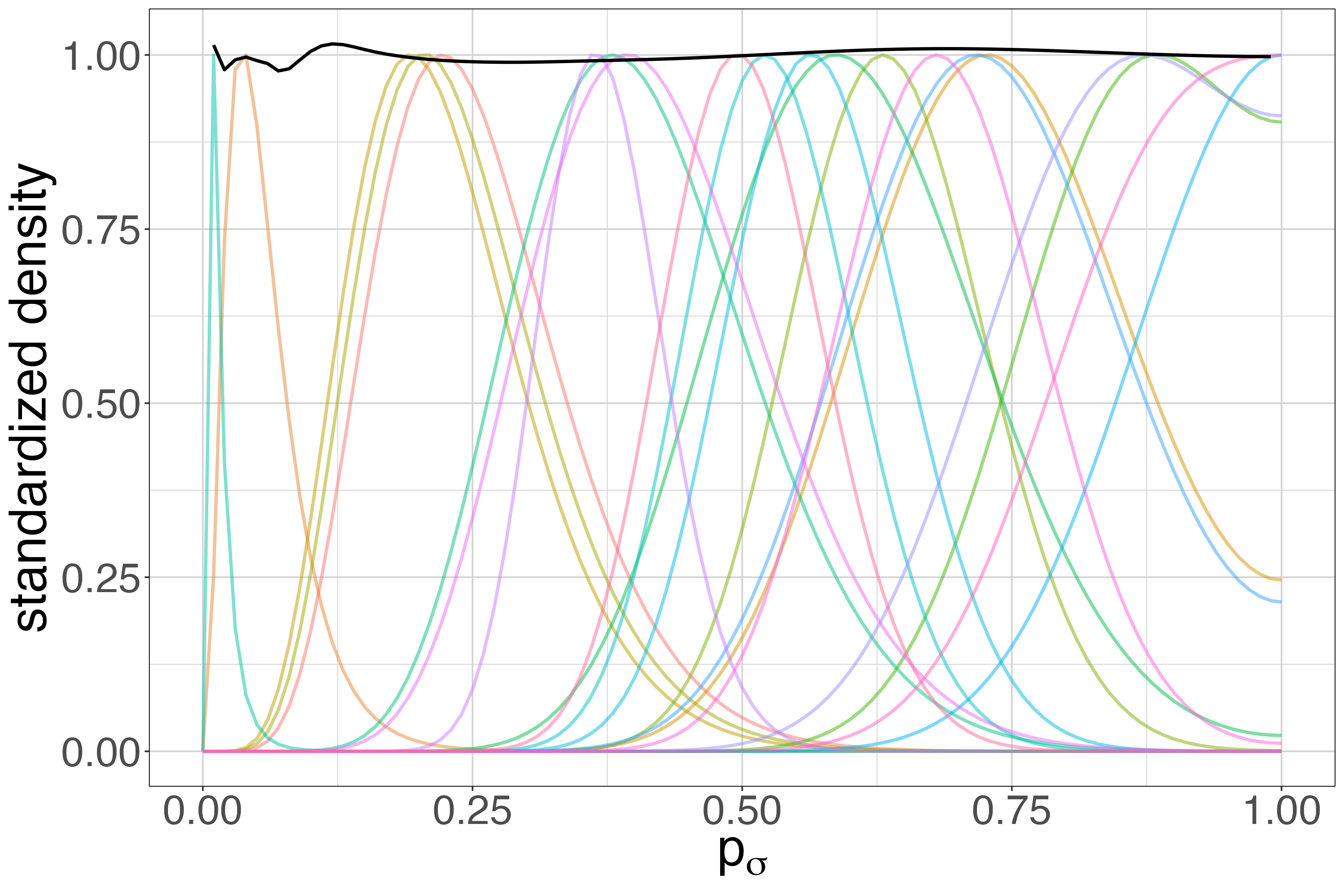}
\hfill
\includegraphics[width=0.3 \textwidth]{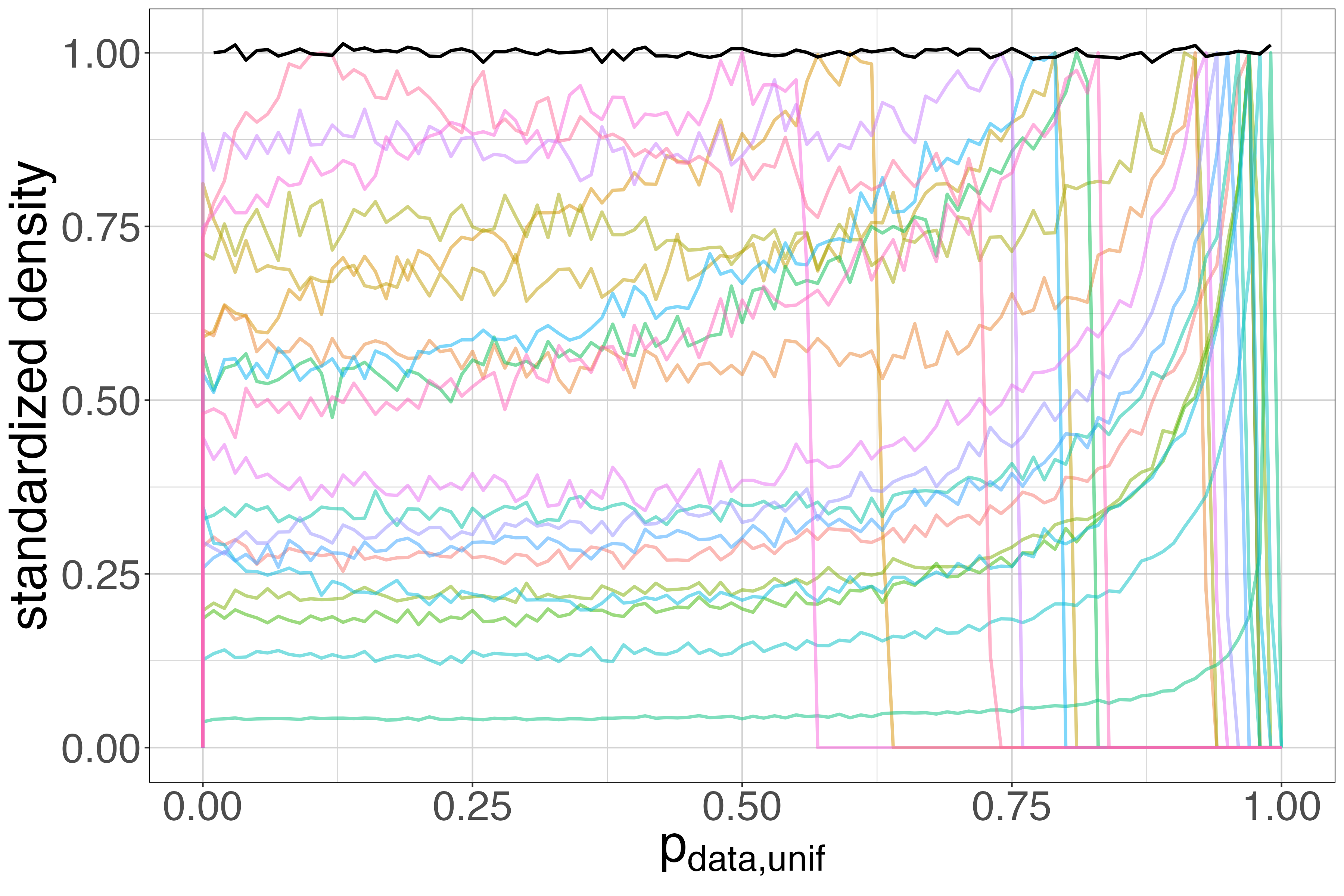}

\caption{
Results using UPCs on the Newcomb data.  (Top) Posterior densities of $p_\mu$, $p_\sigma$, and $p_\mathrm{data,unif}$ given the Newcomb data.
(Bottom) Samples of the posterior densities of $p_\mu$, $p_\sigma$, and $p_\mathrm{data,unif}$ given simulated datasets from the hypothesized model. To aid visualization, each density is standardized to have a maximum of $1$, so that they are all visible in a single plot.
The black lines show the average of these (unstandardized) densities over 100{,}000 simulated datasets.
}
\label{fig:newcomb-upc}
\end{figure}

To visualize what is happening, \cref{fig:newcomb-upc} (top) shows the posterior distribution of each of these p-values given the observed data.
\cref{fig:newcomb-upc} (bottom) shows several simulated null-distributed posteriors -- specifically, each curve is produced by sampling $\mu,\sigma^2$ and $Y_1,\ldots,Y_n\mid\mu,\sigma^2$ from the hypothesized model, and computing the resulting posterior.
The solid black lines in \cref{fig:newcomb-upc} (bottom) show the average of the null-distributed posteriors over $100{,}000$ simulated datasets,
which we know from \cref{theorem:self-consistency} is uniform in expectation.
We see that the observed posteriors for $p_\mu$ and $p_\sigma$ are representative of the null-distributed posteriors, 
but the observed posterior for $p_\mathrm{data,unif}$ is much more concentrated near zero,
which explains the small value of $p_\mathrm{data,unif}^*$.

\begin{figure}
\centering

\includegraphics[width=0.3 \textwidth]{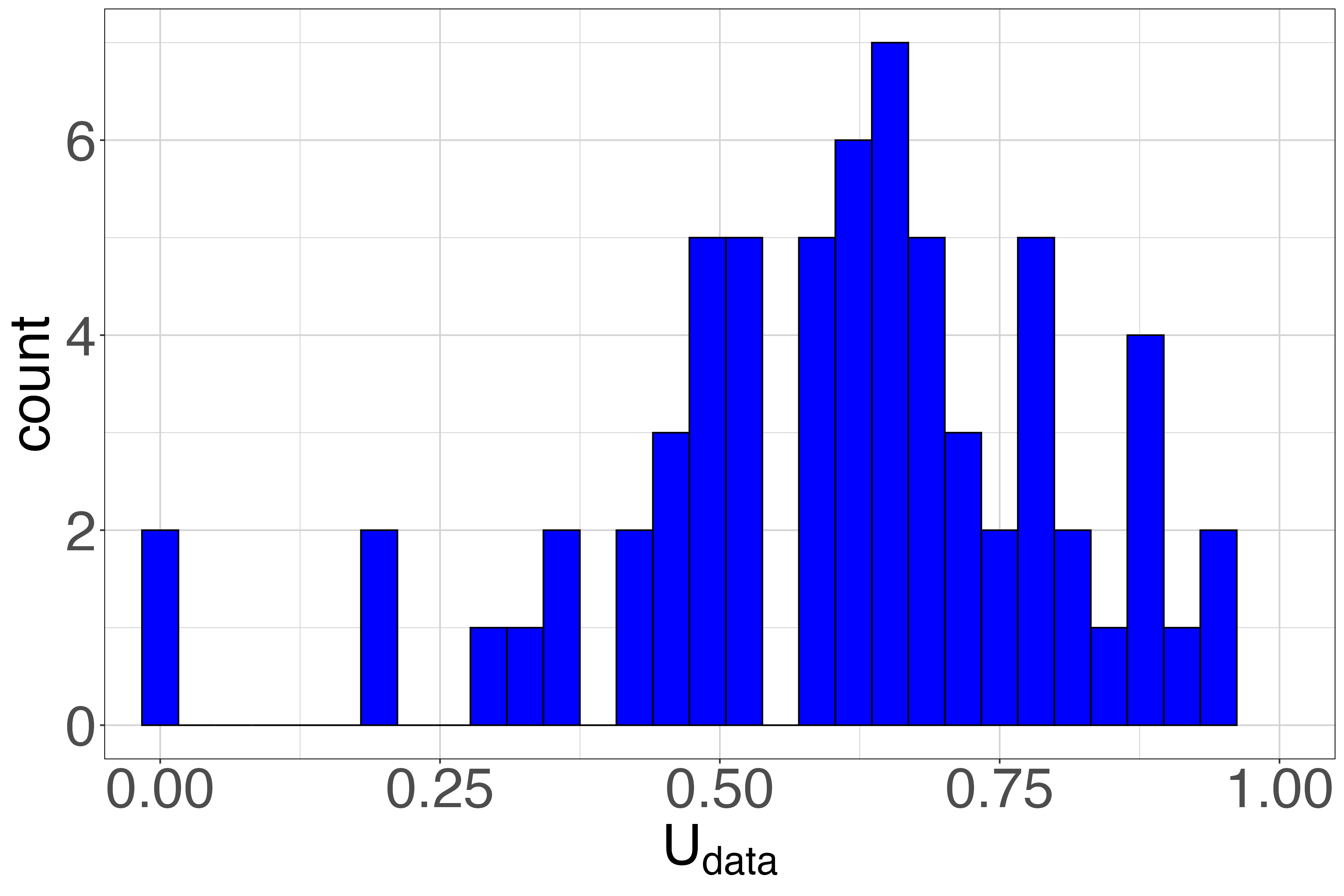}
\hfill
\includegraphics[width=0.3 \textwidth]{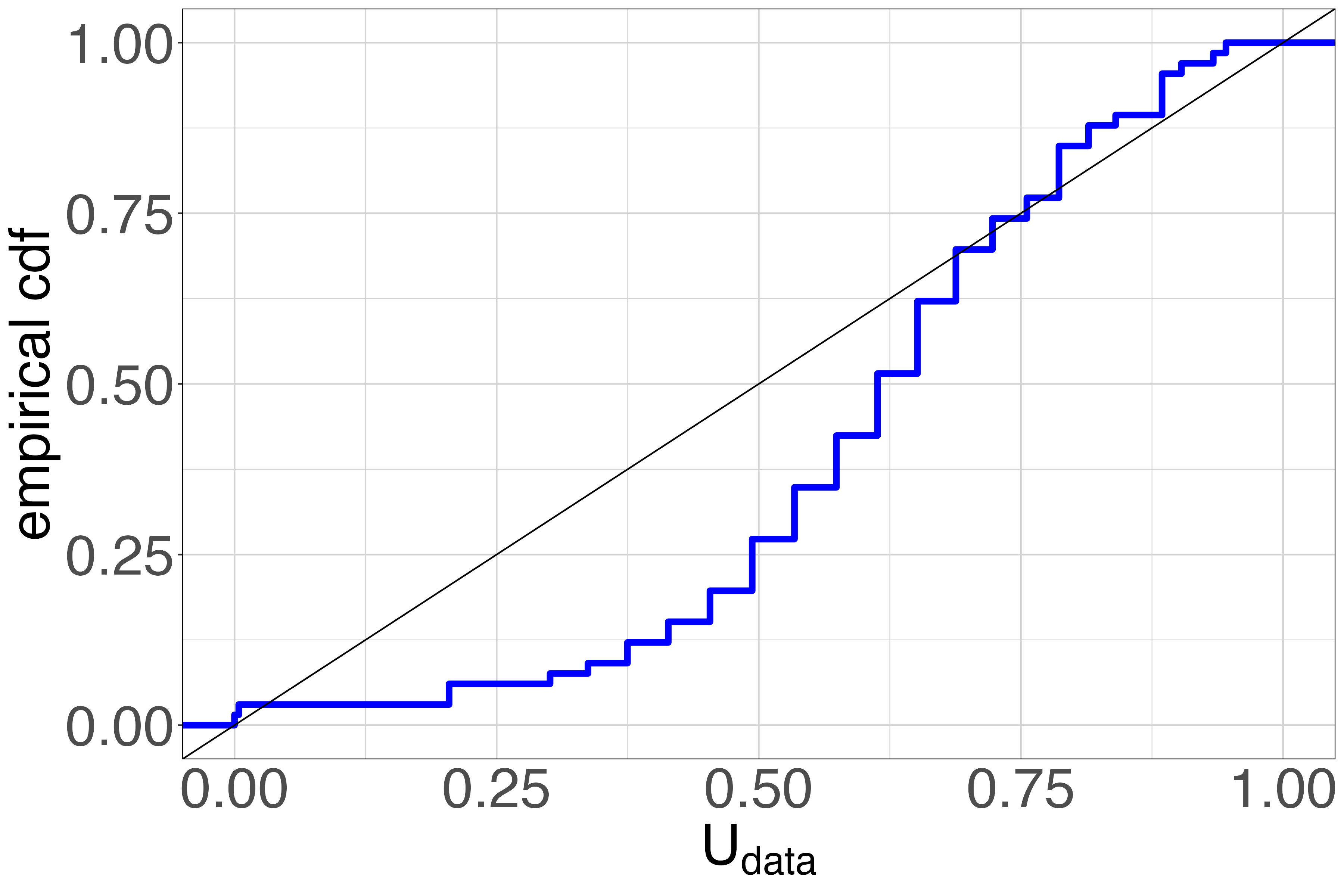}
\hfill 
\includegraphics[width=0.3 \textwidth]{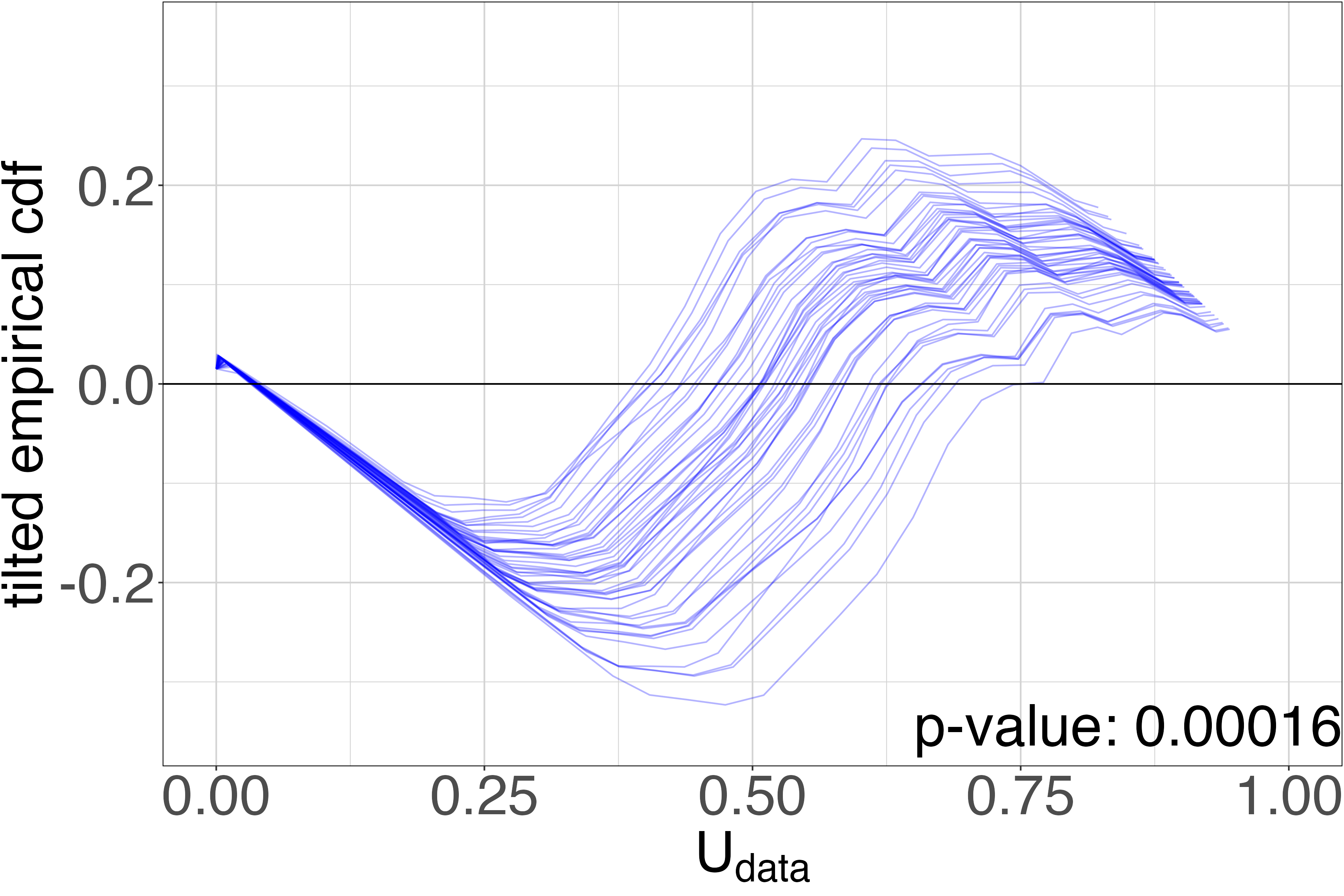}

\includegraphics[width=0.3 \textwidth]{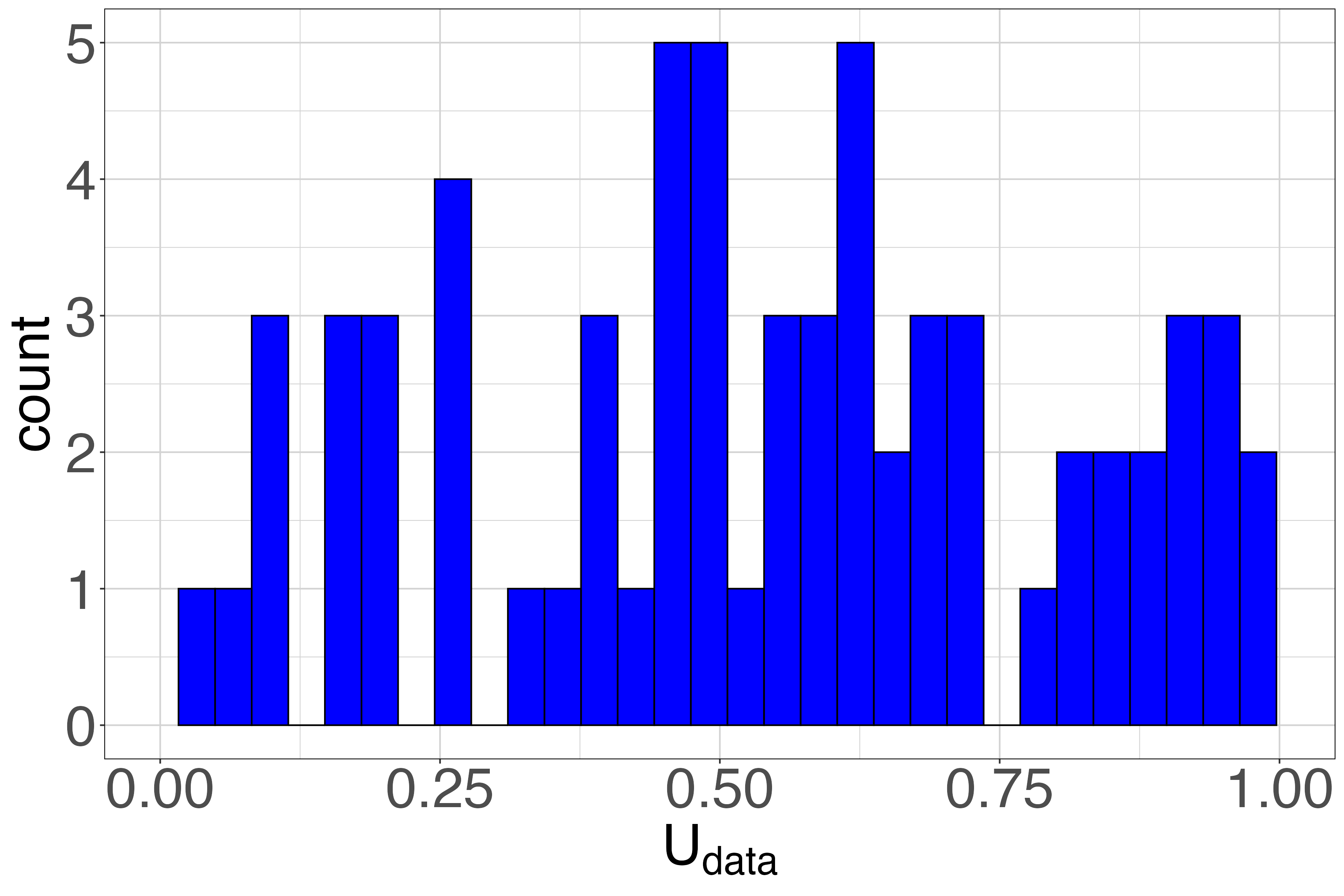}
\hfill 
\includegraphics[width=0.3 \textwidth]{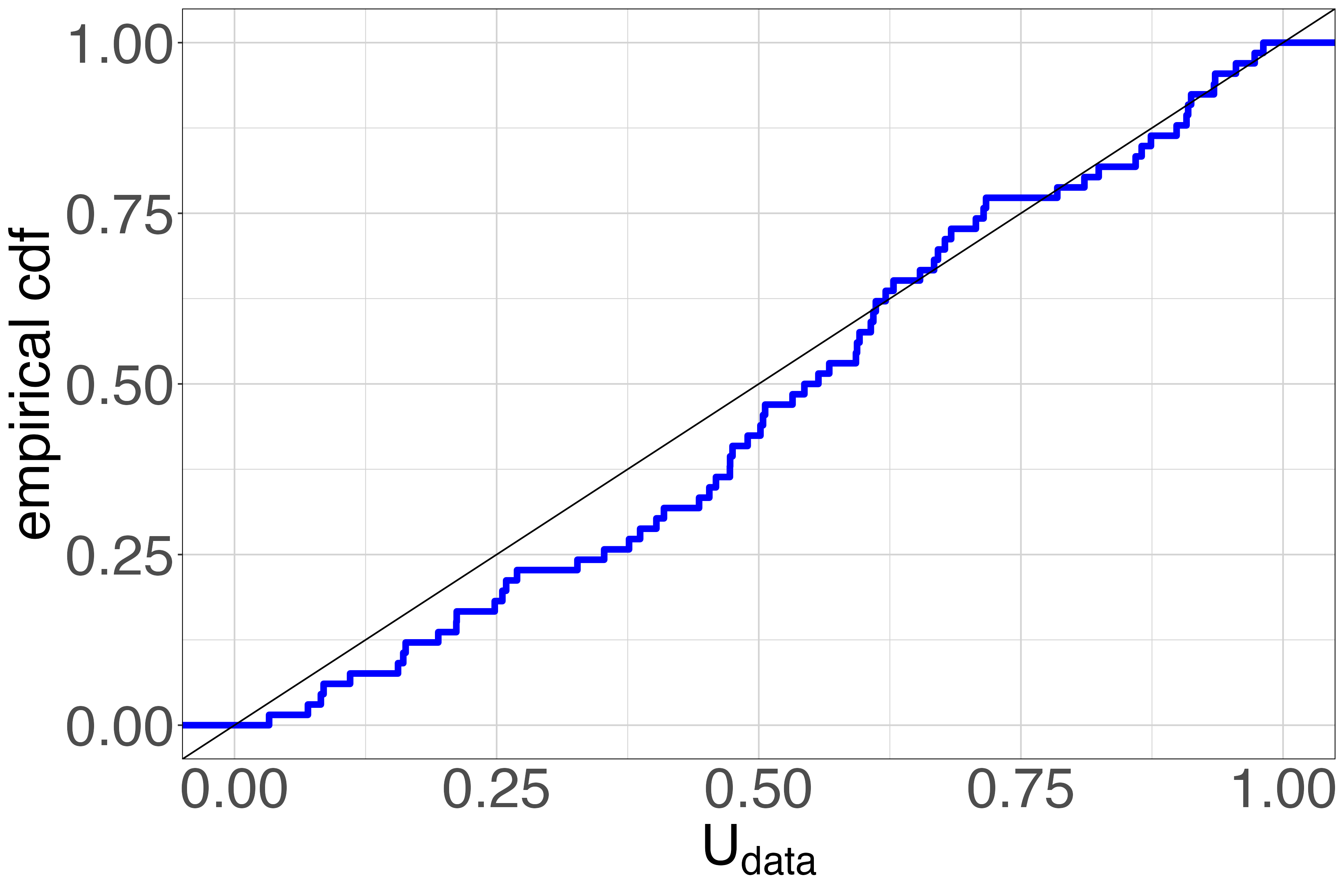}
\hfill
\includegraphics[width=0.3 \textwidth]{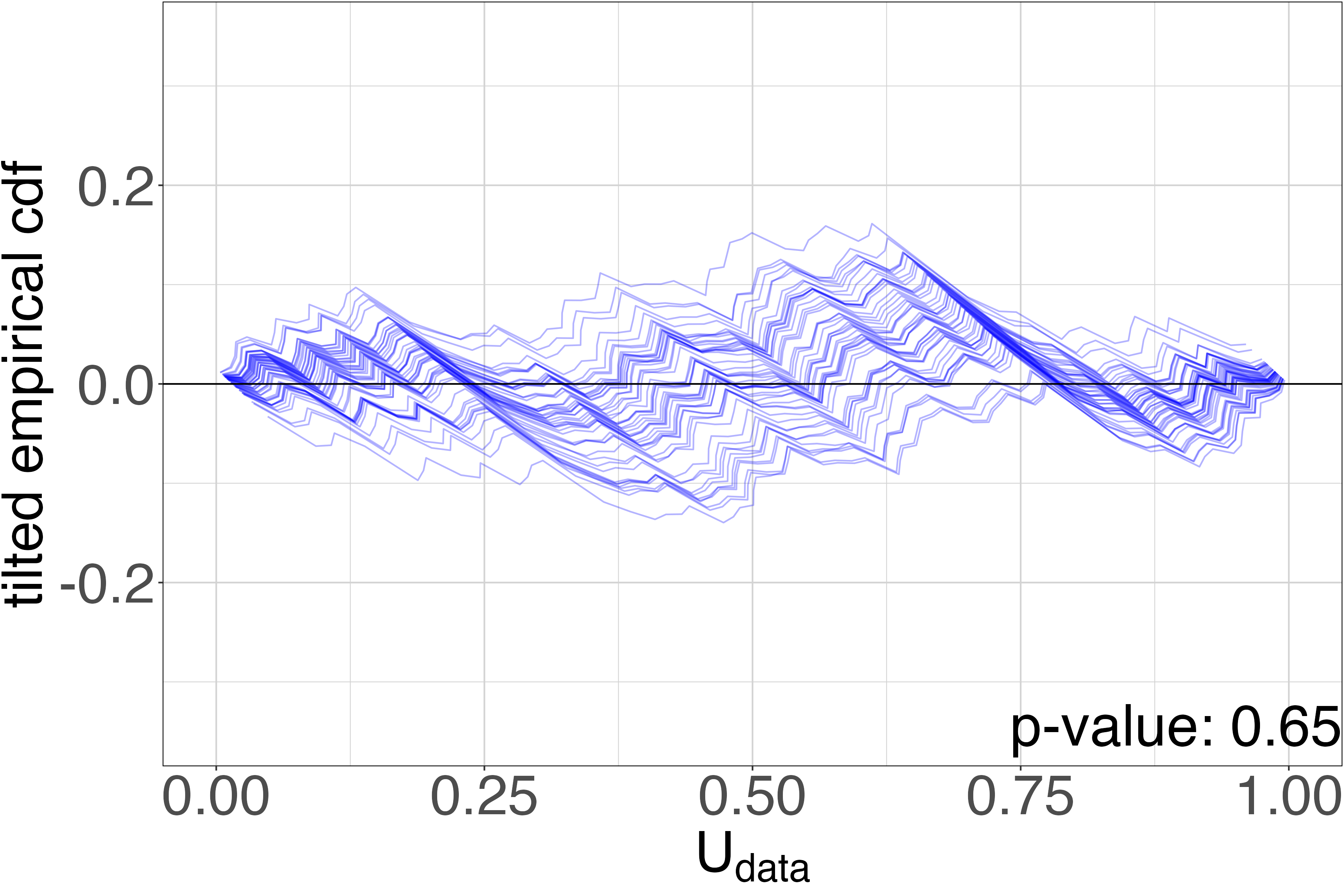}

\caption{Visualizing the data u-values for the Newcomb data.  (Top, left/middle) Histogram and empirical CDF $\hat{F}_{\mathrm{d}}$ of the data u-values $U_{d_1},\ldots,U_{d_n}$ from a single posterior sample, given the Newcomb data. 
(Top, right) Tilted empirical CDF of the data u-values, $\hat{F}_{\mathrm{d}}(u) - u$, for multiple posterior samples given the Newcomb data. 
(Bottom) Same as the top row, but for samples of data u-values from the posterior given a simulated dataset from the hypothesized model.
}
\label{fig:newcomb-data-uvalues}
\end{figure}

To examine the data u-values in more detail, \cref{fig:newcomb-data-uvalues} (top, left/middle) shows a histogram 
and the empirical CDF $\hat{F}_{\mathrm{d}}(u) = \frac{1}{n}\sum_{i=1}^{n} \mathds{1}(\widetilde{U}_{d_i} \leq u)$ of the data u-values
for a single sample of $(\mu,\sigma^2)$ from the posterior given the Newcomb data.
We can see that these data u-values do not appear to be uniformly distributed, particularly in comparison to the corresponding histogram and empirical CDF 
given a simulated dataset from the hypothesized model (\cref{fig:newcomb-data-uvalues}; bottom, left/middle).
To more clearly see the differences between the empirical CDF $\hat{F}_{\mathrm{d}}(u)$ and the uniform CDF $F_{\mathrm{U}(0,1)}(u) = u$, 
in \cref{fig:newcomb-data-uvalues} (right), we plot $\hat{F}_{\mathrm{d}}(u) - u$ 
for multiple posterior samples given the Newcomb data (top) and simulated data from the model (bottom).  For a CDF $F$, we refer to $F(u) - u$ as the corresponding ``tilted CDF.''
This illustrates that the data u-values clearly do not appear to be uniformly distributed,
which explains the small value of $p_{\mathrm{data,unif}}^* = 1.60 \times 10^{-4}$.

To evaluate the effect of the prior on the UPC results, we compare three choices of prior:
(1) the weakly informative prior described above, in which $\mu_0 = 0$, $\kappa_0 = 1/10$, $\alpha_0 = 2$, and $\beta_0 = 300$, 
(2) a data-dependent prior, in which $\mu_0 = \overline{Y}$, $\kappa_0 = n$, $\alpha_0 = n/2$, and $\beta_0 = \hat{\sigma}^2 \alpha_0$,  
where $\overline{Y} = \frac{1}{n} \sum_{i=1}^{n} Y_i$ and $\hat{\sigma}^2 = \frac{1}{n} \sum_{i=1}^{n} (Y_i - \overline{Y})^2$ and 
(3) a poorly chosen informative prior based on previous data, in which $\mu_0 = 179$, $\kappa_0 = n$, $\alpha_0 = n/2$, and $\beta_0 = 42^2 \alpha_0 \kappa_0$.
Prior \#3 is based on experiment of Foucault in 1862, who only twenty years prior to Newcomb estimated the speed of light to be $2.98\times 10^8$ m/s with an error of $\pm 500{,}000$ m/s \citep{froome1971velocity}. 
This corresponds to a value of $24{,}979 \pm 42$ ns in Newcomb's measurements, which represent the time required to travel $7.44373$ km.
Relative to $24{,}800$ ns, this translates to $179 \pm 42$ ns.

\begin{table}[t]
    \centering
    \begin{tabular}{|c|c||c|c|c|}
    \hline
    \textbf{Prior} & \textbf{Data} & $p^*_{\mu}$ & $p^*_{\sigma}$ & $p^*_{\text{data,unif}}$ \\
    \hline
    Weakly informative prior & Newcomb data & $0.45$ & $0.83$ & $1.60  \times 10^{-4}$ \\
                             & Normal data & $0.37$ & $0.48$ & $0.98$  \\
    \hline
    Data-dependent prior & Newcomb data & $0.96$ & $0.93$ & $4.44  \times 10^{-4}$ \\
                                     & Normal data & $0.93$ & $0.95$ & $0.50$ \\
    \hline
    Poorly chosen informative prior & Newcomb data & $2.41  \times 10^{-4}$ & $3.81 \times 10^{-10}$ & $9.09  \times 10^{-6}$ \\
                        & Normal data & $2.25  \times 10^{-4}$ & $2.01 \times 10^{-10}$ & $9.09 \times 10^{-6}$ \\
    \hline
    \end{tabular}
    \caption{Results on Newcomb's speed of light data: Aggregated $p$-values from UPC tests.}
    \label{table:newcomb_uniformity_test}
\end{table}

\cref{table:newcomb_uniformity_test} shows the aggregated p-values for the UPC results for all three choices of prior, on the Newcomb data.
\cref{table:newcomb_uniformity_test} also reports the aggregated p-values on data simulated from $\mathcal{N}(\overline{Y},\hat{\sigma}^2)$, that is, a normal distribution with mean and variance equal to the sample mean and sample variance of the Newcomb data.
Under the weakly informative and data-dependent priors, we see exactly what we would hope for, no evidence 
against the model in the correctly specified case (Normal data) and evidence against the uniformity of the residuals in the misspecified case (Newcomb data). 
Under the poorly chosen prior, we see evidence of misspecification in all three tests, on both 
the Newcomb data and the simulated normal data.
We see this theme throughout our examples: Although 
UPCs do not actively distinguish between misspecification of the prior and the likelihood, the choice of prior does not  
generally affect UPCs meaningfully unless the prior is chosen very poorly.
See \cref{appendix:subsection:example_1} for plots of results using the data-dependent prior and poorly chosen informative prior.

\subsection{Dependent Bernoulli trials}
\label{sec:bernoulli}

Next, we consider a simulation like the dependent Bernoulli trials example presented by~\cite{gelman2013bayesian}, but with a larger sample size.
We generate the following simulated dataset by drawing $Y_1 \sim \mathrm{Bernoulli}(0.5)$
and for $i = 2,\ldots,100$, setting $Y_i = Y_{i-1}$ with probability $0.8$, and drawing $Y_i \sim \mathrm{Bernoulli}(0.5)$ otherwise:
\begin{align}\label{eq:bernoulli_data}
\begin{split}
0~0~1~1~1~1~1~1~1~1~1~1~1~0~0~0~0~0~0~0~0~0~1~1~1~1~1~1~1~0~0~0~0~0~0~0~0~0~0~0~0~0~0~0~1~1~1~1~1~1\phantom{.} \\
~0~0~0~0~0~0~0~0~0~0~0~0~0~0~0~0~0~0~0~0~0~0~0~0~0~0~0~0~0~0~0~0~0~0~0~0~0~0~0~0~0~0~0~0~0~0~1~1~1~1.
\end{split}
\end{align}
Following \citet[Section 6.3]{gelman2013bayesian}, we consider modeling these data as $Y_1,\ldots,Y_n$ i.i.d.\ $\sim \textrm{Bernoulli}(\theta)$, where $n = 100$.  For the prior, we consider $\theta\sim\mathrm{Beta}(1,1)$. We draw $10^6$ samples of $\theta$ from the posterior, $\theta\mid Y_{1:n} \,\sim\, \mathrm{Beta}(1 + \sum_i Y_i,\, 1 + n - \sum_i Y_i)$.

\begin{figure}
\centering
\includegraphics[width=0.49 \textwidth]{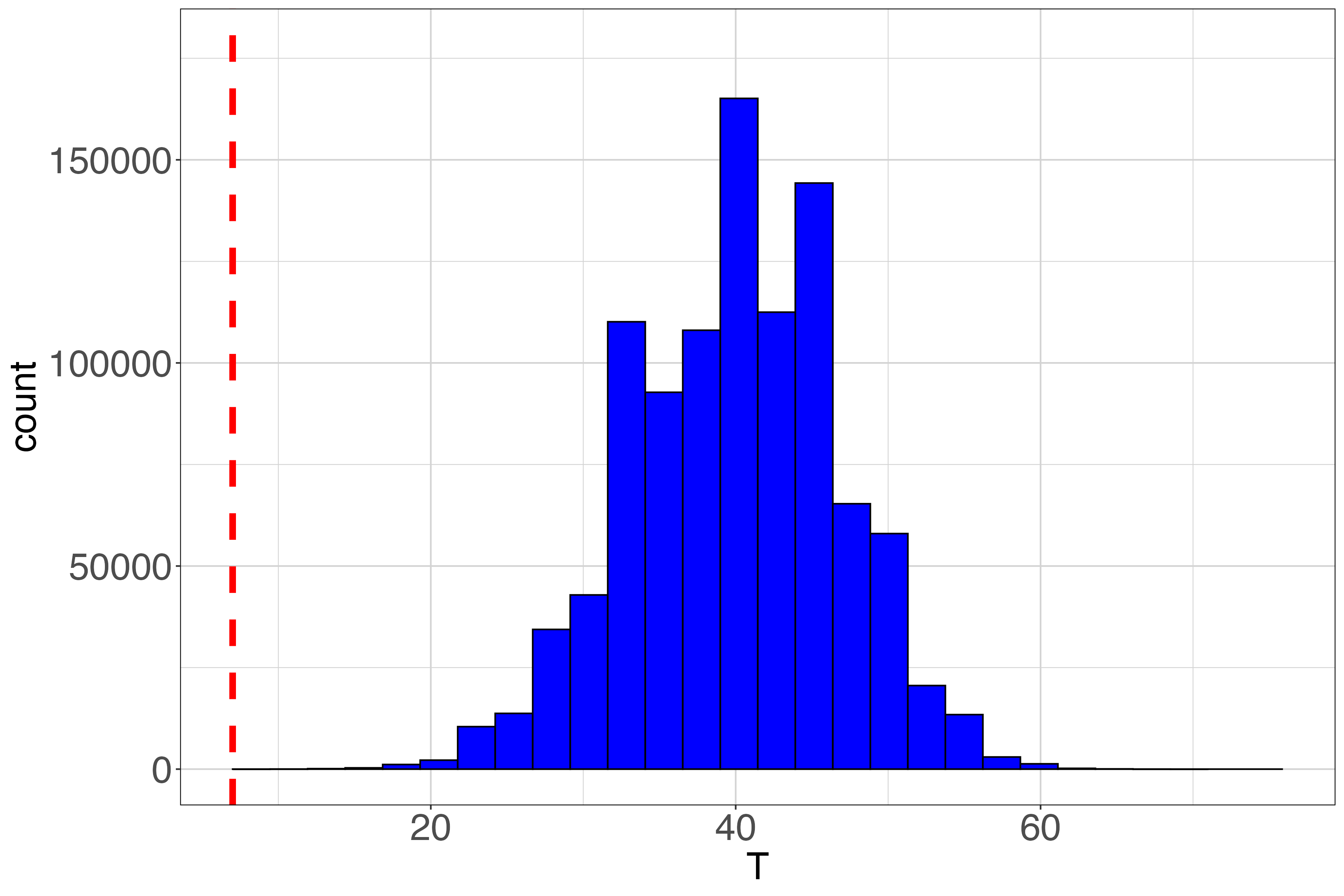}
\hfill
\includegraphics[width=0.49 \textwidth]{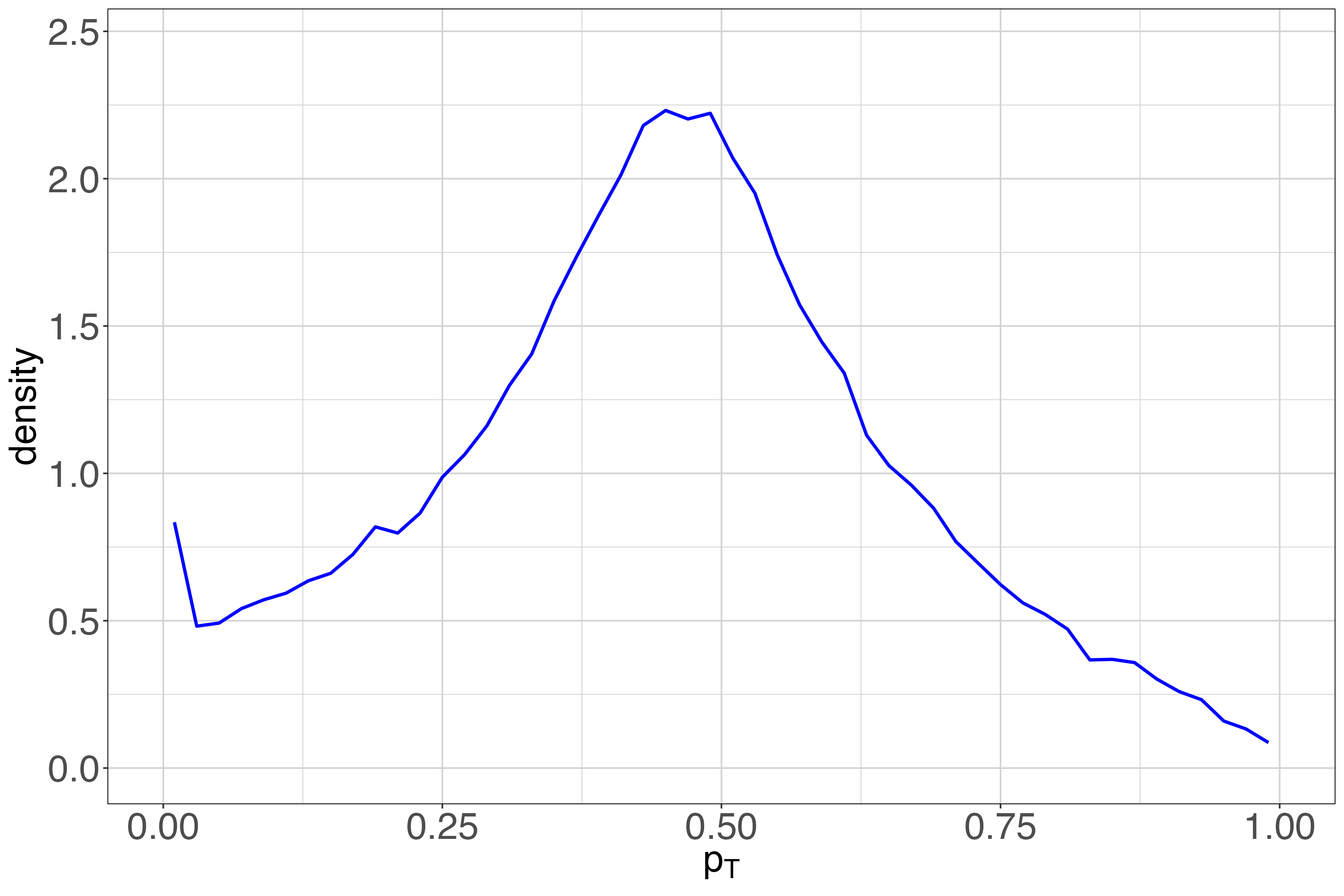}
\caption{Results using PPCs on dependent Bernoulli example. (a) Histogram of the distribution of $T = \sum_{i=1}^{n-1} \mathds{1}(Y_i \neq Y_{i+1})$ under the posterior predictive, along with the observed value of $T$ on the data (vertical red dashed line).  (b) Density of the PPC p-values when sampling from the hypothesized model.}
\label{fig:bernoulli_ppc}
\end{figure}

As a PPC test statistic, \citet{gelman2013bayesian} suggest using the number of switches between $0$ and $1$, that is, $T = \sum_{i=1}^{n-1} \mathds{1}(Y_i \neq Y_{i+1})$.
\cref{fig:bernoulli_ppc}(a) shows that the observed value of $T$ on the data in \cref{eq:bernoulli_data} is extremely unlikely under the posterior predictive, so this PPC successfully detects the fact that the hypothesized model is misspecified. However, again, this is not a well-calibrated test, as we can see from the non-uniformity of the PPC p-values in \cref{fig:bernoulli_ppc}(b) under simulated datasets from the hypothesized model.

For the UPC approach, we reparametrize as $\theta = F_\theta^{-1}(U_1) = U_1$ and $Y_i = \mathds{1}(U_{i+1} \geq 1-\theta)$ for $i=1,\ldots,n$.
For each posterior sample of $\theta$, we sample the u-values by setting $\widetilde{U}_1 = \theta$ and drawing $\widetilde{U}_{i+1} \sim \mathrm{Uniform}(0,\,1-\theta)$ if $Y_i = 0$, or $\widetilde{U}_{i+1} \sim \mathrm{Uniform}(1-\theta,\,1)$ if $Y_i = 1$, independently for $i=1,\ldots,n$,
as described in \cref{sec:computation-of-data-u-values}.

Since any binary variables must necessarily be Bernoulli, the empirical distribution of the data u-values $\widetilde{U}_2,\ldots,\widetilde{U}_{n+1}$ will be close to uniform when the inferred value of $\theta$ is close to the sample mean of the $Y_i$ values.
To verify this empirically, we test for non-uniformity of the data u-values using an Anderson--Darling test to compute $p_{\mathrm{data,unif}}$.  This yields an aggregated p-value of $p_{\mathrm{data,unif}}^* = 0.74$, correctly indicating no issues with the Bernoulli aspect of the model.

Under the null, the u-values are not just marginally $\mathrm{Uniform}(0,1)$, they are also independent.  To test for dependence, we compute a 
Hoeffding test of independence~\citep{Hoeff48} between $\tilde{U}_i$ and $\tilde{U}_{i+1}$ over $i=2,\ldots,n$, 
which produces an aggregated p-value of $p_{\mathrm{data,indep}}^* = 4.61 \times 10^{-6}$. This provides strong evidence of dependence, correctly detecting the form of misspecification present in the hypothesized model relative to the true data generating process.
As before, we also test for extreme values of $\theta$ using p-value $p_\theta = 2\min\{\widetilde{U}_1, 1 - \widetilde{U}_1\}$, 
which yields $p_\theta^* = 0.58$, correctly indicating no issues with the prior.

\begin{figure}
\centering

\includegraphics[width=0.3 \textwidth]{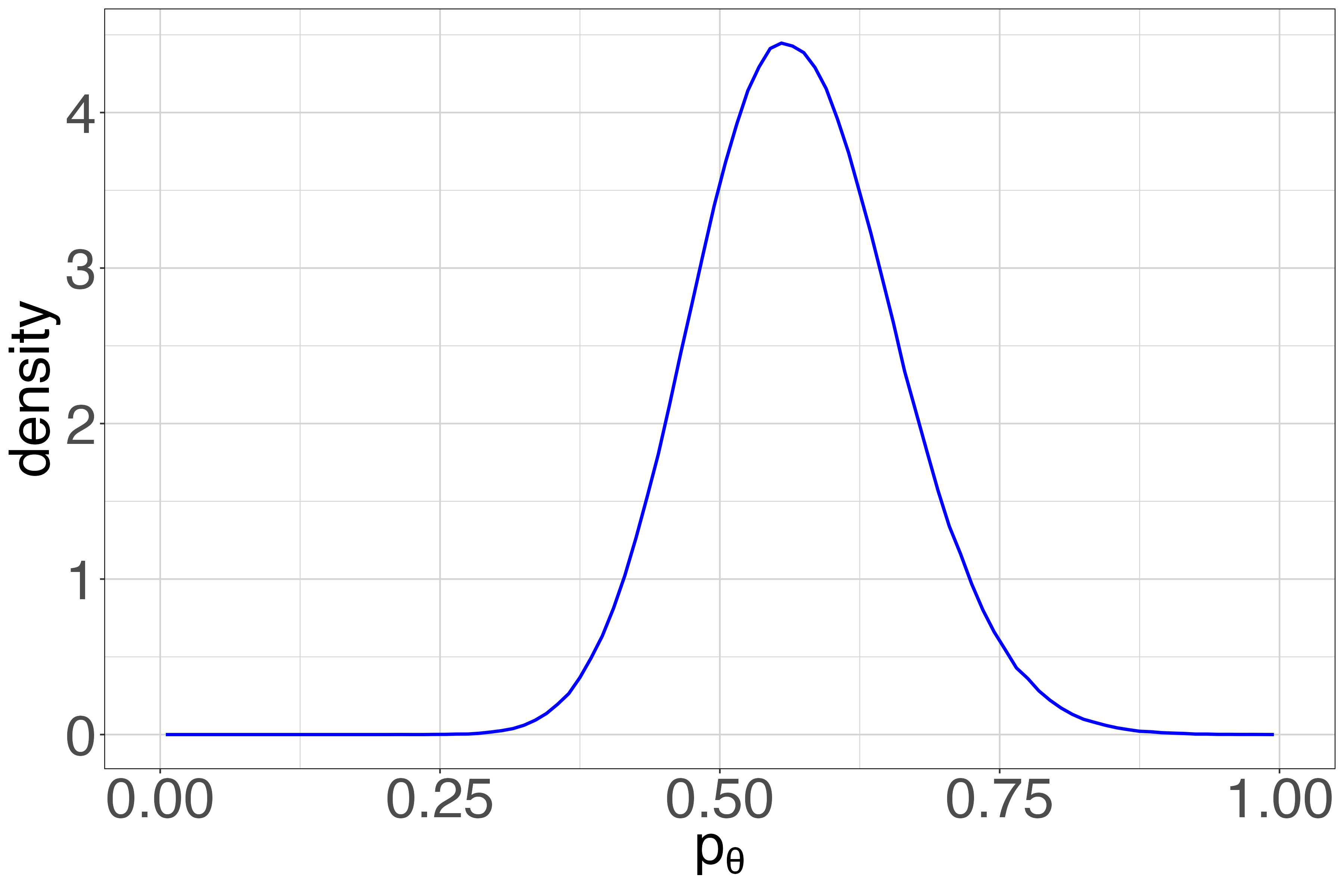}
\hfill 
\includegraphics[width=0.3 \textwidth]{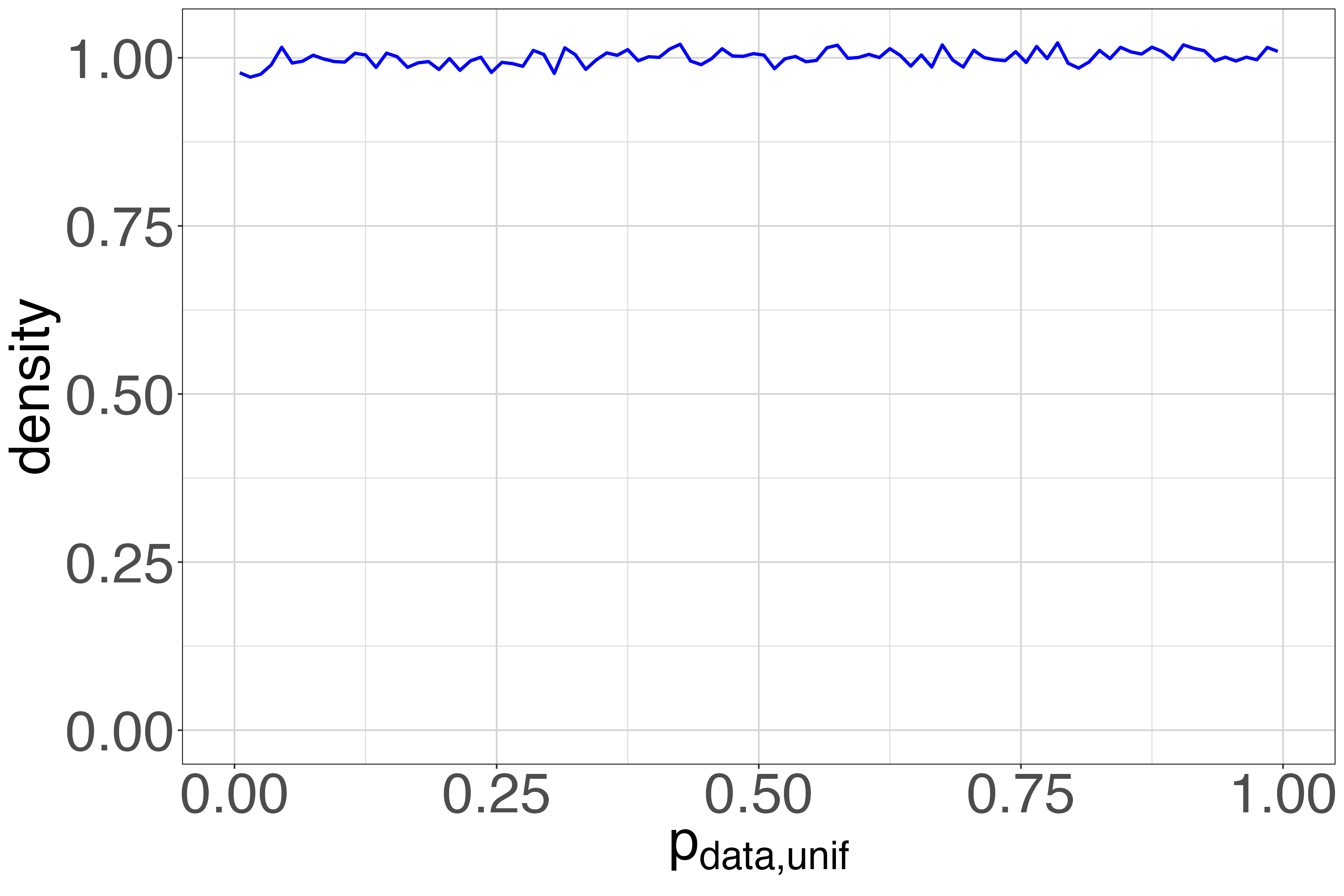}
\hfill
\includegraphics[width=0.3 \textwidth]{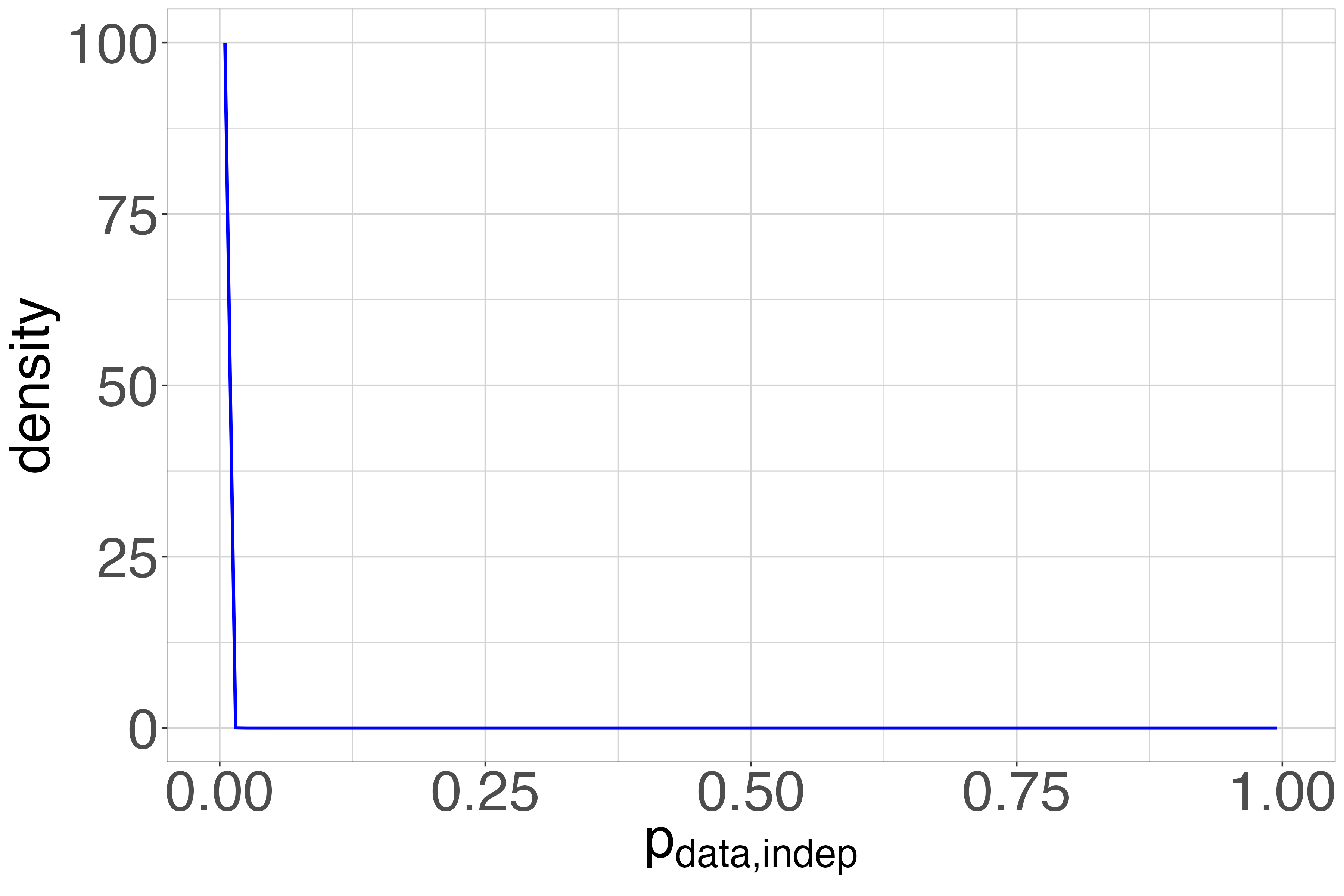}

\includegraphics[width=0.3 \textwidth]{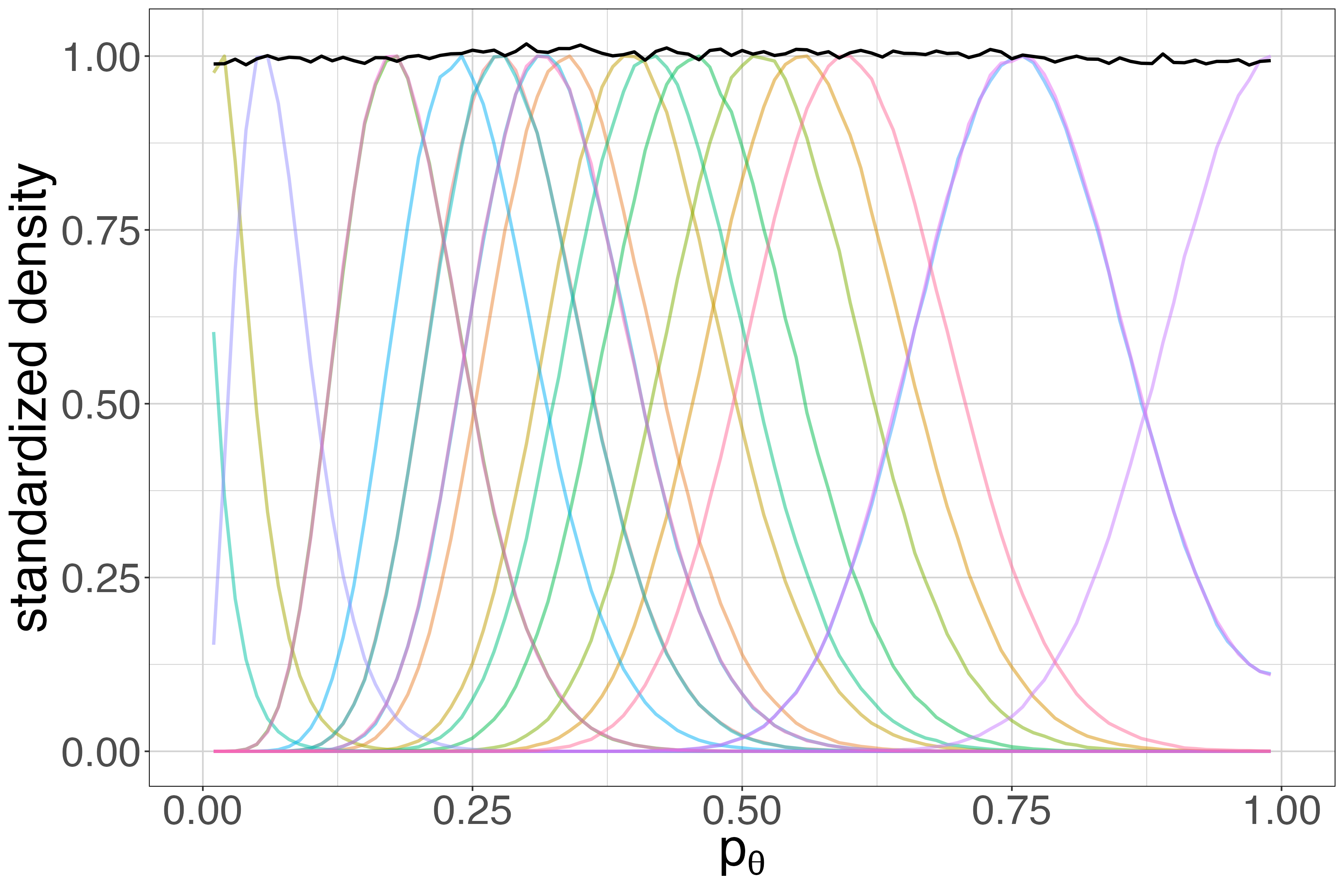}
\hfill 
\includegraphics[width=0.3 \textwidth]{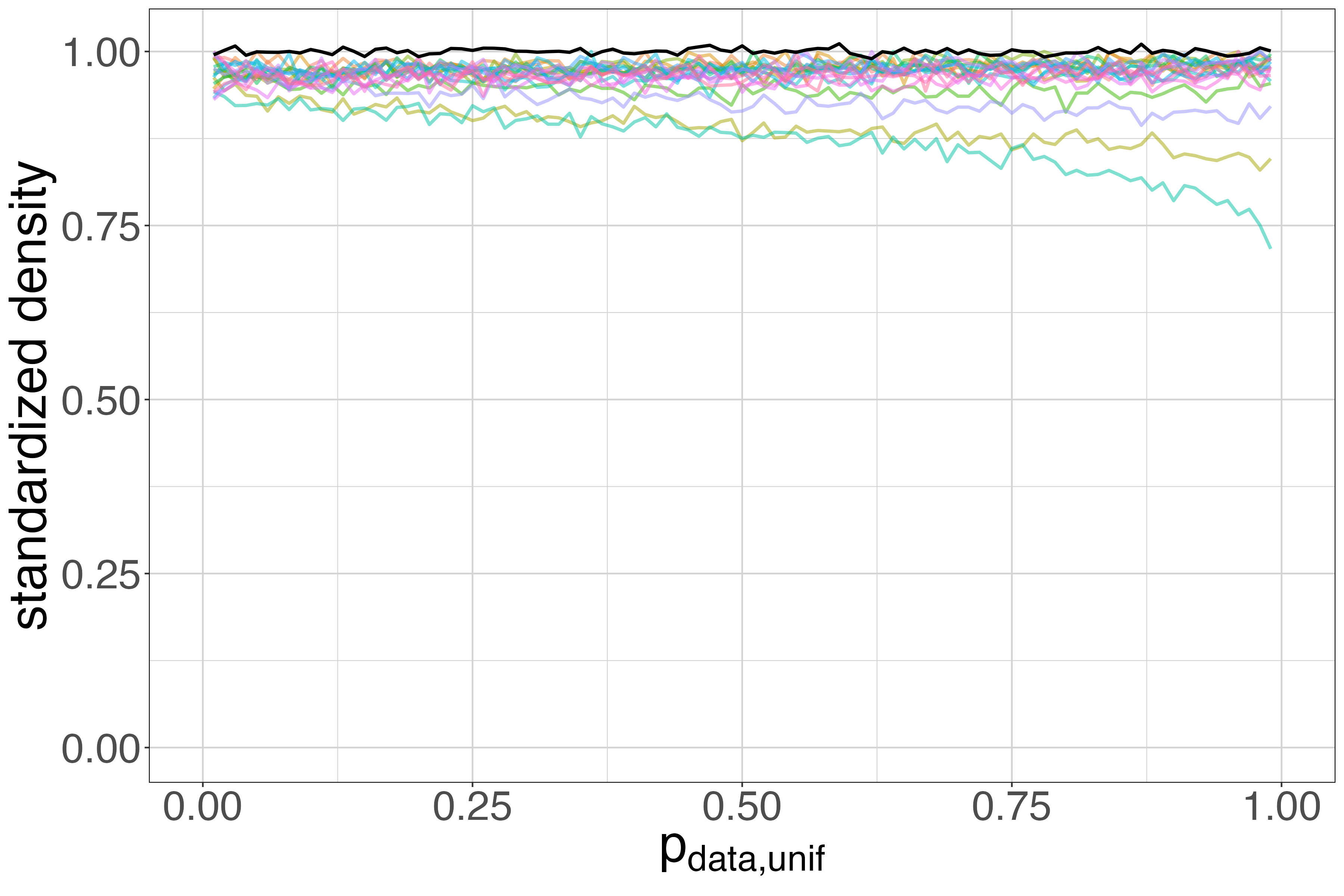}
\hfill
\includegraphics[width=0.3 \textwidth]{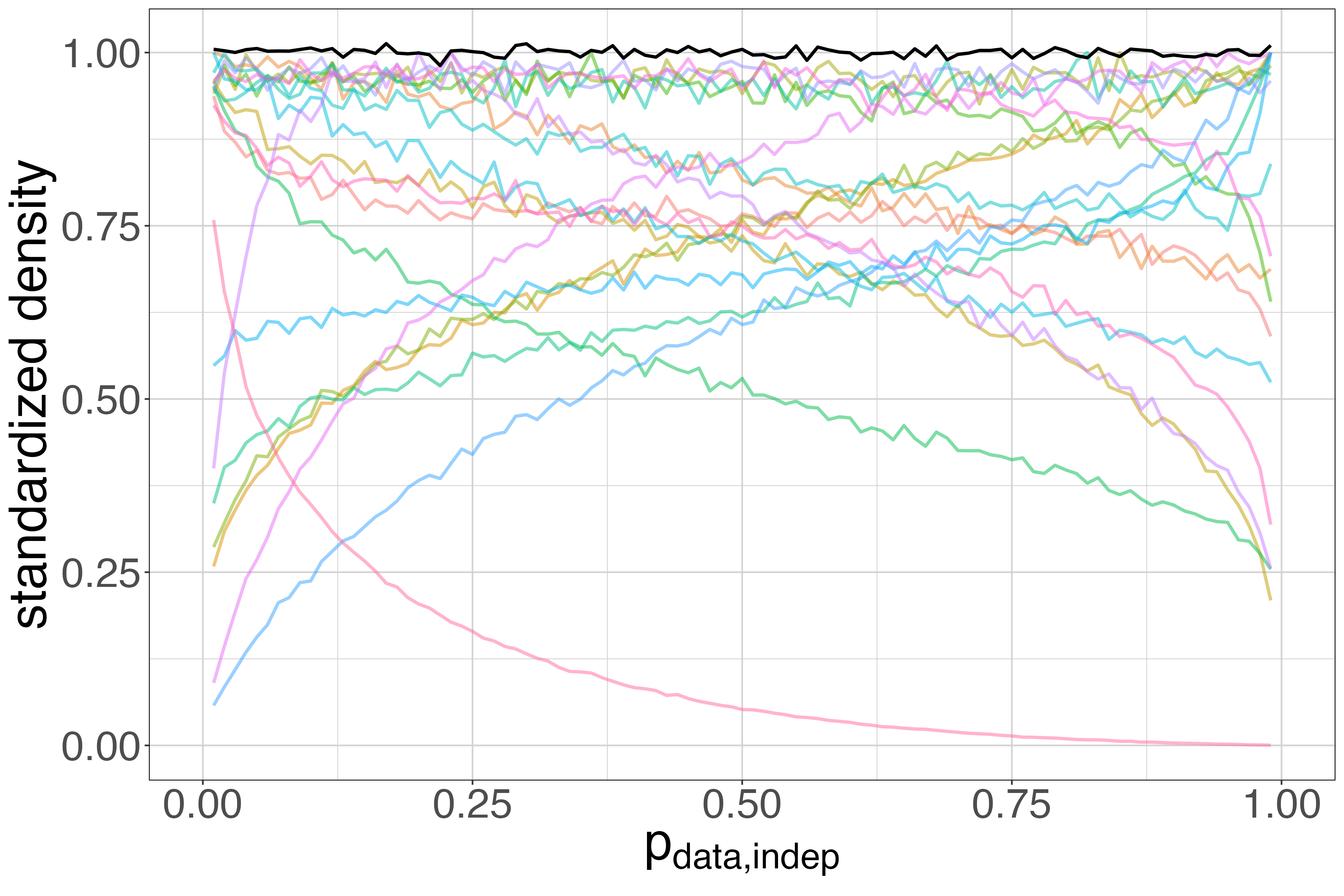}

\caption{Results using UPCs on dependent Bernoulli example. (Top) Posterior densities of the p-values $p_\theta$, $p_{\mathrm{data,unif}}$, and $p_{\mathrm{data,indep}}$ given the observed data in \cref{eq:bernoulli_data}.
(Bottom) Several samples of the same posterior densities, given simulated datasets generated by sampling $\theta$ and $Y_1,\ldots,Y_n$ from the hypothesized model. To aid visualization, each density is standardized to have a maximum of $1$, so they are all visible in a single plot. The black solid lines show the average of these posterior densities over $100{,}000$ simulations, which is exactly uniform in expectation.
}
\label{fig:bernoulli_upc}
\end{figure}

\cref{fig:bernoulli_upc} (top) shows the posterior densities of $p_\theta$, $p_{\mathrm{data,unif}}$, and $p_{\mathrm{data,indep}}$ given the data in \cref{eq:bernoulli_data}.
\cref{fig:bernoulli_upc} (bottom) shows samples of these same posteriors under the null, that is, given simulated datasets from the hypothesized model.
As in \cref{fig:newcomb-upc}, the black lines show the average of these null-distributed posteriors over 100{,}000 simulated datasets,
which we know from the theory is uniform in expectation. 
We can see that the observed posterior for $p_{\mathrm{data,indep}}$ is concentrated near zero, 
which visually illustrates why $p_{\mathrm{data,indep}}^*$ is small.

\begin{table}
    \centering
    \begin{tabular}{|c|c|c|c|}
        \hline
        \textbf{prior} & \textbf{$p_\theta^*$} & \textbf{$p_{\mathrm{data,unif}}^*$} & \textbf{$p_{\mathrm{data,indep}}^*$} \\
        \hline
        Uniform prior, $\mathrm{Beta}(1,1)$ & $0.58$ 
                      & $0.74$ 
                      & $4.61 \times 10^{-6}$ \\
        \hline
        Jeffreys prior, $\mathrm{Beta}(1/2, 1/2)$ & $1$ 
                            & $0.72$
                            & $4.61 \times 10^{-6}$\\
        \hline
        Poorly chosen prior, $\mathrm{Beta}(1,50)$ 
                            & $1.89 \times 10^{-4}$ 
                            & $2.08 \times 10^{-4}$ 
                            & $4.61 \times 10^{-6}$ \\
        \hline
    \end{tabular}
    \caption{Dependent Bernoulli trials: Aggregated $p$-values from UPC tests for extreme $\theta$ values ($p_\theta^*$), non-uniform data u-values ($p_{\mathrm{data,unif}}^*$), and dependent data u-values ($p_{\mathrm{data,indep}}^*$).}
    \label{table:bernoulli_tests}
\end{table}

\begin{figure}
    \begin{subfigure}[t]{0.3\textwidth}
        \centering
        \includegraphics[width=\textwidth]{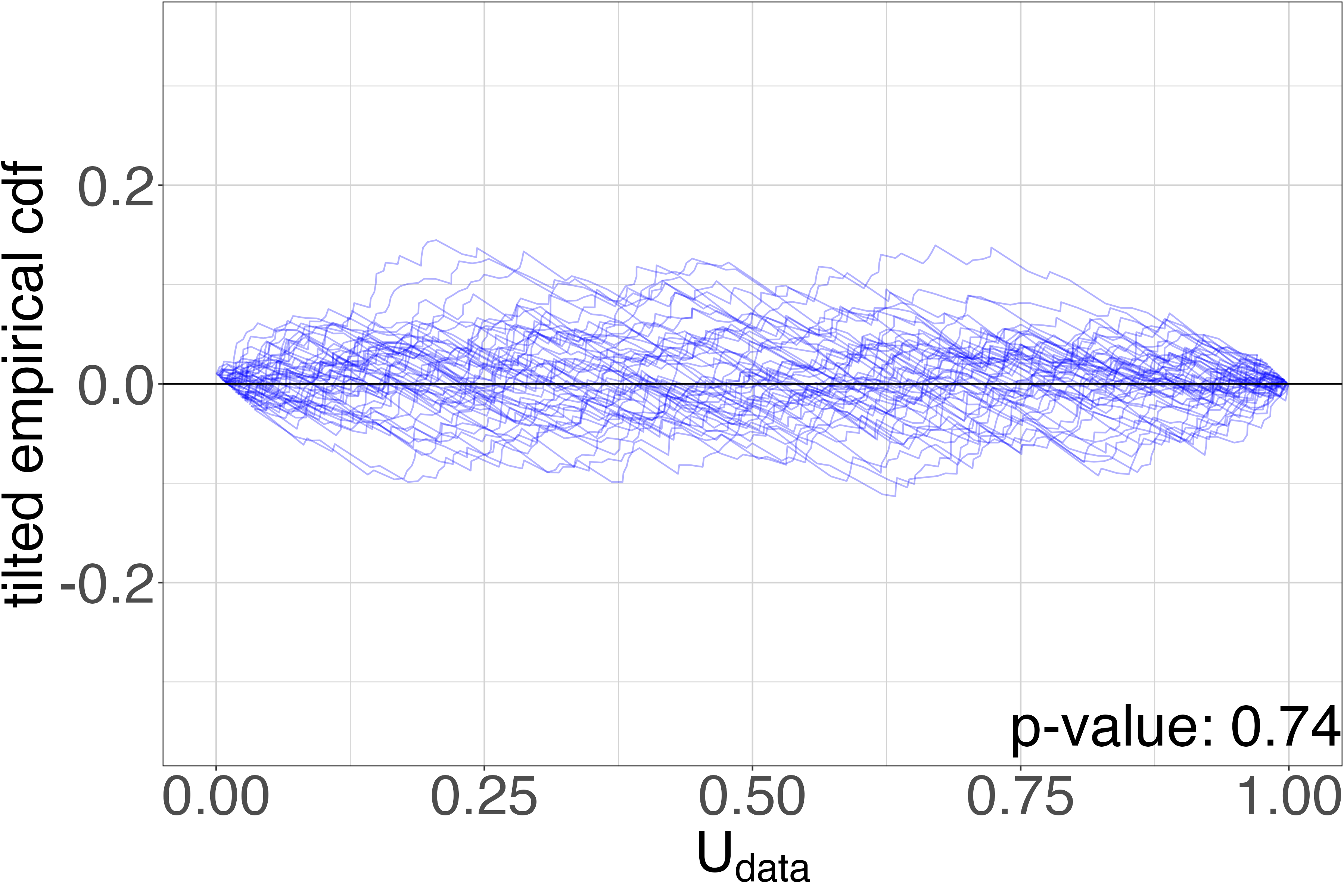}
        \caption{Uniform Prior}
        \label{fig:bernoulli_sim_residual_ecdf_uniform_prior}
    \end{subfigure}
    \hfill
    \begin{subfigure}[t]{0.3\textwidth}
        \centering
        \includegraphics[width=\textwidth]{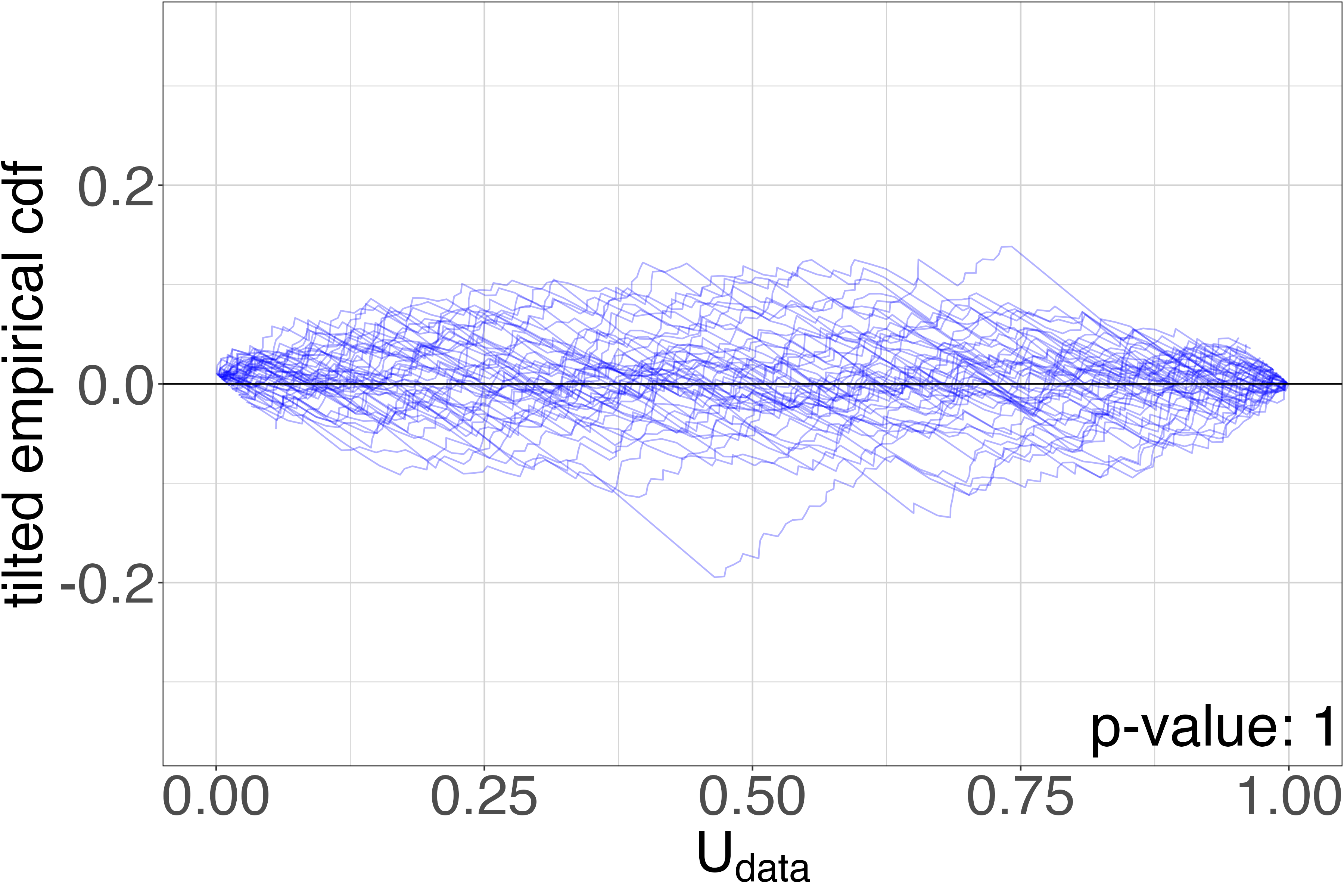}
        \caption{Jeffreys Prior}
        \label{fig:bernoulli_sim_residual_ecdf_uninformative_prior}
    \end{subfigure}
    \hfill
    \begin{subfigure}[t]{0.3\textwidth}
        \centering
        \includegraphics[width=\textwidth]{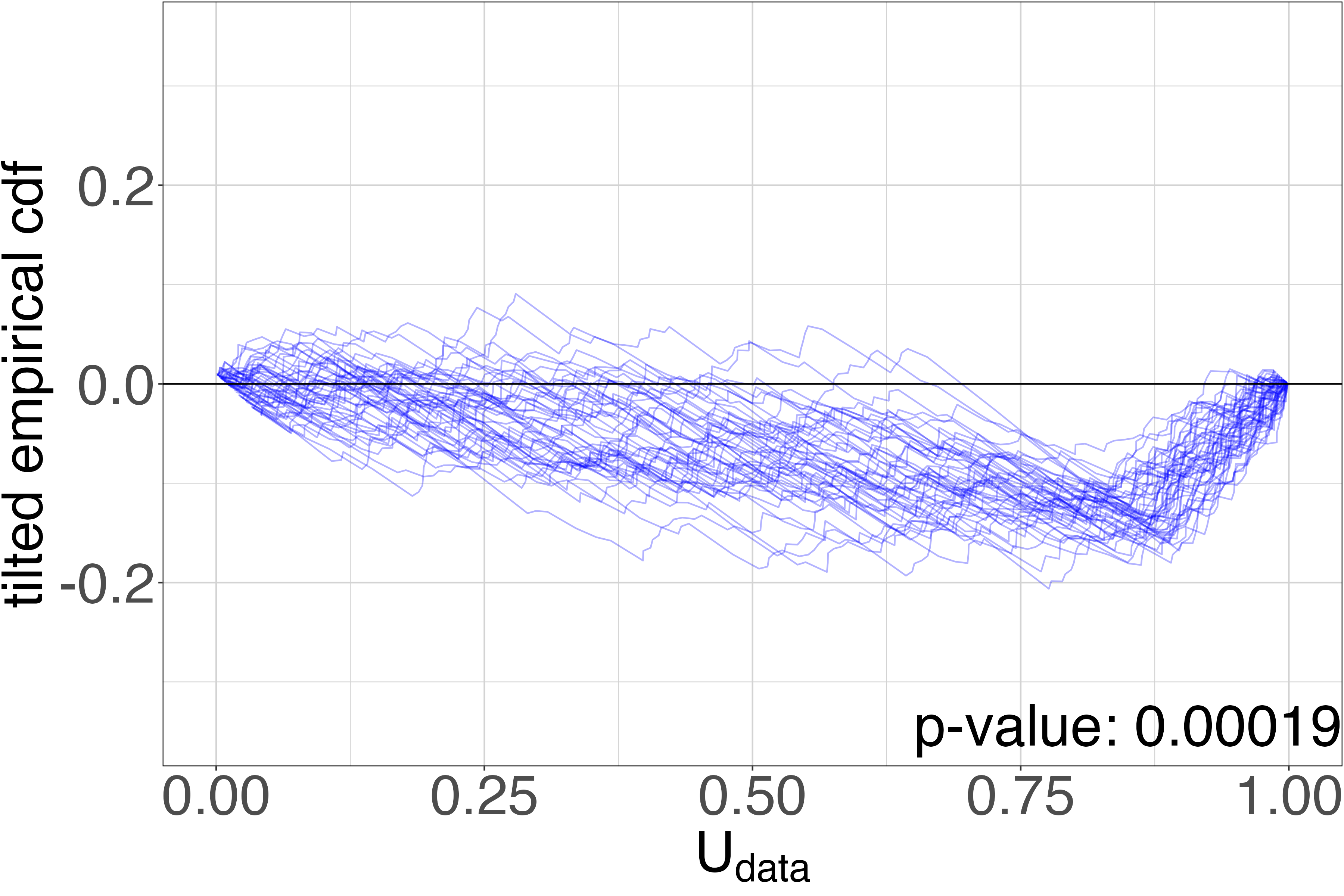}
        \caption{Poorly Chosen Prior}
        \label{fig:bernoulli_sim_residual_ecdf_very_bad_prior}
    \end{subfigure}
    \hfill
    \caption{Tilted CDF of the data u-values, $\hat{F}_\mathrm{d}(u) 
- u$, for multiple posterior samples given the Bernoulli data, under each prior. The departure from uniformity is visibly clear under the poorly chosen prior.}
    \label{fig:bernoulli_sim_residual_ecdf}
\end{figure}

To assess the effect of the choice of prior, we compare (1) the uniform prior as described above, $\theta\sim\mathrm{Beta}(1,1)$, (2) the Jeffreys prior, $\theta\sim\mathrm{Beta}(1/2,1/2)$, and (3) a poorly chosen prior, $\theta\sim\mathrm{Beta}(1,50)$. \cref{table:bernoulli_tests} shows the aggregated p-values for each combination of prior and test.
Under the two reasonable priors (uniform and Jeffreys), we see no evidence against uniformity of the parameter u-value and data u-values (as expected),
and we do see evidence against the independence assumption (as desired, since the data exhibit dependence). 
Under the poorly chosen prior, the null of model correctness is rejected in all three tests; thus, while it is not as clear which aspect of the model is problematic under this prior, the extreme $\theta$ value suggests that improving the prior is a good place to start.
See \cref{appendix:fig:bernoulli_upc_other_priors} for plots of the posteriors under the Jeffreys prior and poorly chosen prior.
For each prior, \cref{fig:bernoulli_sim_residual_ecdf} shows the tilted empirical CDF of the data u-values, $\hat{F}_\mathrm{d}(u) 
- u$, for multiple posterior samples given the Bernoulli data.
In contrast to the PPC approach, we simply use default choices of UPC test, without having to design test statistics.

\subsection{Logistic regression model for adolescent smoking data}
\label{sec:logistic}

\begin{figure}
\centering 
\begin{minipage}{0.45\textwidth}
    \centering
    \includegraphics[width=\textwidth]{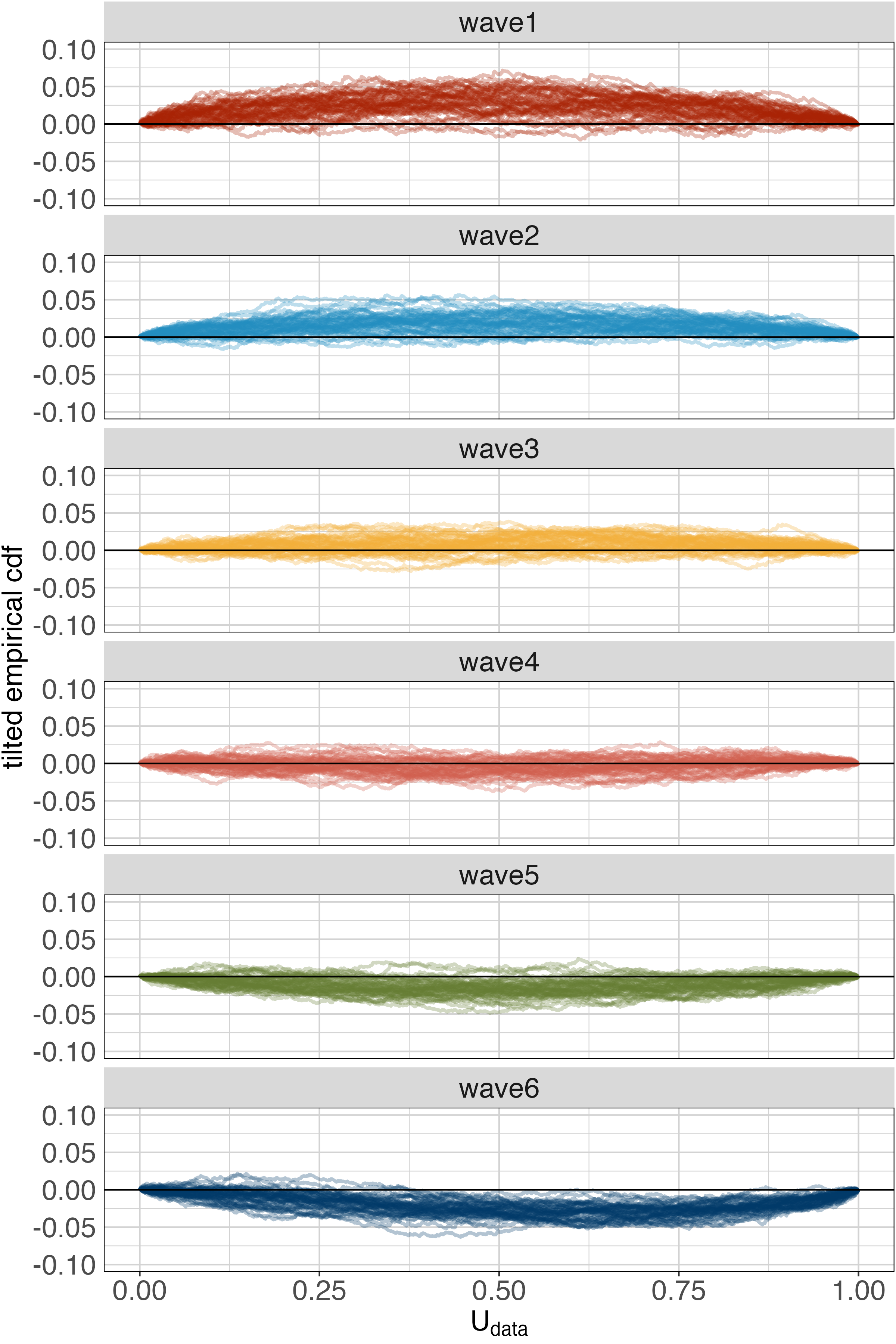}
\end{minipage}
\hfill 
\begin{minipage}{0.54\textwidth}
    \includegraphics[width=0.49\textwidth]{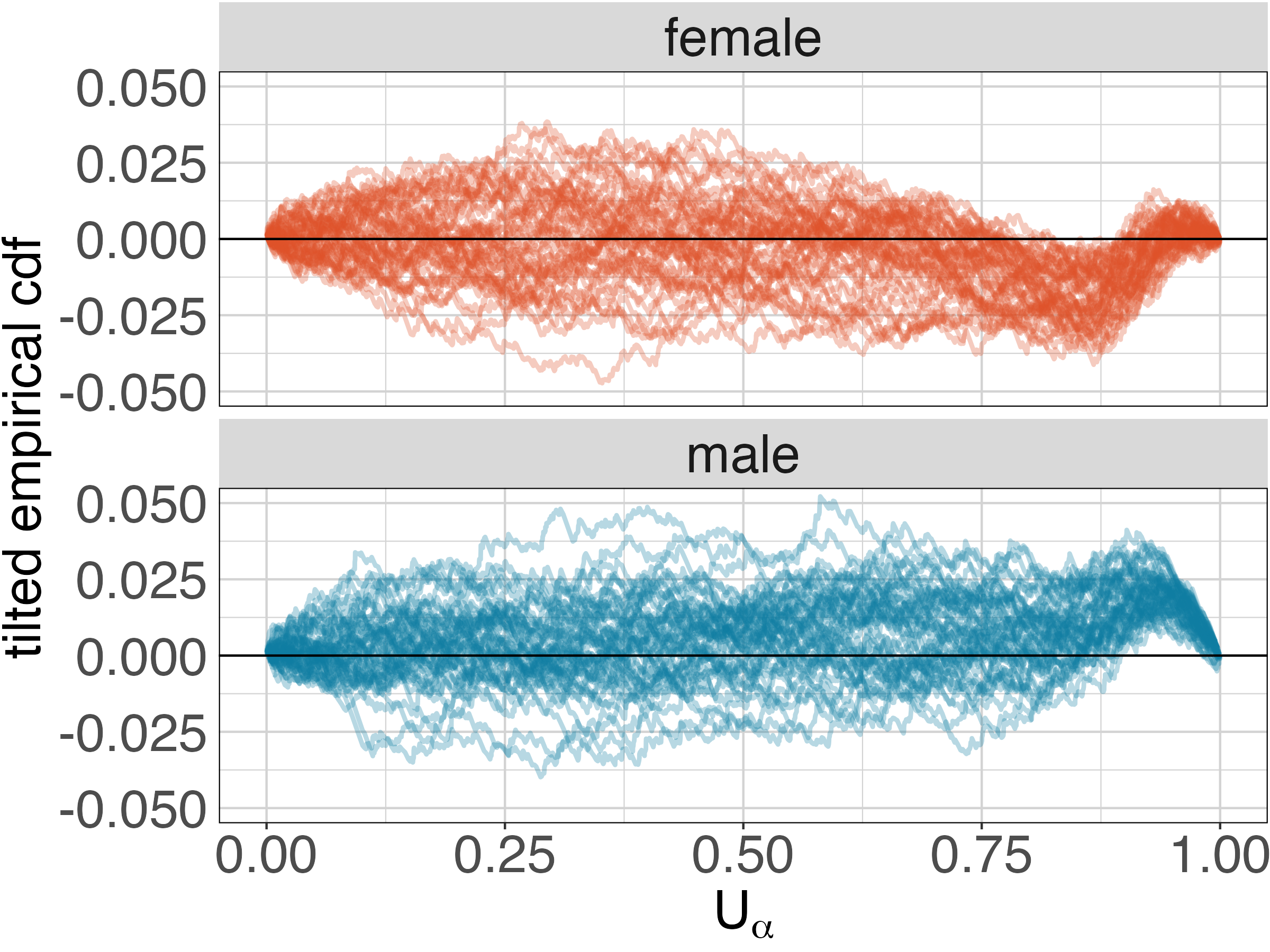}
    \hfill
    \includegraphics[width=0.49\textwidth]{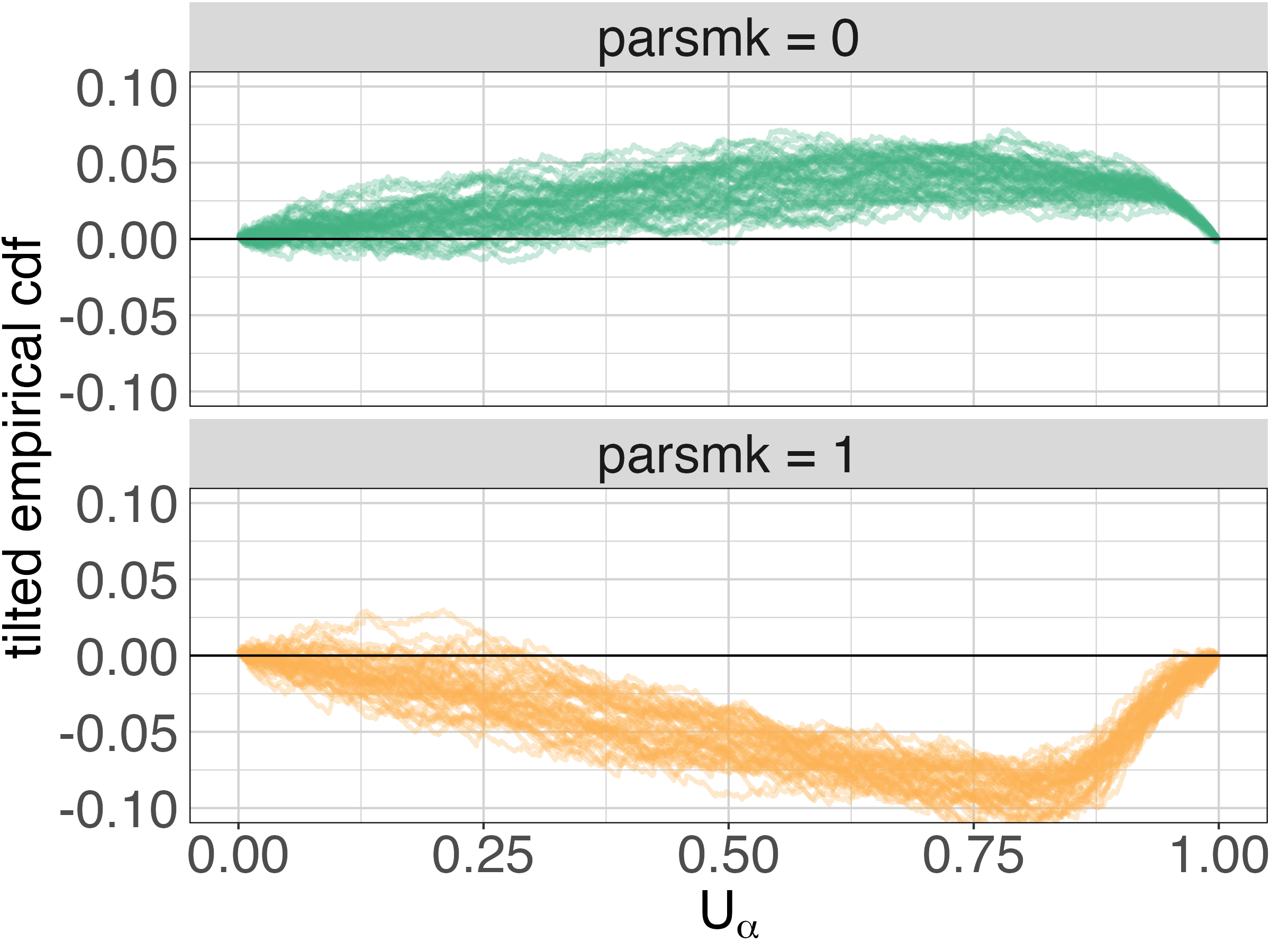}
    \vspace{0.5em}
    \includegraphics[width=0.49\textwidth]{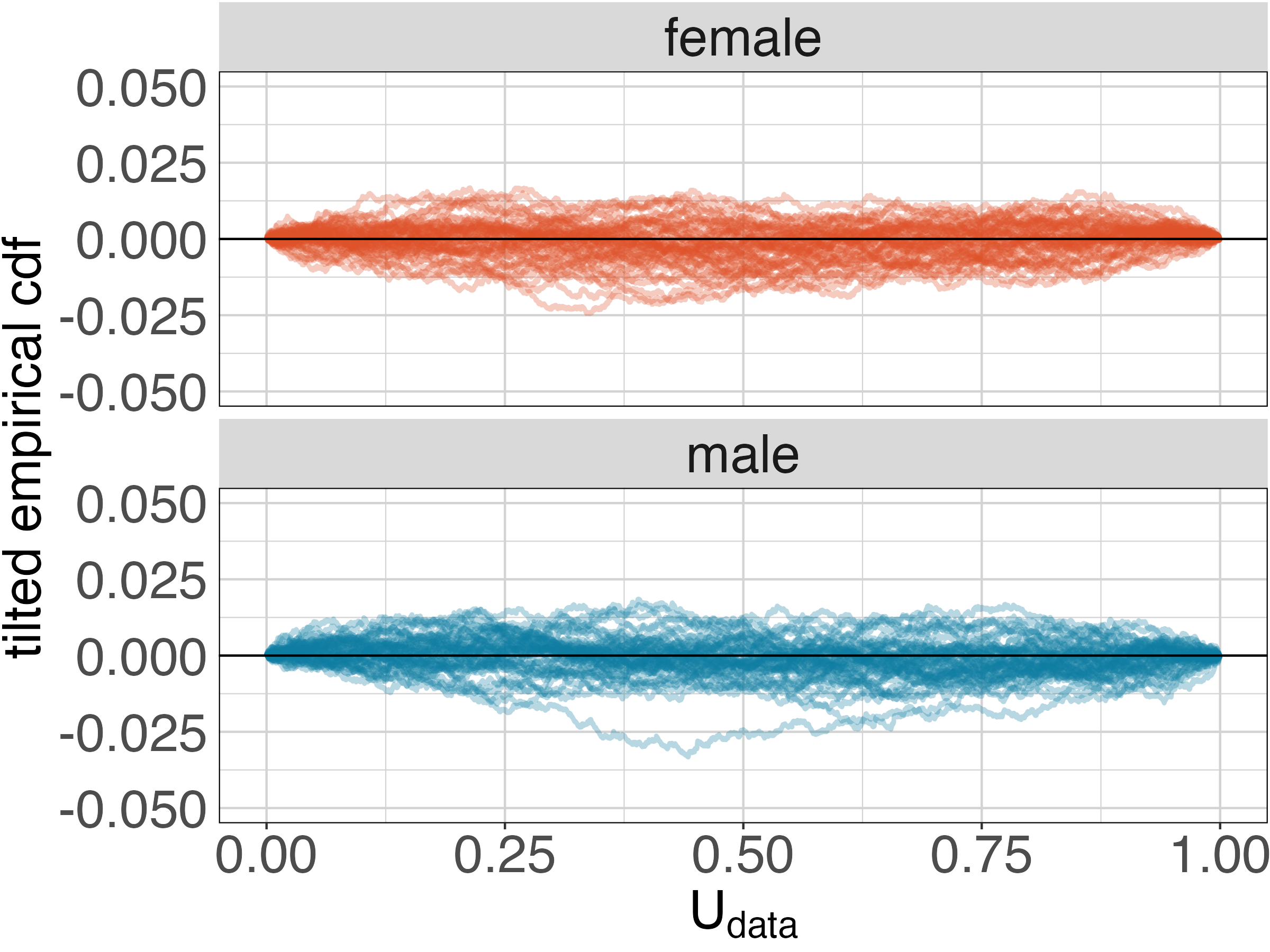}
    \hfill
    \includegraphics[width=0.49\textwidth]{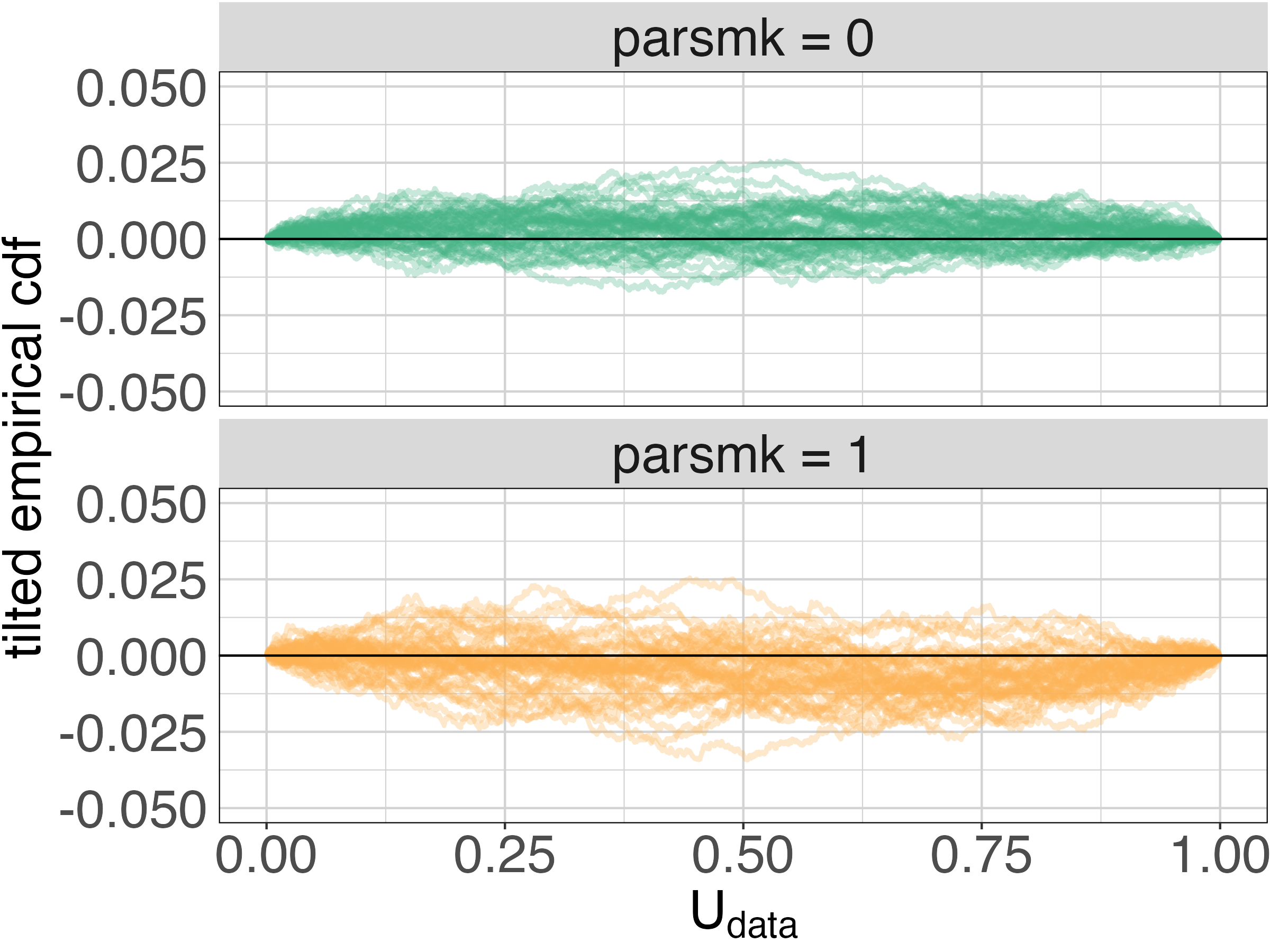}
    \vfill
    \centering
    \includegraphics[width=0.75\textwidth]{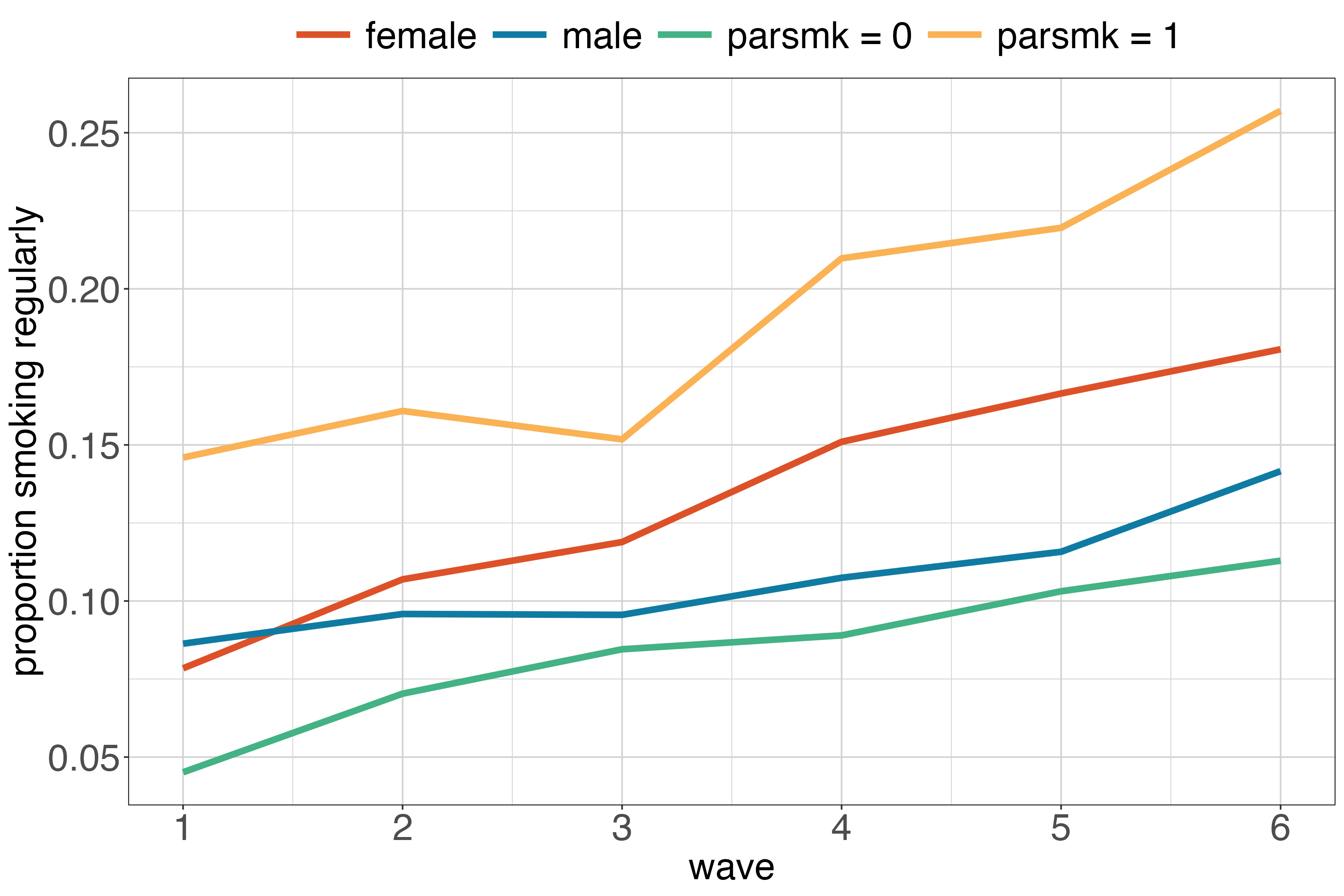}
\end{minipage}

\caption{UPC results for Model \#1 on the logistic regression example. (Left) Posterior samples of the tilted empirical CDFs of the data u-values, stratified by \texttt{wave}. (Top right) Posterior samples of the tilted 
empirical CDFs of the $\alpha$ u-values and the data u-values, stratified by \texttt{sex} and \texttt{parsmk}. (Bottom right) Proportion of individuals that smoke regularly as a function of \texttt{wave}, stratified by \texttt{sex} and \texttt{parsmk}.}
\label{fig:smoking_1}
\end{figure}

In this section, we apply the UPC methodology to a logistic regression example used by \citet[Section 6.3]{gelman2013bayesian} to illustrate PPCs.
Each of the $n = 8{,}730$ observations comes from one of $m = 1{,}760$ individuals (\texttt{newid}) who 
were surveyed at up to six time points (\texttt{wave}) and asked whether or not they smoked regularly (\texttt{smkreg}).
Individuals' sex (\texttt{sex}) and parental smoking status (\texttt{parsmk}) were also recorded as potential covariates.
\cref{fig:smoking_1} (bottom right) shows the proportion of individuals that smoked regularly versus \texttt{wave}, stratified by \texttt{sex} and \texttt{parsmk}.
We treat \texttt{smkreg} as the outcome and \texttt{sex}, \texttt{parsmk}, and \texttt{wave} as covariates, 
standardizing each covariate to have zero mean and unit variance across all observations.

We show how to use the UPC framework for iterative model criticism. 
In particular, we build up from a simple model to a more complex model, 
using UPCs to reject certain model assumptions in each round. 
At the outset, we commit to performing at most two rounds of model criticism (producing three models in total), 
with a cumulative Type I error rate of $\alpha = 0.2$, which we split evenly between the rounds.

\paragraph{Model \#1.}
We start with a simple random effects model with weakly informative priors.
The outcome $Y_{j k}\in\{0,1\}$ represents whether individual $j$ smokes regularly at wave $k$ (\texttt{smkreg}), which is modeled as

\begin{align*}
    (Y_{j k}\mid\alpha) &\sim \mathrm{Bernoulli}\big(\mathrm{logit}^{-1}(\alpha_j)\big), \\
    \alpha_j &\sim \mathcal{N}(\mu,5^2), \\
    \mu &\sim \mathcal{N}(0,5^2)
\end{align*}
for $j\in\{1,\ldots,m\}$, $k\in\{1,\ldots,6\}$.
Since some individuals were not observed at some waves, we handle missing entries by assuming they are missing completely at random; thus, they can be marginalized out of the model and contribute nothing to the likelihood. Indexing the observations $i = 1,\ldots,n$, we define $i(j,k)$ to be the index of the observation for individual $j$ at wave $k$ when it is nonmissing, otherwise $i(j,k)$ is undefined.
We use JAGS~\citep{Plum03} for posterior inference with MCMC, drawing $100{,}000$ posterior samples after a burn-in of $5{,}000$ iterations. We thin the posterior samples, keeping only every $100$th sample, leaving $1{,}000$ samples.
To implement the UPC approach, we write $\mu = 5\Phi^{-1}(U_1)$, $\alpha_j = \mu + 5\Phi^{-1}(U_{j+1})$, and $Y_{j k} = \mathds{1}(U_{i(j,k)+m+1} \geq 1-q_{j k})$,
where $q_{j k} = \mathrm{logit}^{-1}(\alpha_j)$. Inverting these as before, for each posterior sample, we sample the u-values by setting $\widetilde{U}_1 = \Phi(\mu/5)$, $\widetilde{U}_{j+1} = \Phi\big((\alpha_j-\mu)/5\big)$, $\widetilde{U}_{i(j,k)+m+1} \sim \mathrm{Uniform}(0, 1-q_{j k})$ if $Y_{j k} = 0$, and $\widetilde{U}_{i(j,k)+m+1} \sim \mathrm{Uniform}(1-q_{j k},1)$ if $Y_{j k} = 1$.

For this initial model, a primary question is whether the exclusion of all covariates is problematic.  
Thus, as discussed in \cref{section:choice-of-tests}, a natural first step is to test for external dependence between u-values and covariates.
By \cref{theorem:external}, if the true distribution of the outcome is independent of a covariate, then the u-values will independent of that covariate as well.
Therefore, we test for dependence between the data u-values and (1) \texttt{wave}, (2) \texttt{sex}, and (3) \texttt{parsmk}.  We also test for dependence between the $\alpha$ u-values and (4) \texttt{sex} and (5) \texttt{parsmk}.  Testing for dependence between $\alpha$ and \texttt{wave} is excluded since $\alpha_j$ is a subject-level parameter and \texttt{wave} varies within subject.
We use the Mann--Whitney U test for \texttt{sex} and \texttt{parsmk}, and Hoeffding's test for \texttt{wave}; see \cref{section:choice-of-tests}.

For each posterior sample, we compute the p-values for each of these five tests, and aggregate across posterior samples using the Cauchy combination method for each test, obtaining
$p_{\mathrm{data},\texttt{wave}}^* = 1.67 \times 10^{-7}$,
$p_{\mathrm{data},\texttt{sex}}^* = 0.72$,
$p_{\mathrm{data},\texttt{parsmk}}^* = 8.47 \times 10^{-3}$,
$p_{\alpha,\texttt{sex}}^* = 0.68$, and
$p_{\alpha,\texttt{parsmk}}^* = 1.81 \times 10^{-11}$.
To control Type~I error under multiple testing, we apply the Holm--Bonferroni correction to these five aggregated p-values, after which the p-values for \texttt{wave} and \texttt{parsmk} remain significant at the $0.1$ level; see \cref{appendix:section:logistic_regression} for details.

This provides strong evidence of dependence with \texttt{wave} and \texttt{parsmk}, but not \texttt{sex}. 
To visually confirm that these formal results make sense, \cref{fig:smoking_1} contrasts the empirical CDFs of u-values across strata defined by covariate values. 
For instance, \cref{fig:smoking_1} (left) shows $\hat{F}_\mathrm{data}^{\texttt{wave}=k}(u) - u$ for $k\in\{1,\ldots,6\}$ for several posterior samples, where $\hat{F}_\mathrm{data}^{\texttt{wave}=k}$ is the empirical CDF of the data u-values for data points at $\texttt{wave}=k$ for a given posterior sample.
The clear trend as \texttt{wave} goes from $1$ to $6$ reflects the very small value of $p_{\mathrm{data},\texttt{wave}}^*$.

\begin{figure}
\centering 
\begin{minipage}{0.45\textwidth}
    \centering
    \includegraphics[width=\textwidth]{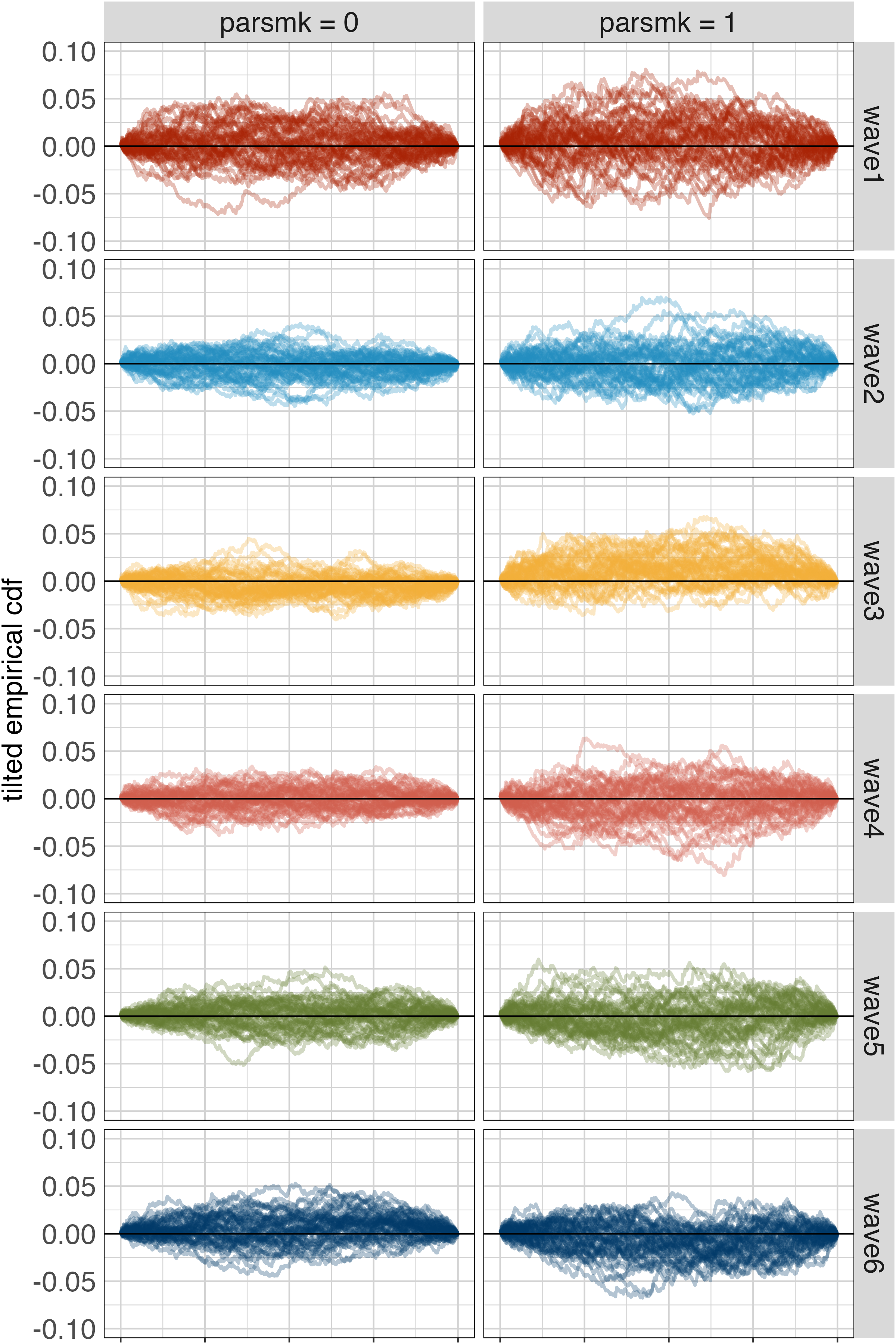}
\end{minipage}
\hfill 
\begin{minipage}{0.54\textwidth}
    \includegraphics[width=0.49\textwidth]{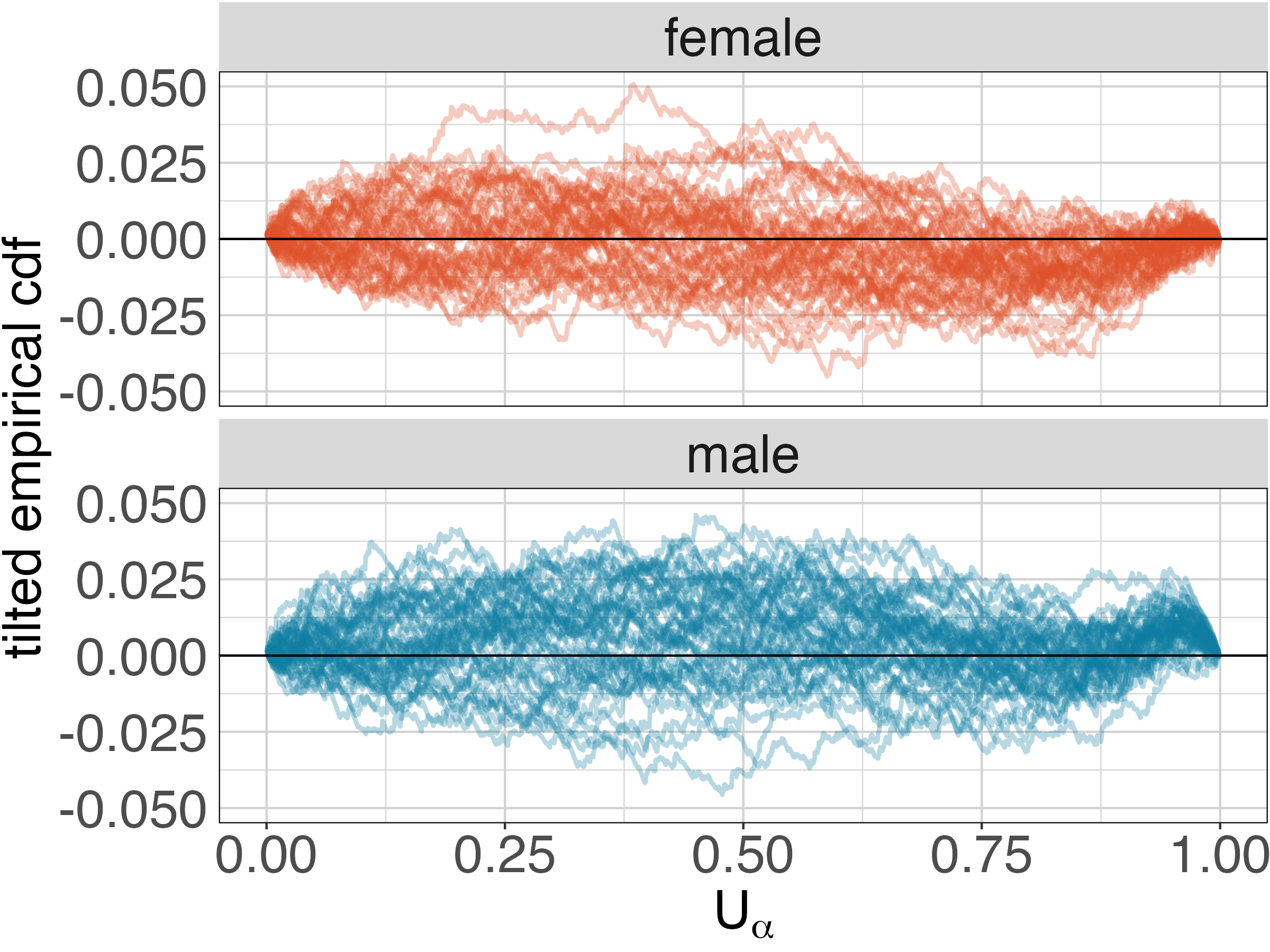}
    \hfill
    \includegraphics[width=0.49\textwidth]{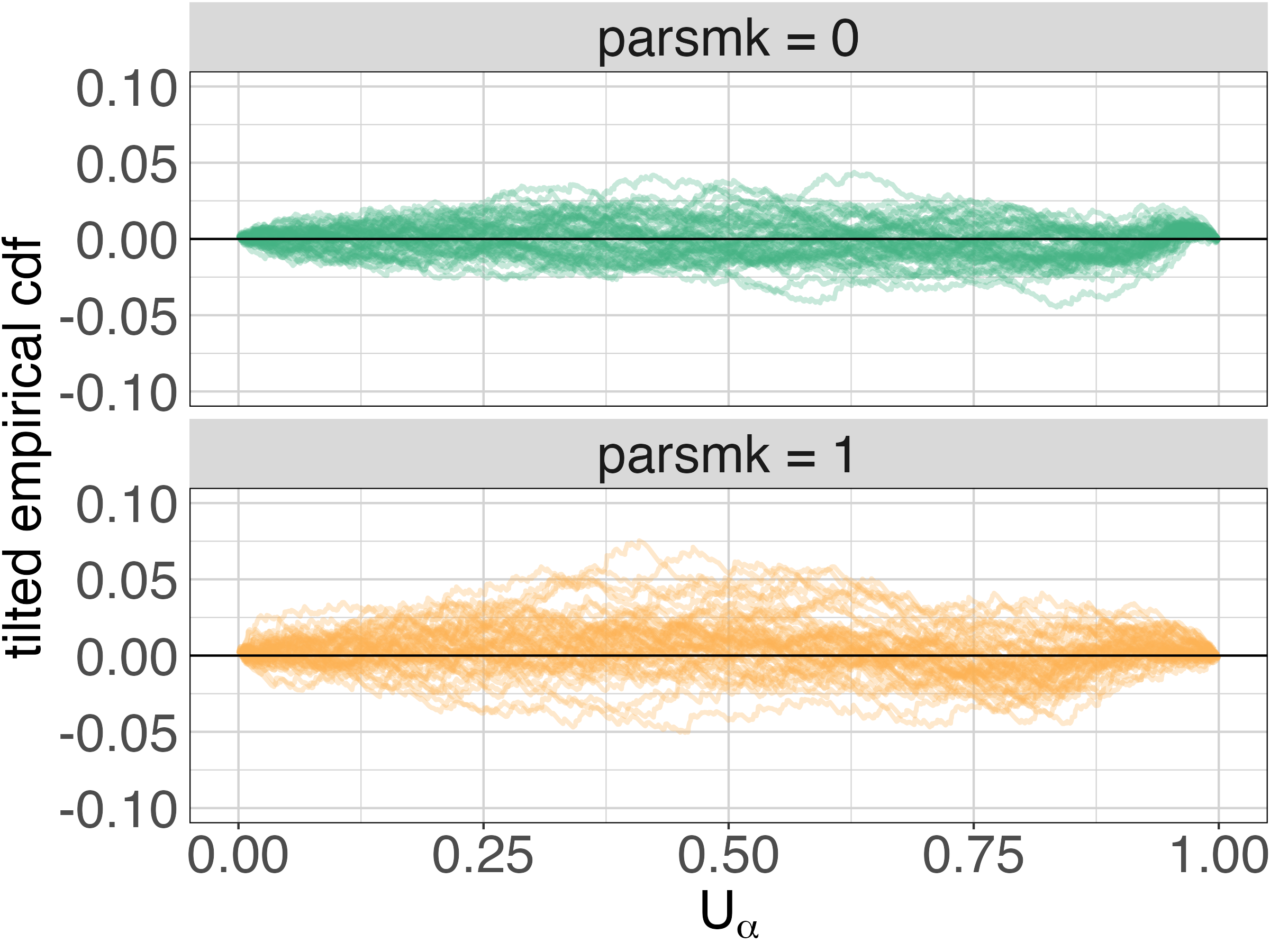}
    \vspace{0.5em}
    \includegraphics[width=0.49\textwidth]{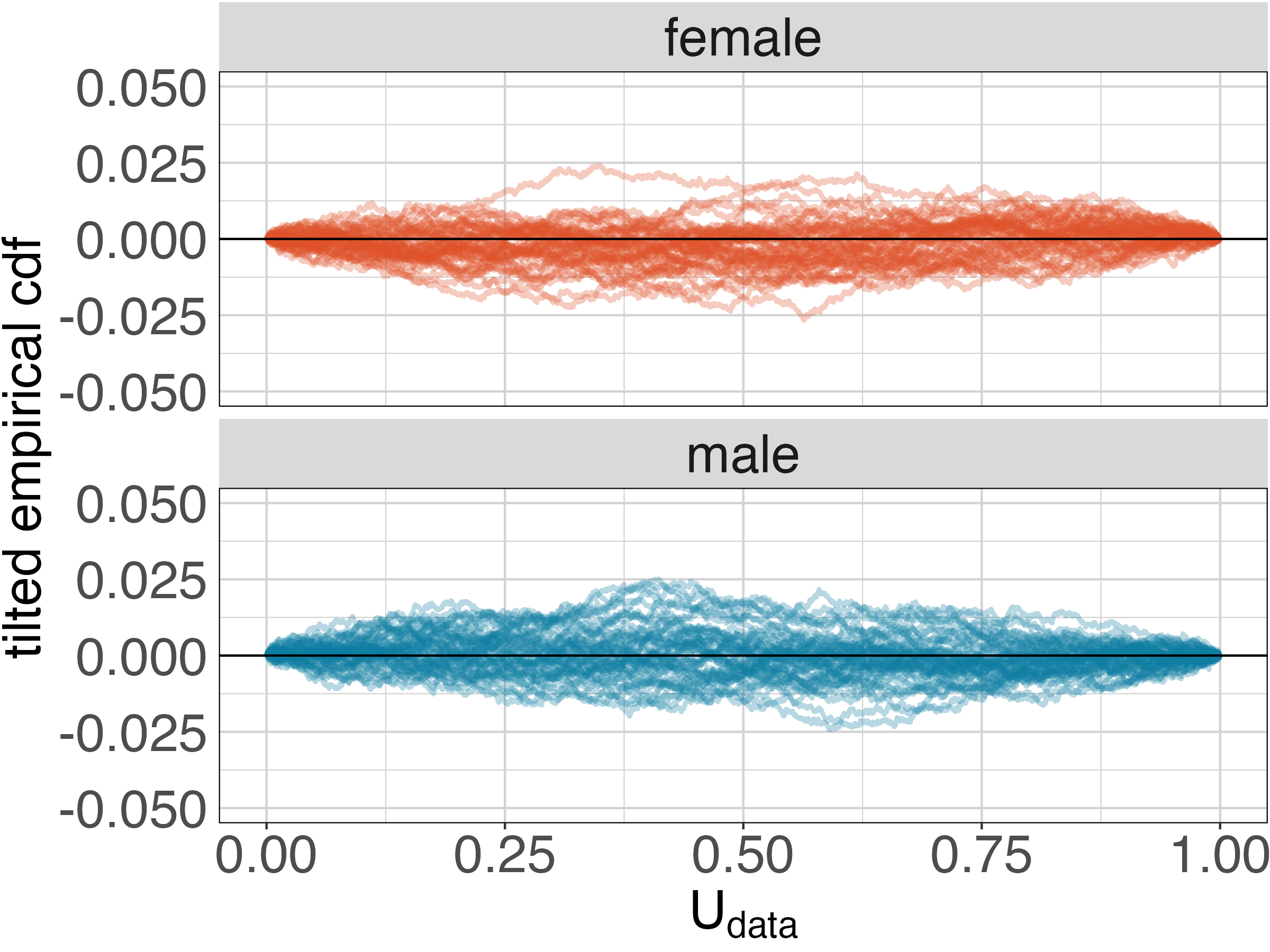}
    \hfill
    \includegraphics[width=0.49\textwidth]{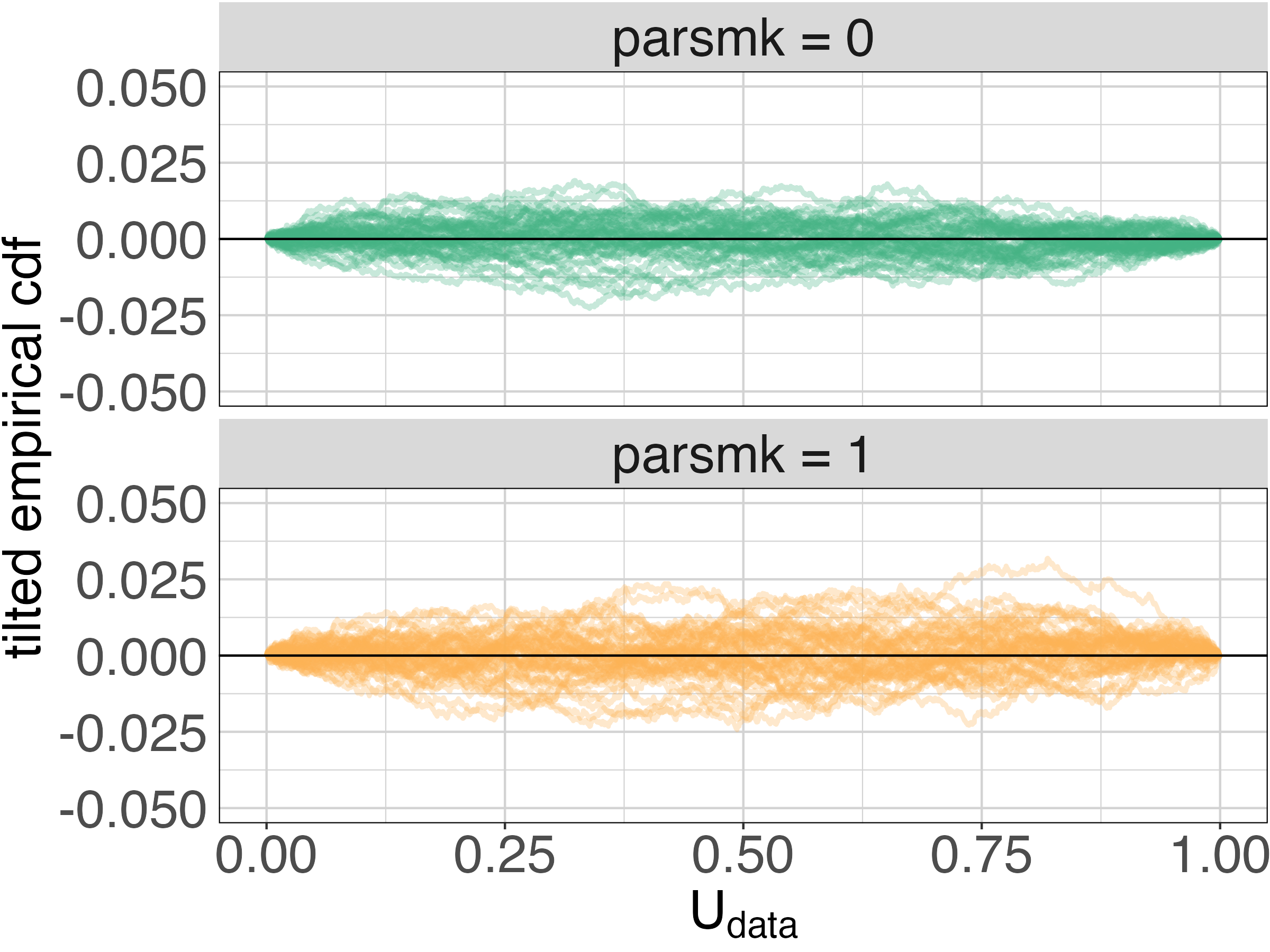}
    \vfill
    \centering
    \includegraphics[width=0.65\textwidth]{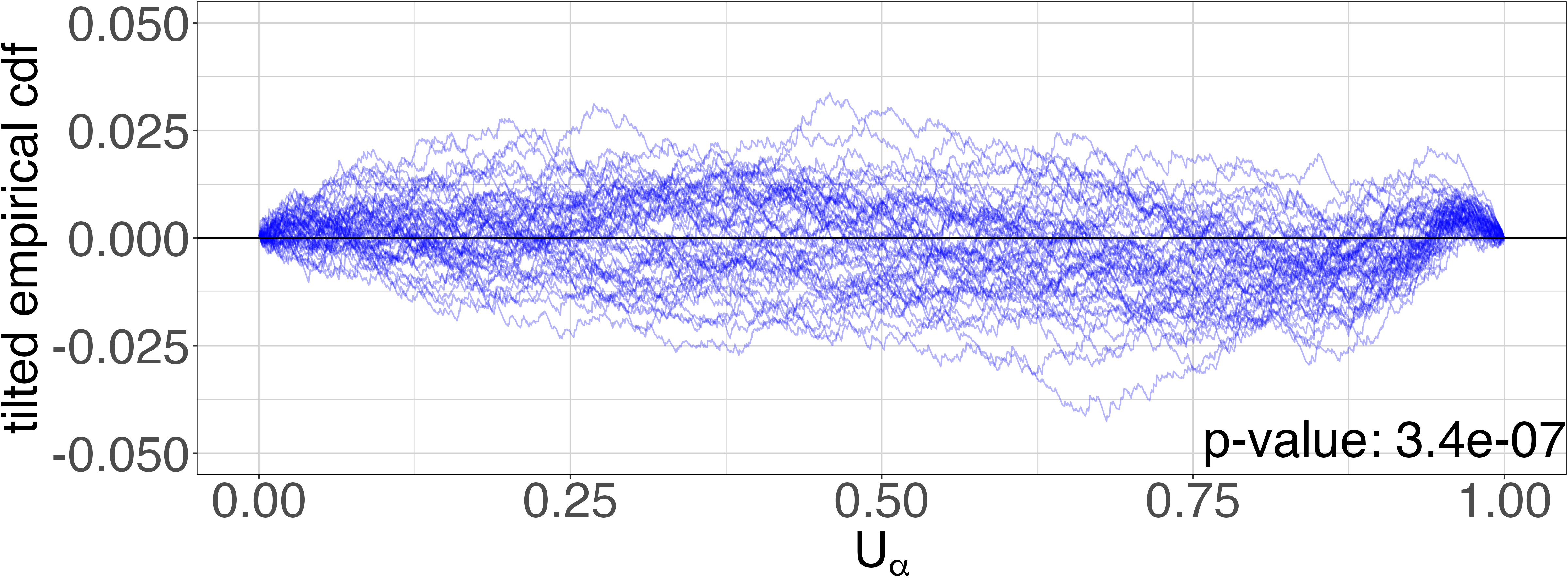}
    \vspace{0.5em}
    \includegraphics[width=0.65\textwidth]{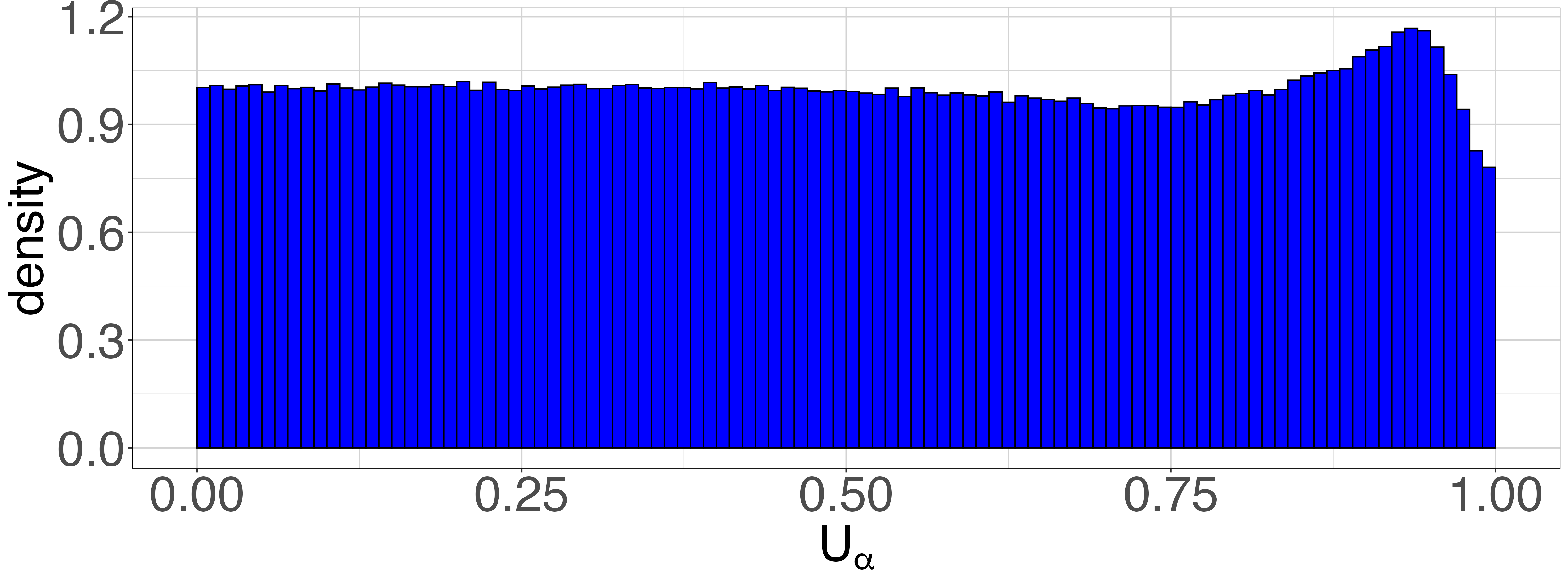}
    
\end{minipage}

    \caption{UPC results for Model \#2 on the logistic regression example. (Left) Tilted empirical CDFs of the data u-values, stratified by $\texttt{wave}\times\texttt{parsmk}$.  (Top right) Same as the corresponding plots in \cref{fig:smoking_1}, but for Model \#2. (Bottom right) Tilted empirical CDFs of the $\alpha$ u-values for multiple posterior samples, and a density histogram of the $\alpha$ u-values over all posterior samples.} 

    \label{fig:smoking_2}
\end{figure}

\paragraph{Model \#2.}
Based on the Model \#1 results, we augment the model to include \texttt{wave} and \texttt{parsmk}:
\begin{align*}
    (Y_{j k}\mid \beta,\alpha) &\sim \mathrm{Bernoulli}\big(\mathrm{logit}^{-1}(\alpha_j + \beta_1\texttt{wave}_{j k} + \beta_2\texttt{parsmk}_{j k})\big), \\
    \beta_1,\beta_2 &\sim \mathcal{N}(0,5^2), \\
    \alpha_j &\sim \mathcal{N}(\mu,5^2), \\
    \mu&\sim \mathcal{N}(0,5^2)
\end{align*}
for all $j,k$.
As before, we use JAGS to draw $100{,}000$ posterior samples after $5{,}000$ burn-in iterations, we thin to $1{,}000$ samples, and we compute the u-values.
To illustrate the effectiveness of the UPC method, we first verify that dependence with \texttt{wave} and \texttt{parsmk} is no longer detected. Indeed, running the same tests as before yields
$p_{\mathrm{data},\texttt{wave}}^* \approx 1$,
$p_{\mathrm{data},\texttt{sex}}^* = 0.55$,
$p_{\mathrm{data},\texttt{parsmk}}^* \approx 1$,
$p_{\alpha,\texttt{sex}}^* = 0.75$, and
$p_{\alpha,\texttt{parsmk}}^* = 0.20$.
\cref{fig:smoking_2} provides visual confirmation of these results.

To explore possible deficiencies in Model \#2, we consider two tests: (1) we test for dependence between the data u-values and $\texttt{wave}\times\texttt{parsmk}$ using Hoeffding's test, to assess whether this interaction is needed, and (2) we test for uniformity of the $\alpha$ u-values using Anderson--Darling, to evaluate the choice of prior on the $\alpha$ parameters.
These tests yield aggregated p-values of $p_{\mathrm{data},\texttt{wave}\times\texttt{parsmk}}^* \approx 1$ and $p_{\alpha,\mathrm{unif}}^* = 3.41 \times 10^{-7}$. 
This suggests that there is no need to add the interaction to the model, but that the prior on the $\alpha$ values is defective in some way.

To visually confirm and understand these findings, \cref{fig:smoking_2} shows posterior samples of tilted empirical CDF of (left) the data u-values, stratified by $\texttt{wave}\times\texttt{parsmk}$, and (bottom right) the $\alpha$ u-values, as well as a density histogram of the $\alpha$ u-values across posterior samples.
There is little variation across values of $\texttt{wave}\times\texttt{parsmk}$, which explains the large p-value for the interaction test.  Meanwhile, the $\alpha$ u-values tend to be approximately uniform except for a medium-sized bump at around $0.9$.
This suggests that there may be some additional latent structure that is not accounted for in the measured covariates.

\paragraph{Model \#3.}

\begin{figure}
\centering 

\includegraphics[width=0.45\textwidth]{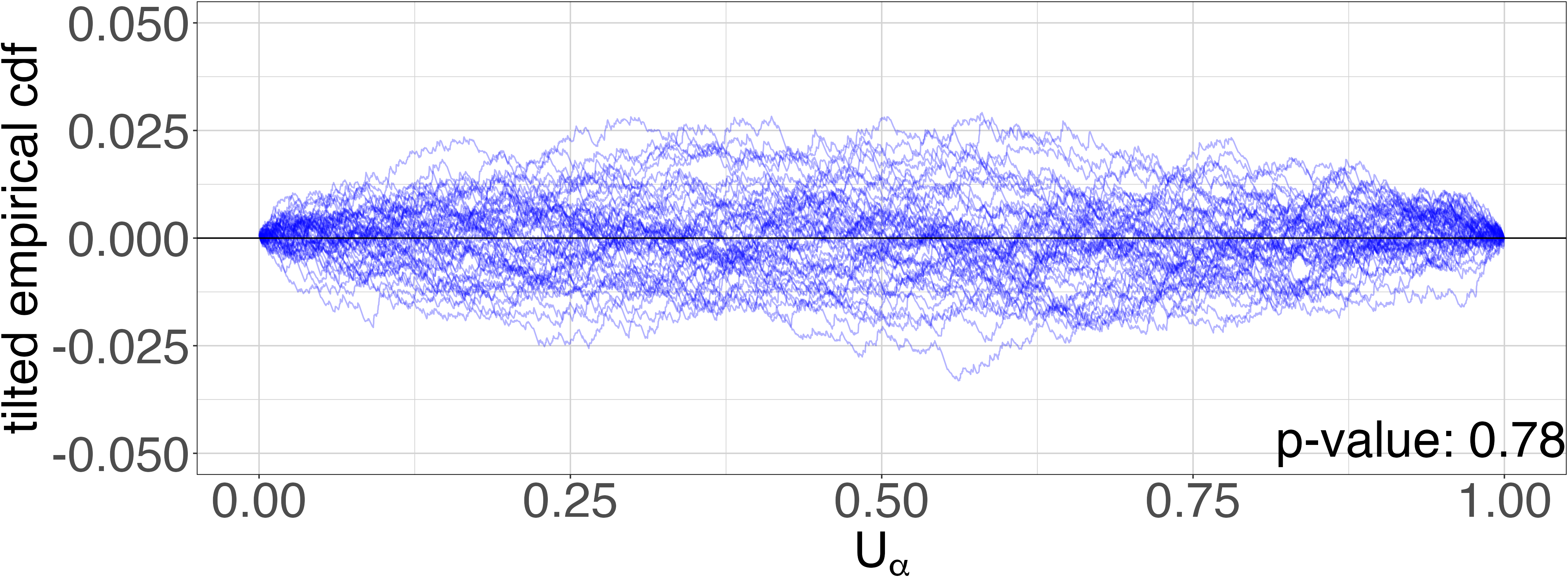}
%
\hfill
\includegraphics[width=0.45\textwidth]{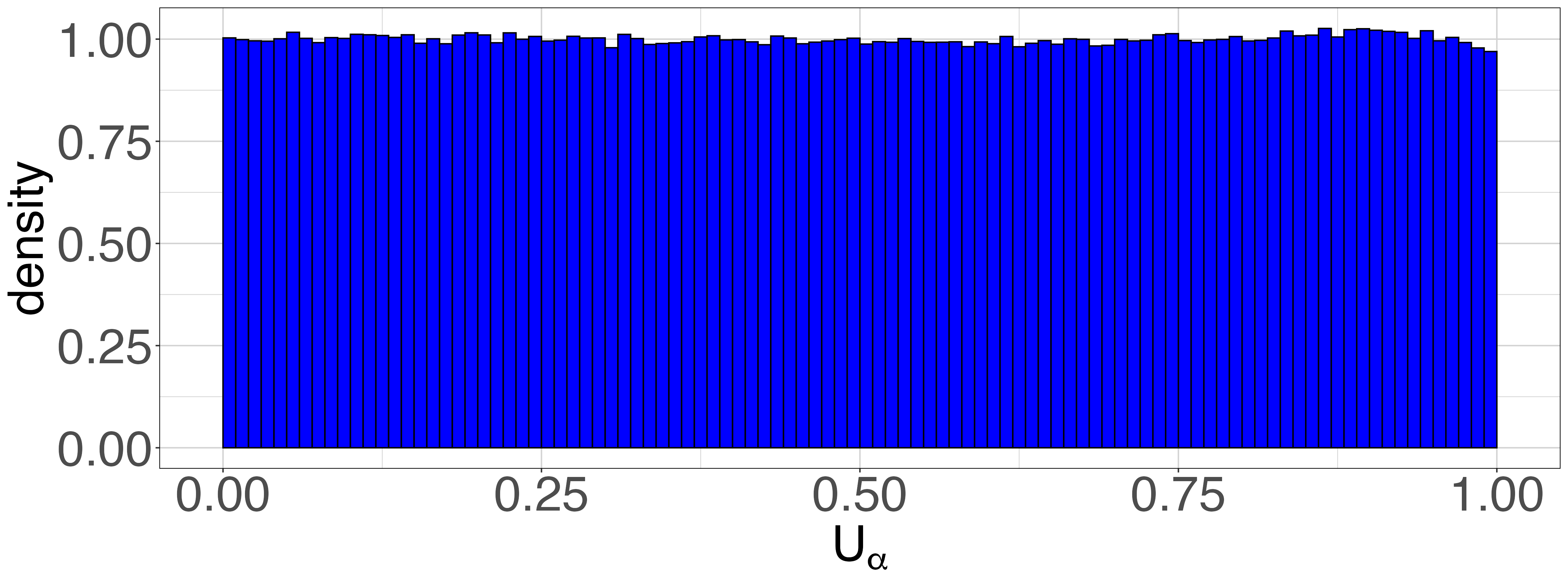}

\hfill

\caption{UPC results for Model \#3 on the logistic regression example.  (Left) Tilted empirical CDFs of the $\alpha$ u-values for multiple posterior samples. (Right) Histogram of the $\alpha$ u-values over all posterior samples. 
}
\label{fig:smoking_3}
\end{figure}

To account for the non-uniformity of the $\alpha$ u-values in Model \#2, we augment the model to employ a two-component mixture model for the prior on the $\alpha$ values, as follows:
\begin{align*}
    (Y_{j k}\mid\beta,\alpha) &\sim \mathrm{Bernoulli}\big(\mathrm{logit}^{-1}(\alpha_j + \beta_1\texttt{wave}_{j k} + \beta_2\texttt{parsmk}_{j k})\big), \\
    \beta_1,\beta_2 &\sim \mathcal{N}(0,5^2), \\
    (\alpha_j\mid \mu,\tau,Z) &\sim \mathcal{N}(\mu_{Z_j},\tau_{Z_j}^{-1}), \\
    (Z_j\mid\pi) &\sim \mathrm{Bernoulli}(\pi), \\
    \pi &\sim \mathrm{Beta}(1,1), \\
    \mu_1,\mu_2 &\sim \mathcal{N}(0,5^2), \\
    \tau_1,\tau_2 &\sim \mathrm{Gamma}(1, 0.2)
\end{align*}
for all $j,k$, where $\mathrm{Gamma}(a,b)$ denotes the gamma distribution with shape $a$ and rate $b$.  Again, we use JAGS to draw $100{,}000$ posterior samples after $5{,}000$ burn-in iterations, we thin to $1{,}000$ samples, and we compute the u-values.
Again, to demonstrate the effectiveness of UPCs, we verify that this addresses the non-uniformity of the $\alpha$ u-values that was seen in Model \#2.  Computing the same tests as done for Model \#2, we obtain $p_{\mathrm{data},\texttt{wave}\times\texttt{parsmk}}^* \approx 1$ and $p_{\alpha,\mathrm{unif}}^* = 0.78$ for Model \#3, which suggests that the inadequacy of the $\alpha$ prior in Model \#2 has been sufficiently addressed.
\cref{fig:smoking_3} provides visual confirmation that the $\alpha$ u-values do indeed appear to be close to uniform in Model \#3.  
In fact, for Model \#3, the null of model correctness is not rejected under any of the UPC tests considered in this section, indicating that this model is at least providing a reasonable fit to these data. 
Additionally, \cref{tab:smoking_3} shows that the mixture component assignments $Z_j$ have a clear interpretation, since nearly all individuals with $Z_j = 0$ never smoke regularly throughout the study.

\subsection{Autoregressive model - Simulation study}
\label{sec:AR1}

We present a simulation study using an autoregressive model.
In the other examples, we consider only the observed dataset and simulated datasets from the hypothesized model.
In this example, we simulate datasets from various perturbations of the hypothesized model in order to see the effect on the UPCs.
This demonstrates which UPCs are affected by different types of perturbations, and the power we have to detect each type of perturbation.
Suppose the hypothesized model is an autoregressive AR(1) model:
\begin{equation}
\begin{split}
    & Y_i = \phi Y_{i-1} + \sigma \epsilon_i, 
\label{ex:basic_AR1_model}\\
    & \phi \sim \pi(\phi), \quad \sigma\sim\pi(\sigma)
\end{split}
\end{equation}
for $i = 1,\ldots,n$, with $Y_0 = 0$ and $\epsilon_1,\ldots,\epsilon_n$ i.i.d.\ $\sim \mathcal{N}(0,1)$.
We will consider various choices for the priors $\pi(\phi)$, $\pi(\sigma)$
and for the true data generating process (DGP).  
In each scenario, we simulate $100{,}000$ datasets from the true DGP with $n = 500$, and for each dataset, we use Stan~\citep{CGH+17} to draw one sample from the posterior after 1{,}000 burn-in iterations.

To implement UPCs, we reparametrize in terms of u-values via $\phi = F_\phi^{-1}(U_1)$, $\sigma = F_\sigma^{-1}(U_2)$, and $Y_i = \phi Y_{i-1} + \sigma\Phi^{-1}(U_{d_i})$, where $d_i = i+2$.
For each posterior sample of $\phi$ and $\sigma$, we compute the u-values via
$\widetilde{U}_1 = F_\phi(\phi)$, 
$\widetilde{U}_2 = F_\sigma(\sigma)$, and 
$\widetilde{U}_{d_i} = \Phi((Y_i - \phi Y_{i-1})/\sigma)$. 
Consider the following UPCs:
(1) test for extreme values of $\phi$ and $\sigma$ via the p-values $p_\phi = 2\min\{\widetilde{U}_1, 1 - \widetilde{U}_1\}$ and $p_\sigma = 2\min\{\widetilde{U}_2, 1 - \widetilde{U}_2\}$,
(2) test for non-uniformity of the data u-values using Anderson--Darling, yielding $p_{\mathrm{data,unif}}$,
(3) test for dependence between the data u-values $\tilde{U}_{d_i}$ and the index $i$ (for $i=1,\ldots,n$) using the Hoeffding independence test, yielding $p_{\mathrm{data,index}}$,
(4) test for first-order serial correlation by using a Hoeffding independence test between $\tilde{U}_{d_i}$ and $\tilde{U}_{d_i+1}$ (for $i=1,\ldots,n-1)$, yielding $p_{\mathrm{data,lag1}}$, and
(5) test for second-order serial correlation by using a Hoeffding independence test between $\tilde{U}_{d_i}$ and $\tilde{U}_{d_i+2}$ (for $i=1,\ldots,n-2$), yielding $p_{\mathrm{data,lag2}}$.

\begin{figure}
    \centering
    \includegraphics[width=\textwidth]{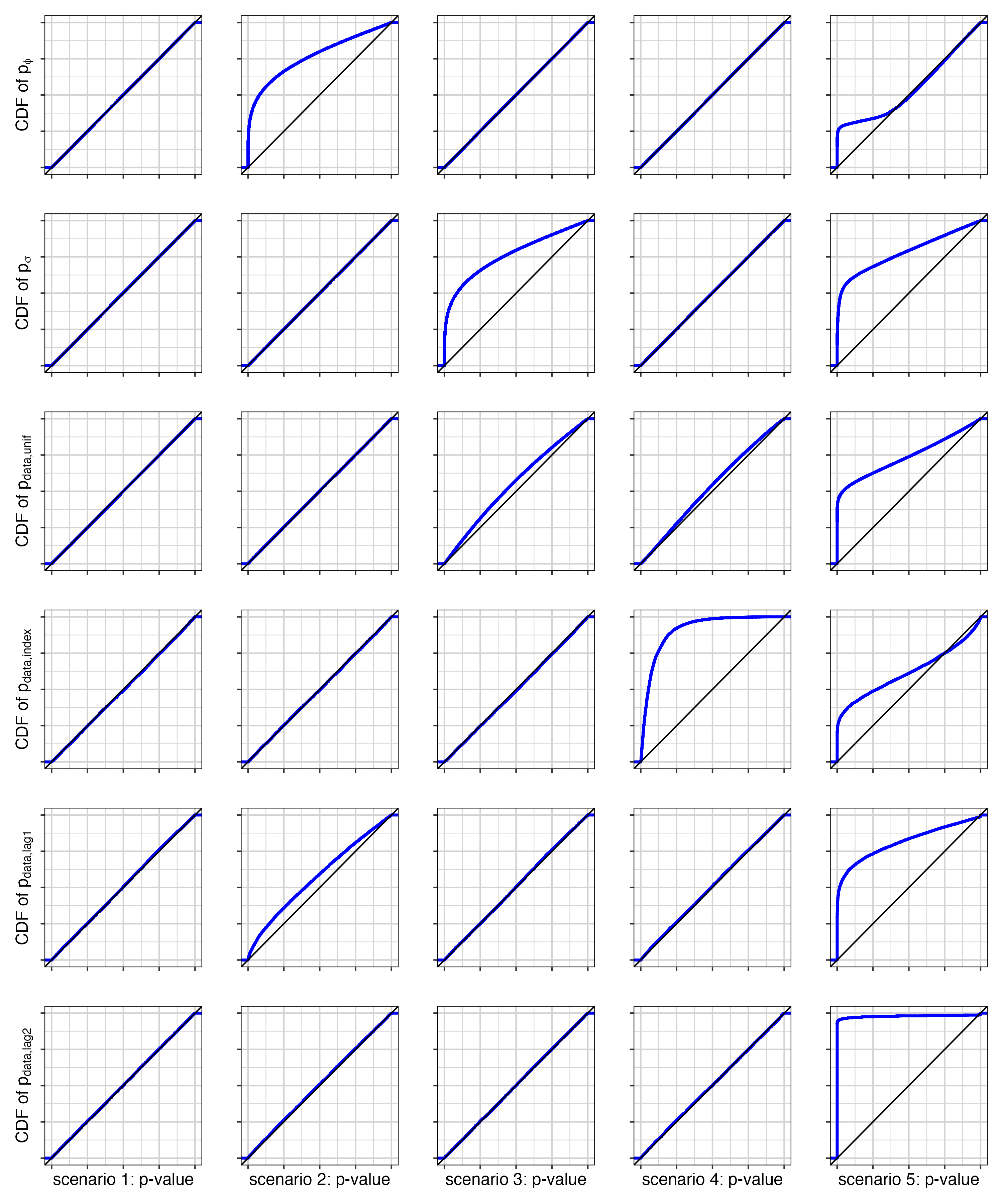}

    \caption{Expected CDFs of p-values of each test in the AR simulation study scenarios, where the expectation is taken with respect to datasets drawn from the true DGP: Scenario 1 (correct model), Scenario 2 (narrow prior for $\phi$),  Scenario 3 (narrow prior for $\sigma$), Scenario 4 (heteroskedastic errors), and Scenario 5 (second-order dependence).}
    \label{fig:AR_omnibus_ecdf}
\end{figure}

\paragraph{Scenario \#1: Correct model.}
As a sanity check, we begin with the case where the hypothesized model is identical to the true DGP.
Let $\mathcal{TN}(\mu,\sigma^2,[a,b])$ denote the truncated normal distribution obtained by restricting the support of $\mathcal{N}(\mu,\sigma^2)$ to the interval $[a,b]$.
Suppose that under both the true DGP and the hypothesized model, a dataset is generated by drawing parameters
$\phi\sim\mathcal{TN}(0, 0.4^2, [-0.5, 0.5])$ and $\sigma\sim\mathcal{TN}(1.5, 0.4^2, [1, 2])$ independently, and generating $Y_1,\ldots,Y_n$ as in \cref{ex:basic_AR1_model}.
\cref{fig:AR_omnibus_ecdf} shows the expected posterior CDFs of $p_\phi$, $p_\sigma$, $p_{\mathrm{data,unif}}$, $p_{\mathrm{data,index}}$, $p_{\mathrm{data,lag1}}$, and $p_{\mathrm{data,lag2}}$, where by ``expected'' we are referring to averaging over datasets drawn from the true DGP; this is called the ``data-averaged posterior'' by \citet{talts2018validating}.
In \cref{fig:AR_omnibus_ecdf}, we see that the p-values are all uniformly distributed, as expected based on our theory.

\paragraph{Scenario \#2: Narrow prior for $\phi$.}
Next, we consider a case where the hypothesized likelihood is correct but the prior on $\phi$ is too concentrated.
Suppose the true DGP and hypothesized model are the same as in Scenario \#1, except that $\phi\sim\mathcal{TN}(0, 0.1^2, [-0.5, 0.5])$ under the hypothesized model.
In \cref{fig:AR_omnibus_ecdf}, we see that $p_\phi$ is relatively concentrated near zero. Thus, the UPCs correctly detect that the $\phi$ values needed to explain the data are extreme, relative to the hypothesized model prior.
Meanwhile, the rest of the p-values are approximately uniform, except that $p_{\mathrm{data,lag1}}$ exhibits a slight departure from uniformity. This is appealing since the prior on $\phi$ is in fact the only aspect of the model that does not match the true DGP.

\paragraph{Scenario \#3: Narrow prior for $\sigma$.}
Now, we consider a case where the hypothesized prior on $\sigma$ is too concentrated.  Again, suppose the same true DGP and hypothesized model as in  Scenario \#1, except that  $\sigma\sim\mathcal{TN}(1.5,0.1^2,[1,2])$ under the hypothesized model.
Thus, in this case the prior on $\sigma$ is the one that does not match the true DGP.
As one might hope, the UPCs correctly detect extreme values of $\sigma$ relative to the prior, and detect no other serious issues.

\paragraph{Scenario \#4: Heteroskedastic errors.}
In this case, we suppose the true DGP and hypothesized model are the same as in Scenario \#1, except under the true DGP the data are generated according to $Y_i = \phi Y_{i-1} + \sqrt{c_i} \sigma \varepsilon_i$ where $c_i = 1 + (2 i - n - 1)/n$ for $i=1,\ldots,n$. Meanwhile, the hypothesized model still uses \cref{ex:basic_AR1_model}.
Thus, the true DGP exhibits heteroskedasticity, whereas the model assumes homoskedasticity.
In \cref{fig:AR_omnibus_ecdf}, we see extreme non-uniformity in $p_{\mathrm{data,index}}$, correctly reflecting that the model does not capture the index dependence in the data. Again, as desired, the UPCs detect no other serious issues.

\paragraph{Scenario \#5: Second-order dependence.}
Finally, we suppose the true DGP and hypothesized model are again the same as in Scenario \#1, except the true DGP follows an AR(2) structure with $Y_i = \phi Y_{i-1} + \mathrm{sgn}(\phi)|\phi|^{1/4} Y_{i-2} + \sigma \varepsilon_i$.
Thus, the true DGP exhibits second-order dependence, whereas the hypothesized model assumes only first-order dependence.
In \cref{fig:AR_omnibus_ecdf}, we see that $p_{\mathrm{data,lag2}}$ is highly concentrated near zero,  providing a clear indication that there is an issue pertaining to higher-order dependence.
Interestingly, this perturbation of the model has a more global impact than in the other scenarios, affecting the other tests as well. None of the other p-value distributions as extreme as $p_{\mathrm{data,lag2}}$, but all of them put significant mass on p-values very near zero.
\section{Discussion}
\label{section:discussion}

Uniform parametrization checks (UPCs) provide a 
general-purpose technique for Bayesian model criticism.
As demonstrated in the examples, UPCs are easy-to-use and provide insight into which aspects of a model are misspecified.
As shown in the theory, UPCs are guaranteed to yield valid p-values under the null of model correctness and---in many cases---the u-values are uniform if and only if the model is correct.
Compared to posterior predictive checks (PPCs), a key advantage of UPCs is that there is a default set of UPC tests that can be used on any model, rather than having to design PPC test quantities in a model-specific way.

There are several interesting directions for future work on UPCs.
First, as observed in the autoregressive model example (\cref{sec:AR1}), misspecification of one aspect of a model can affect UPCs pertaining to other aspects of the model.
A more complete theory is needed for understanding the relationship between misspecification of each part of a model and the resulting posterior distribution of u-values of other parts.
Relatedly, we have observed that some parametrizations are preferable, in that the effect of perturbing one part of a model is more isolated to the corresponding subset of u-values.
Characterizing the role of model parametrization in this correspondence would make it possible to implement models in a way that is conducive to model criticism.
Another important direction is to develop a standard ``best practices'' workflow for iterative model improvement, generalizing and formalizing the process illustrated in the logistic regression example (\cref{sec:logistic}).
Finally, it will be interesting to use the UPC methodology in other models and employ it in practical applications.

\section*{Acknowledgments}
\label{section:acknowledgments}

We would like to thank Aki Vehtari, Andrew Gelman, Yuling Yao, David Dunson, and Ryan Giordano for helpful comments.
C.T.C.\ was supported by National Institutes of Health (NIH) Training Grant T32CA09337. 
J.W.M.\ was supported in part by the National Cancer Institute of the NIH under award number R01CA240299. 
The content is solely the responsibility of the authors and does not necessarily represent the official views of the National Institutes of Health.

\bibliographystyle{chicago}
\bibliography{bib}


\newpage
\setcounter{page}{1}
\setcounter{section}{0}
\setcounter{table}{0}
\setcounter{figure}{0}
\renewcommand{\theHsection}{SIsection.\arabic{section}}
\renewcommand{\theHtable}{SItable.\arabic{section}.\arabic{table}}
\renewcommand{\theHfigure}{SIfigure.\arabic{section}.\arabic{figure}}
\renewcommand{\thepage}{S\arabic{page}}  
\renewcommand{\thesection}{S\arabic{section}}   
\renewcommand{\thetable}{S\arabic{section}.\arabic{table}}   
\renewcommand{\thefigure}{S\arabic{section}.\arabic{figure}}

\begin{center}
{\LARGE Supplementary material for \\``Bayesian model criticism using uniform parametrization checks''}
\end{center}

\label{section:appendix}

\section{Proofs}
\label{subsection:appendix:theory}

\begin{proof}[\bf Proof of \cref{theorem:self-consistency}]
Suppose $Y^0\stackrel{d}{=} Y$.  Since $\widetilde{\bm{\theta}}\mid Y^0$ and $\widetilde{U}\mid \widetilde{\bm{\theta}},Y^0$ are defined according to the conditional distributions $\Pi_{\theta|y}$ and $\Pi_{u|\theta,y}$ under the hypothesized model, it follows that $(\widetilde{U},\widetilde{\bm{\theta}},Y^0) \stackrel{d}{=} (U,\bm{\theta},Y)$, where $(U,\bm{\theta},Y)$ is distributed according to the hypothesized model.
In particular, $\widetilde{\bm{\theta}} \stackrel{d}{=} \bm{\theta}$ and $\widetilde{U} \stackrel{d}{=} U$.
\end{proof}

\begin{proof}[\bf Proof of \cref{theorem:partial-converse}]
Suppose $\widetilde{U} \stackrel{d}{=} U$ and $g$ is a bijection.
Recall that by definition, $\widetilde{U}$ is distributed according to $\Pi_{u|\theta,y}$ given $\widetilde{\bm{\theta}} = \theta$ and $Y^0 = y$, where $\Pi_{u|\theta,y}$ is the conditional distribution under the hypothesized model. Since $(\bm{\theta},Y) = g(U)$ under the hypothesized model and $g$ is a bijection, we have
\begin{align}\label{eq:partial-converse1}
    U = g^{-1}(\bm{\theta},Y).    
\end{align}
Since the conditional distribution of $\widetilde{U}\mid\widetilde{\bm{\theta}},Y^0$ is defined according to the hypothesized model, we have 
\begin{align}\label{eq:partial-converse2}
    \widetilde{U} \stackrel{d}{=} g^{-1}(\widetilde{\bm{\theta}},Y^0).
\end{align}
Since $\widetilde{U} \stackrel{d}{=} U$ and $g$ is a bijection, \cref{eq:partial-converse1,eq:partial-converse2} imply that $(\bm{\theta},Y) \stackrel{d}{=} (\widetilde{\bm{\theta}},Y^0)$.
In particular, $Y \stackrel{d}{=} Y^0$.
\end{proof}

\begin{proof}[\bf Proof of \cref{theorem:external}]
This is a simple consequence of the assumed independence relations.  We claim that for any random elements $X$, $Y$, and $Z$, if $X\indep Y$ and $Z\indep X\mid Y$ then $Z\indep X$.  To see this, observe that for any bounded, measurable real-valued functions $f$ and $g$, 
\begin{align*}
    \E\big(f(Z)g(X)\big) &= \E\Big(\E\big(f(Z)g(X)\mid X,Y\big)\Big)
    = \E\Big(g(X)\E\big(f(Z)\mid X,Y\big)\Big) \\
    &= \E\Big(g(X)\E\big(f(Z)\mid Y\big)\Big)
    = \E\big(g(X)\big)\E\big(\E(f(Z)\mid Y)\big) \\
    &= \E(g(X))\E(f(Z)).
\end{align*}
Therefore, $Z\indep X$.  The result follows by applying this with $(\widetilde{U},\widetilde{\bm{\theta}})$ and $Y^0$ playing the role of $Z$ and $Y$.
\end{proof}

\begin{proof}[\bf Proof of \cref{theorem:identifiability}]
(i) Let $u_{1:K},u'_{1:K}\in(0,1)^K$ such that $u_{1:K}\neq u'_{1:K}$, that is, $u_k \neq u'_k$ for some $k$.  Then, letting $\theta = g_{\mathrm{p}}(u_{1:K})$ and $\theta' = g_{\mathrm{p}}(u'_{1:K})$, we have $\theta\neq\theta'$ since $g_{\mathrm{p}}$ is one-to-one.
Hence, $P_\theta \neq P_{\theta'}$.  This establishes identifiability of $U_{1:K}$.
(ii) Let $u,u'\in(0,1)^D$ such that $u\neq u'$.  Letting $(\theta,y) = g(u)$ and $(\theta',y') = g(u')$, we have that either $\theta\neq\theta'$ or $y\neq y'$, since $g$ is one-to-one.
If $\theta\neq\theta'$ then $P_\theta \neq P_{\theta'}$.
Otherwise, $y\neq y'$.
This proves the claim.
\end{proof}
\section{Additional details for the normal model example}
\label{appendix:subsection:example_1}

This section contains additional empirical results and implementation details for the example from \cref{sec:newcomb} on the normal model for Newcomb's speed of light data.

\subsection{Additional empirical results}

\begin{figure}
    \begin{subfigure}[t]{0.5\textwidth}
        \centering
        \includegraphics[width=0.8\textwidth]{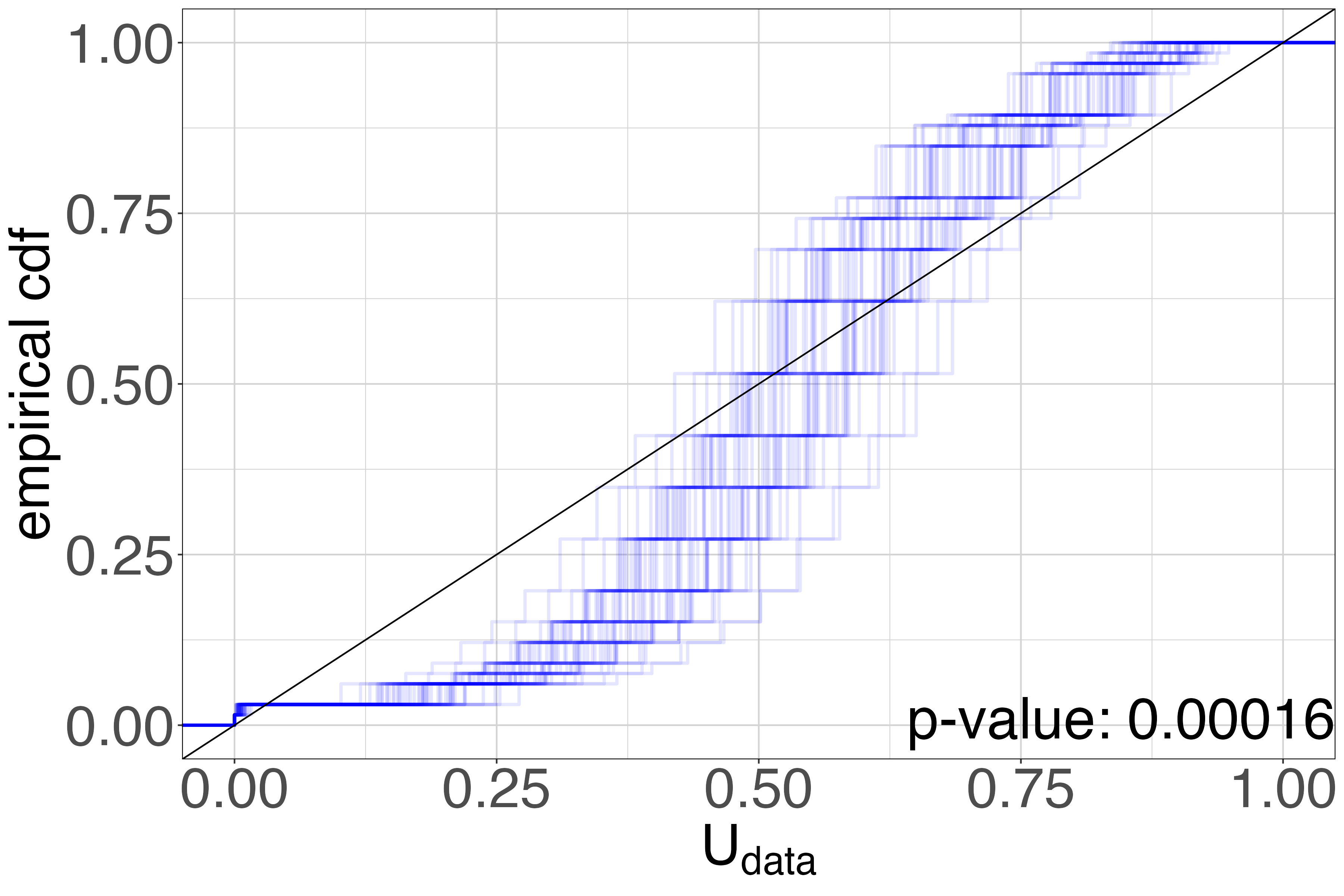}
        \caption{Newcomb data, Weakly informative prior}
        \label{fig:newcomb_resid_ecdf_empirical_prior}
    \end{subfigure}
    \hfill
    \begin{subfigure}[t]{0.5\textwidth}
        \centering
        \includegraphics[width=0.8\textwidth]{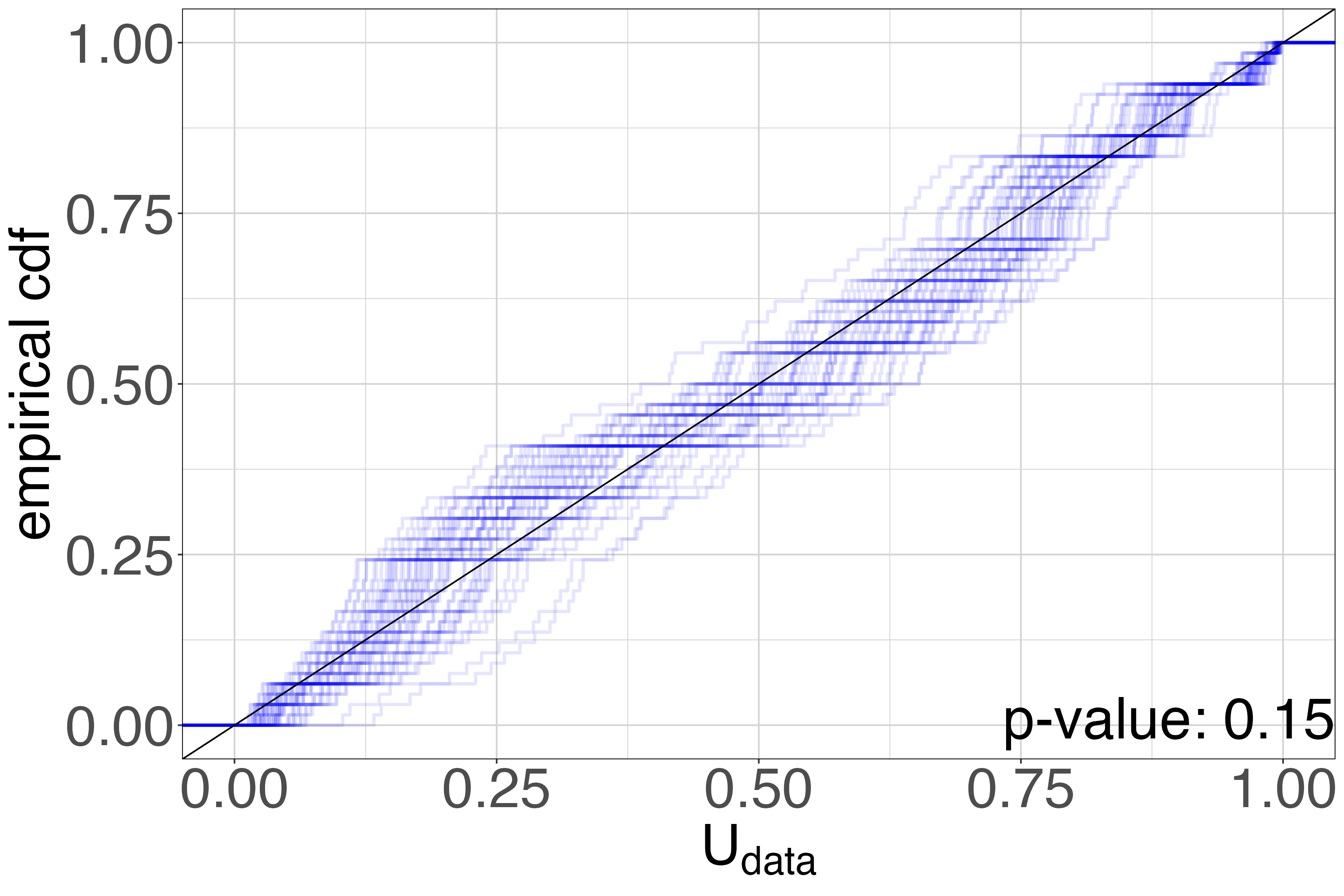}
        \caption{Normal data, Weakly informative prior}
        \label{fig:sim_resid_ecdf_empirical_prior}
    \end{subfigure}
    \par\vspace{2em}
    \begin{subfigure}[t]{0.5\textwidth}
        \centering
        \includegraphics[width=0.8\textwidth]{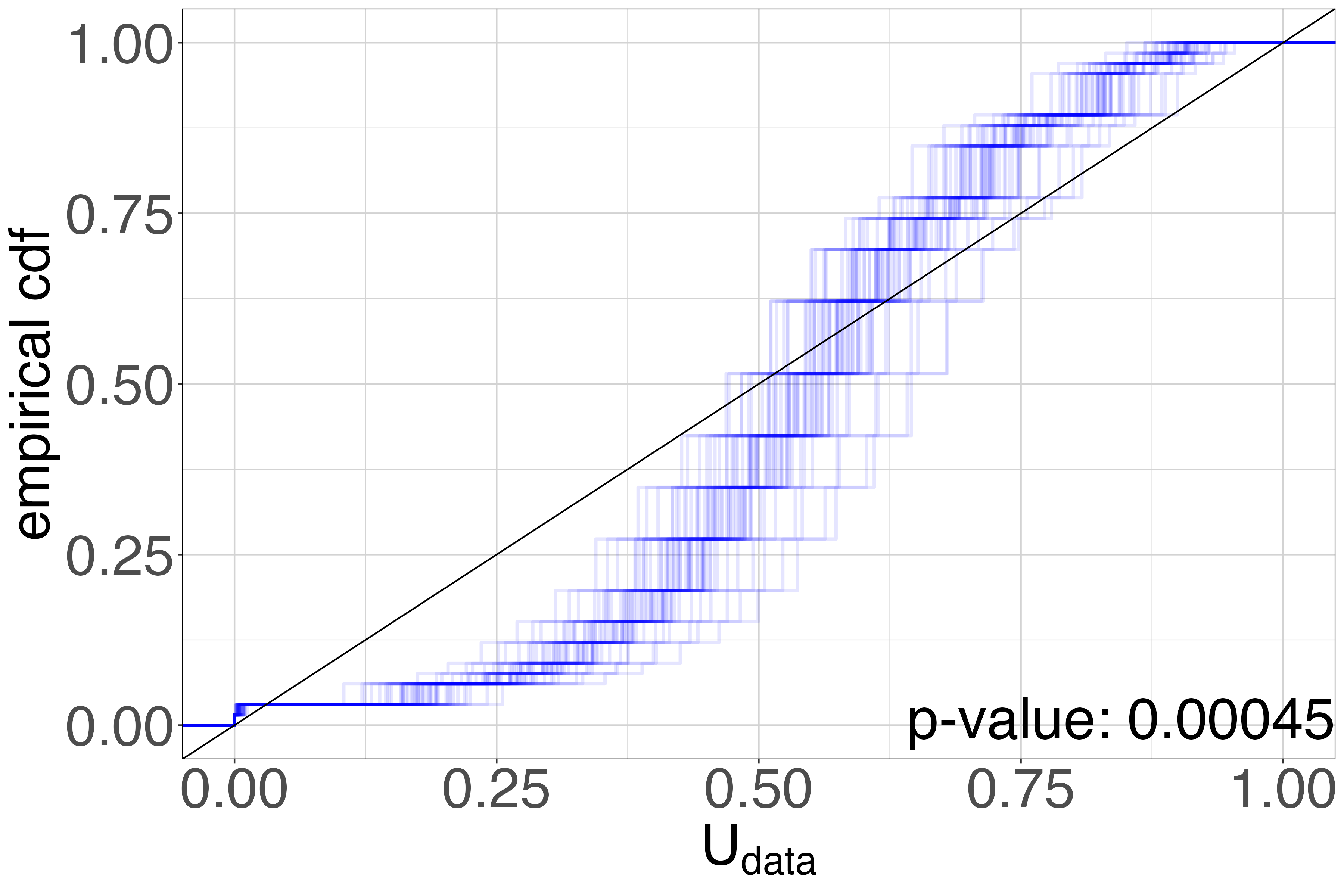}
        \caption{Newcomb data, Data-dependent prior}
        \label{fig:newcomb_resid_ecdf_empirical_prior}
    \end{subfigure}
    \hfill
    \begin{subfigure}[t]{0.5\textwidth}
        \centering
        \includegraphics[width=0.8\textwidth]{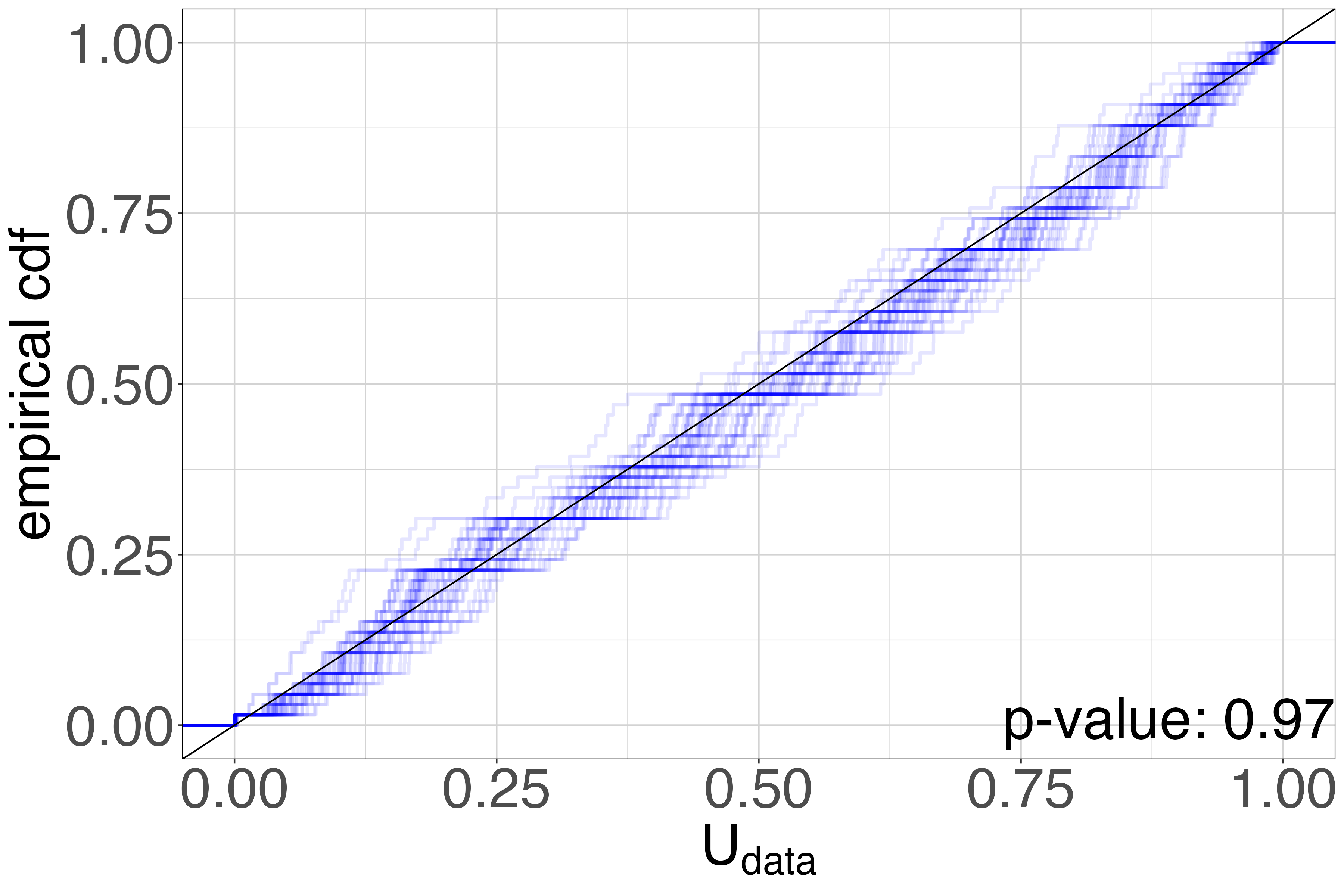}
        \caption{Normal data, Data-dependent prior}
        \label{fig:sim_resid_ecdf_empirical_prior}
    \end{subfigure}
    \par\vspace{2em}
    \begin{subfigure}[t]{0.5\textwidth}
        \centering
        \includegraphics[width=0.8\textwidth]{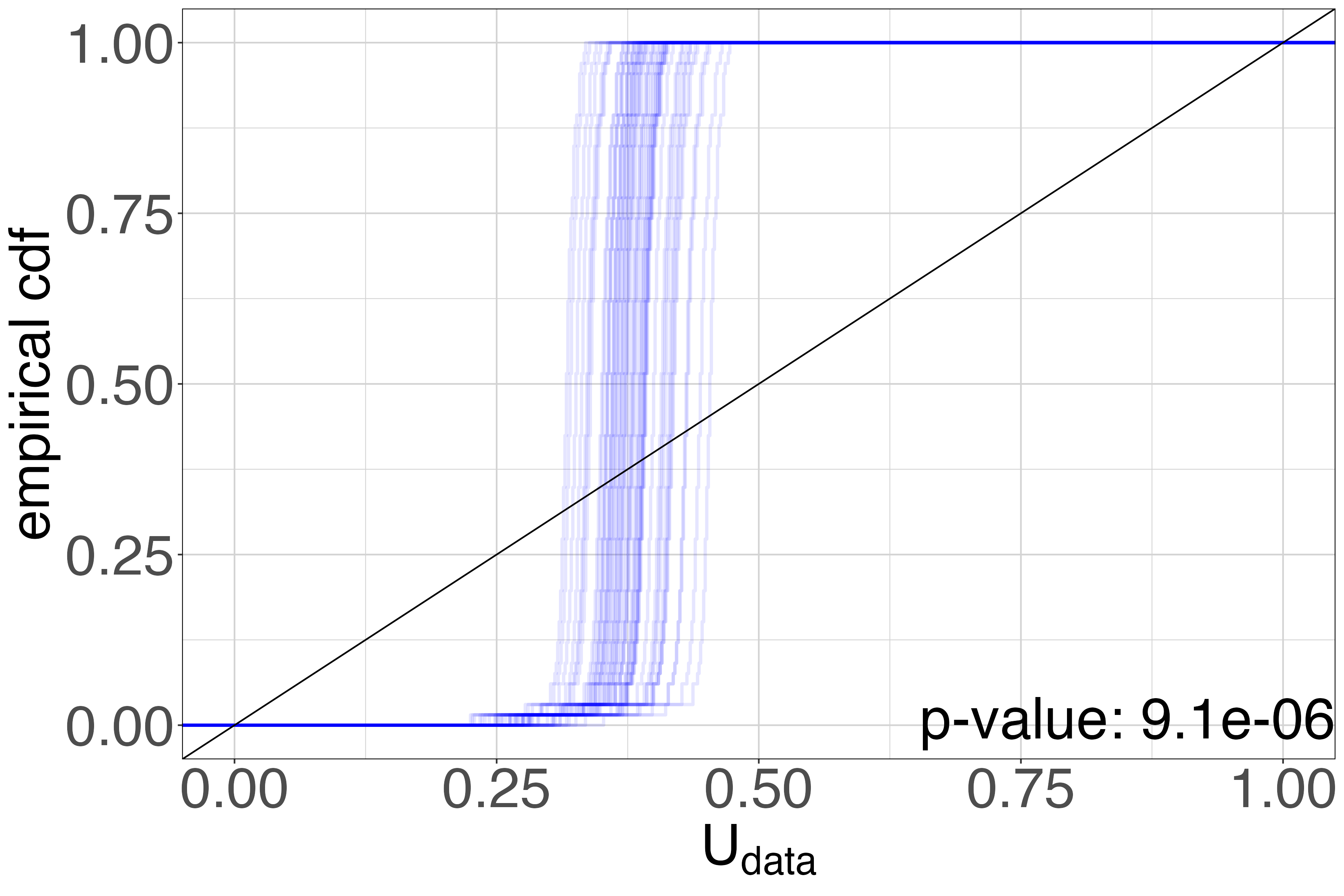}
        \caption{Newcomb data, Poorly chosen prior}
        \label{fig:newcomb_resid_ecdf_very_bad_prior}
    \end{subfigure}
    \hfill
    \begin{subfigure}[t]{0.5\textwidth}
        \centering
        \includegraphics[width=0.8\textwidth]{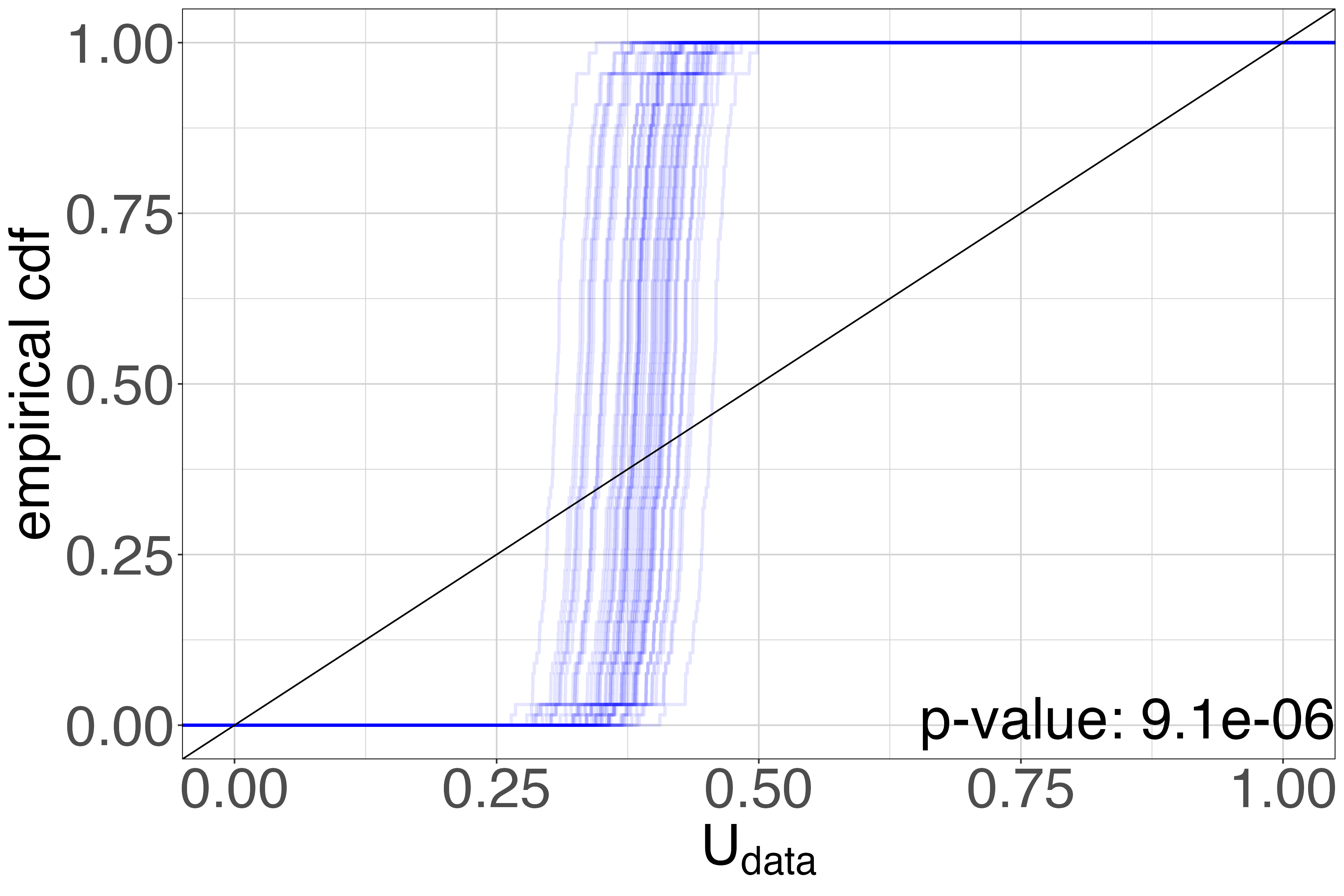}
        \caption{Normal data, Poorly chosen prior}
        \label{fig:sim_resid_ecdf_very_bad_prior}
    \end{subfigure}

    \caption{(Left) Empirical CDF $\hat{F}_\mathrm{d}$ of the data u-values $U_{d_1},\ldots,U_{d_n}$ from multiple posterior samples given the Newcomb data, when using the (a) weakly informative prior, (c) data-dependent prior, and (e) poorly chosen informative prior.  (Right) Same, but given a simulated dataset from a normal distribution with mean and variance matching the Newcomb data, when using the (b) weakly informative prior, (d) data-dependent prior, and (f) poorly chosen informative prior.}
    \label{fig:newcomb_vs_sim_ecdf_dap_residuals_empirical}
\end{figure}

In \cref{fig:newcomb_vs_sim_ecdf_dap_residuals_empirical}, we show the empirical CDF of the data u-values given (i) Newcomb's speed of light data and (ii) a simulated dataset from the normal distribution with mean and variance matching the sample mean and sample variance of the Newcomb data, under each of the three choices of prior considered.
For all three choices of prior, the data u-values for the Newcomb data are clearly non-uniform, which correctly reflects the model misspecification.
For the weakly informative and data-dependent priors, the data u-values for the normal data are approximately uniform.  Meanwhile, the poorly chosen prior has such a detrimental effect that it also makes the data u-values non-uniform even under normally distributed data.

\begin{figure}
    \centering

    \includegraphics[width=0.3 \textwidth]{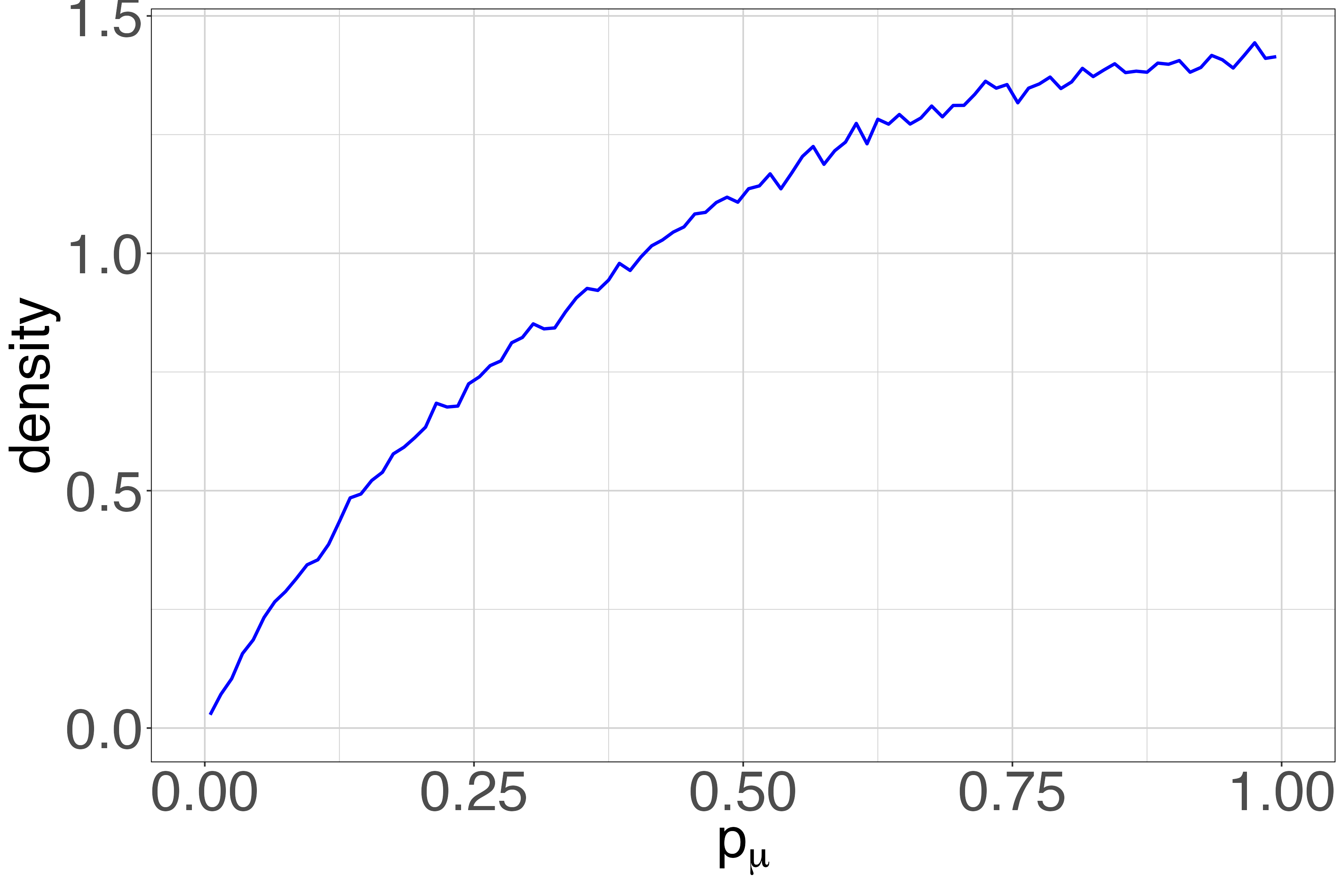}
    \hfill 
    \includegraphics[width=0.3 \textwidth]{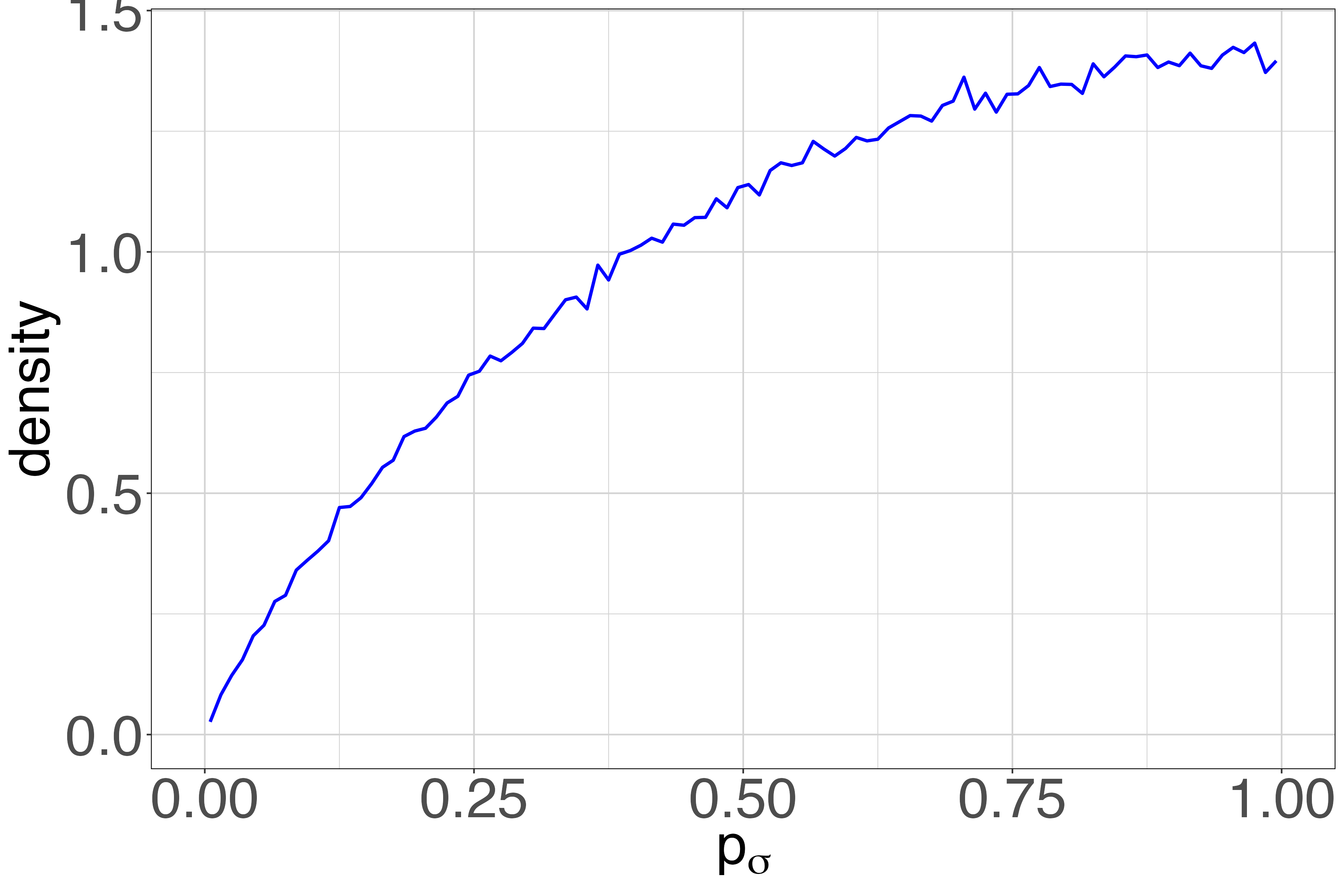}
    \hfill
    \includegraphics[width=0.3 \textwidth]{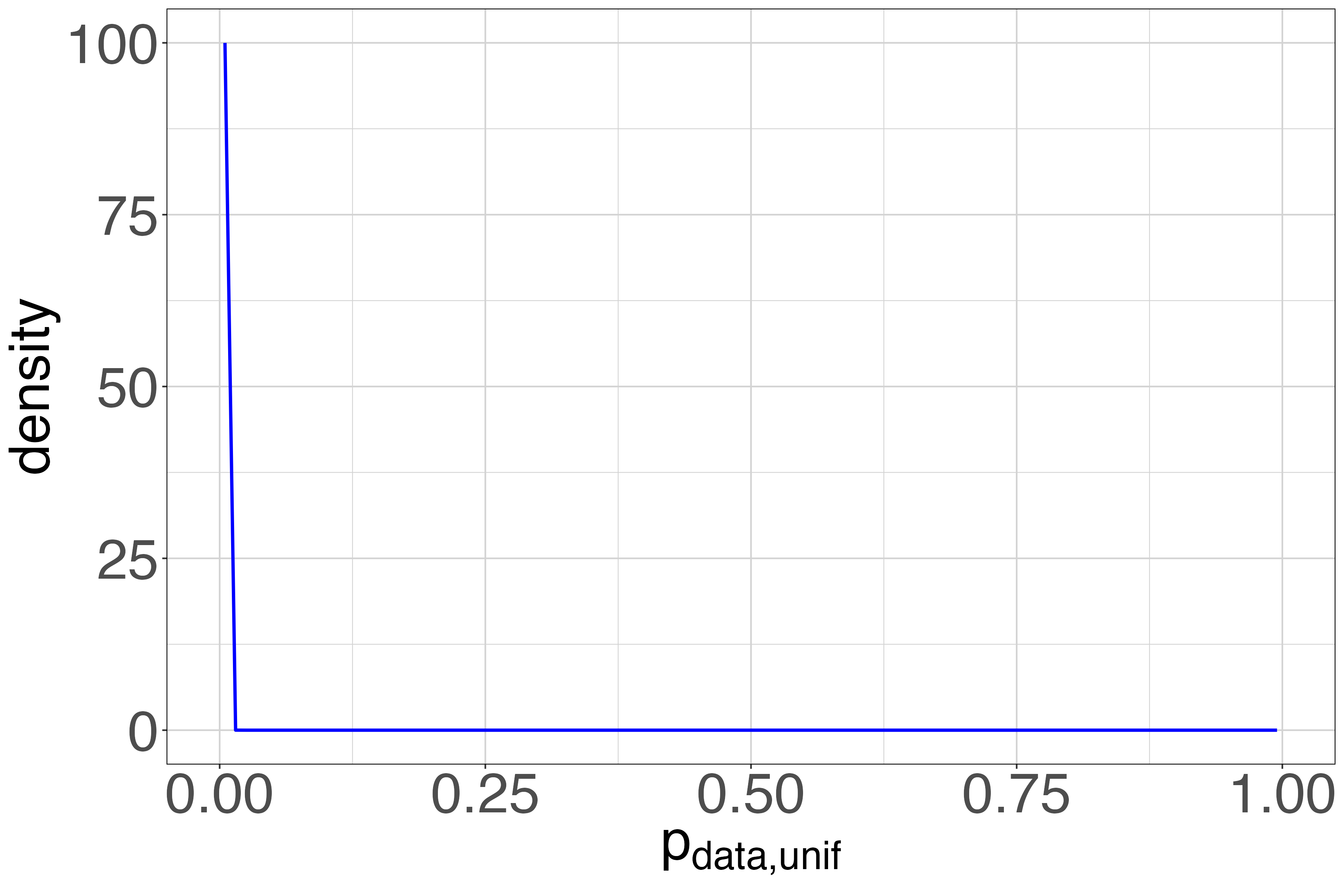}

    \includegraphics[width=0.3 \textwidth]{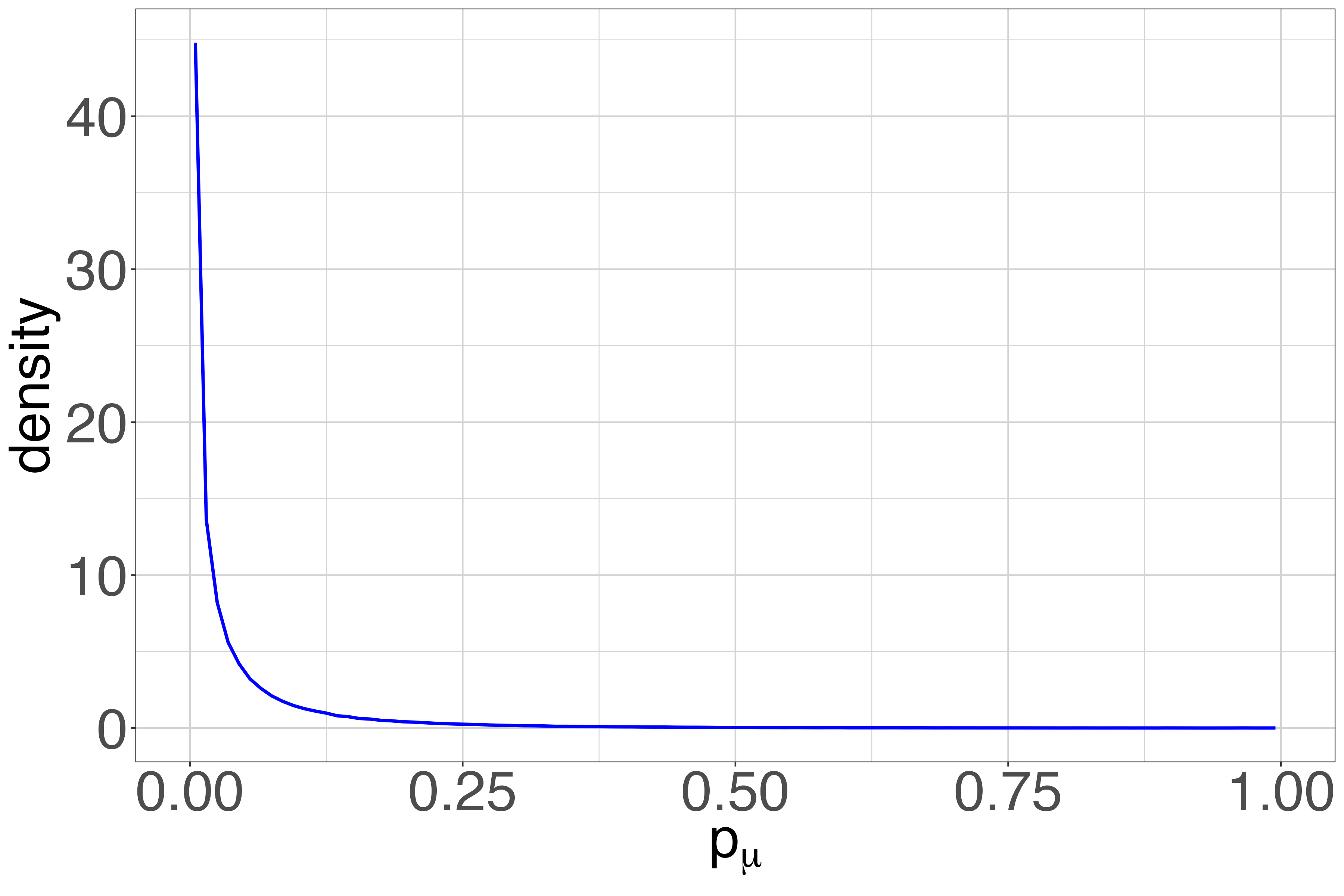}
    \hfill 
    \includegraphics[width=0.3 \textwidth]{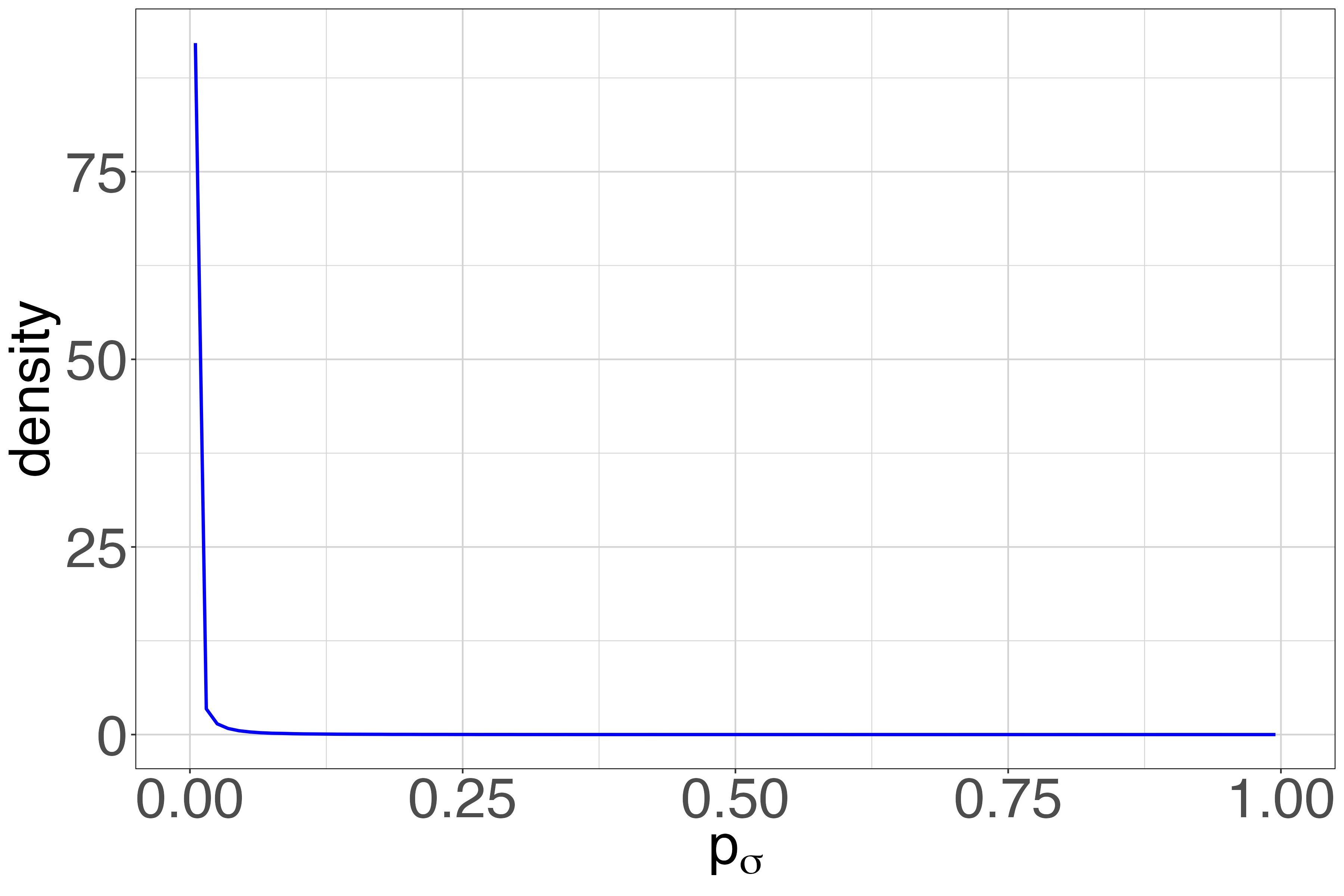}
    \hfill
    \includegraphics[width=0.3 \textwidth]{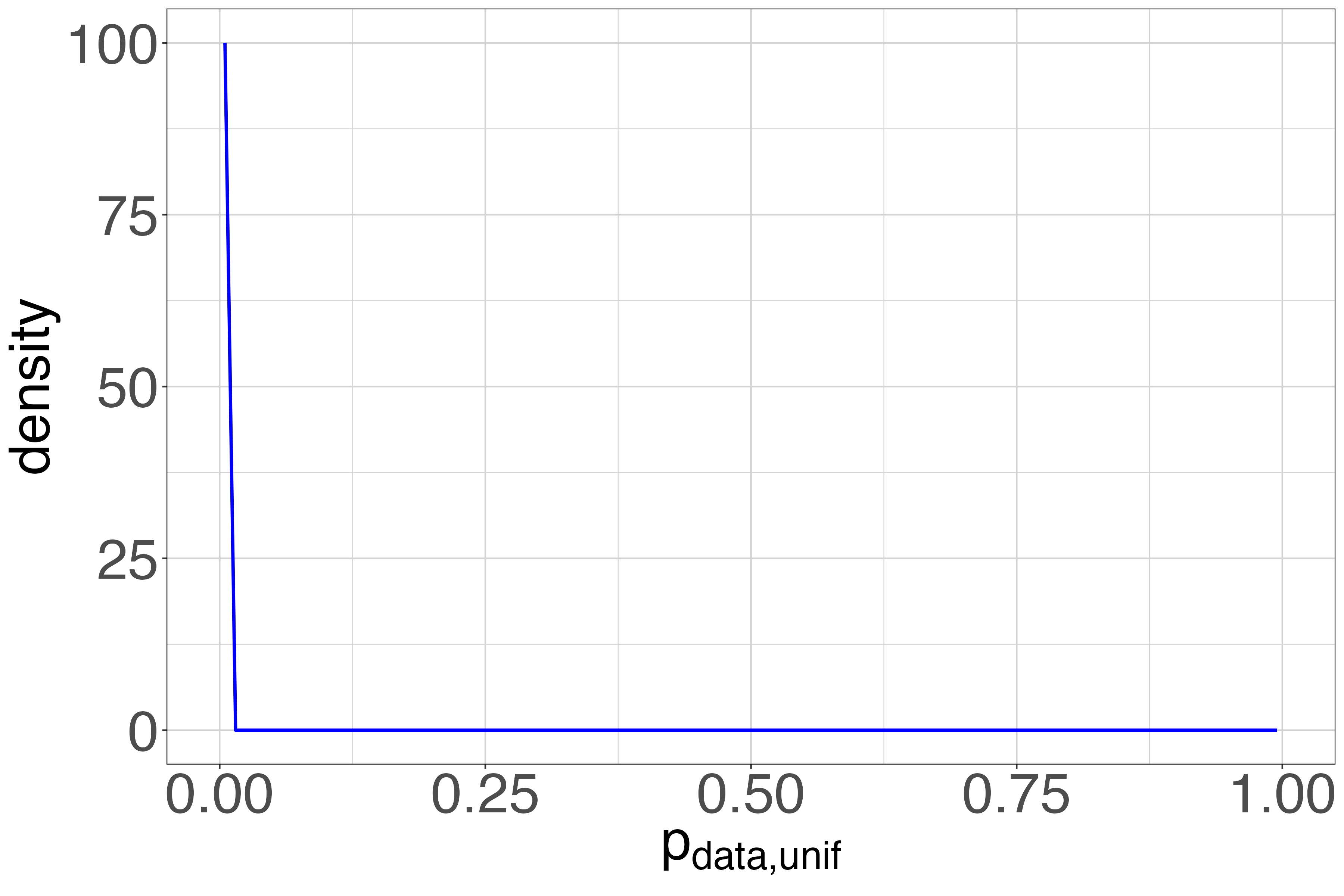}

    \caption{
    Results using UPCs on Newcomb data. Posterior densities of $p_\mu$, $p_\sigma$, and $p_\mathrm{data,unif}$ given the Newcomb data, under the data-dependent prior (top) and poorly chosen informative prior (bottom).  See \cref{fig:newcomb-upc} (top) for the corresponding plots under the weakly informative prior.
    }
    \label{appendix:fig:newcomb_p_densities}
\end{figure}

Recall that in \cref{table:newcomb_uniformity_test}, we report the aggregated p-values from the UPC tests.  To visualize the distributions across which this aggregation occurs, \cref{appendix:fig:newcomb_p_densities} shows the posterior densities of $p_\mu$, $p_\sigma$, and $p_{\mathrm{data,unif}}$ given the Newcomb data, under the data-dependent prior and the poorly chosen informative prior; see \cref{fig:newcomb-upc} for the corresponding plots under the weakly informative prior.  Under all three priors, the distribution of $p_{\mathrm{data,unif}}$ is concentrated near zero, which leads to the small aggregated p-values $p_{\mathrm{data,unif}^*}$ on the Newcomb data.
The distributions of $p_\mu$ and $p_\sigma$ are also concentrated near zero under the poorly chosen prior, which leads to the small aggregated values $p_\mu^*$ and $p_\sigma^*$ for this prior.

\subsection{Calculating the posterior densities of the p-values}

\paragraph{Approximating the PPC p-values for $T(Y) = \min\{Y_1,\ldots,Y_n\}$.}

For any i.i.d.\ random variables $Y_1,\ldots,Y_n$ with common CDF $F_Y$, the CDF of the minimum $T(Y) = \min\{Y_1,\ldots,Y_n\}$ is given by  \begin{align}\label{eq:cdf_of_minimum}
    F_T(t) = 1 - (1 - F_Y(t))^n.
\end{align}
Given the dataset $y = (y_1,\ldots,y_n)$, 
let $Y^* = (Y_1^*,\ldots,Y_n^*)$ be distributed according to the posterior predictive distribution.
Under the hypothesized normal model, the posterior predictive p-value for $T$ given $y$ is 
\begin{align}\label{eq:pp_density}
    \mathbb{P}\big(T(Y^*) \leq T(y)\,\big\vert\, y\big) &= \int \mathbb{P}\big(T(Y^*) \leq T(y)\,\big\vert\, \mu,\sigma, y\big) f(\mu,\sigma\mid y)\, d\mu \,d\sigma \\
    &\approx \frac{1}{S} \sum_{s=1}^S \mathbb{P}\big(T(Y^*) \leq T(y)\,\big\vert\, \mu_s,\sigma_s\big)
\end{align}
where $(\mu_1,\sigma_1),\ldots,(\mu_S,\sigma_S)$ are samples drawn from the posterior of $\mu,\sigma$ given the dataset $y = (y_1,\ldots,y_n)$.
By \cref{eq:cdf_of_minimum}, 
\begin{align}\label{eq:density_of_minimum_of_normals}
    \mathbb{P}\big(T(Y^*) \leq T(y)\,\big\vert\, \mu_s,\sigma_s\big) = 1 - \bigg(1 -  \Phi\bigg(\frac{T(y) - \mu_s}{\sigma_s}\bigg)\bigg)^n,
\end{align}
where $\Phi$ is the standard normal CDF.
By plugging \cref{eq:density_of_minimum_of_normals} into \cref{eq:pp_density}, we can obtain a better approximation to the posterior predictive p-value, compared to sampling $T(Y^*)$ directly.

\paragraph{Approximating the density of the UPC p-values for $\mu$.}

To produce smooth plots of the densities of the p-values in \cref{fig:newcomb-upc}, we derive a partially analytic estimate of the density of $p_\mu$ given $y$.
The following calculations are only for plotting purposes, and not needed to implement the UPC method.

To simplify the notation, we work with the precision $\lambda = 1/\sigma^2$ rather than the variance $\sigma^2$. In terms of $\mu$ and $\lambda$, our prior is $\mu\mid\lambda\sim\mathcal{N}(\mu_0, 1/(\kappa_0\lambda))$ and $\lambda\sim\mathrm{Gamma}(\alpha_0,\beta_0)$.
Let $(\mu_1,\lambda_1),\ldots,(\mu_S,\lambda_S)$ be samples from the posterior given the dataset $y = (y_1,\ldots,y_n)$.
Then we can form a Monte Carlo approximation to the posterior density of $p_\mu$, the UPC p-value for $\mu$, via
\begin{align}
\label{eq:p_mu_monte_carlo}
    f(p_{\mu} \vert y) 
    =
    \int f(p_{\mu} \vert \lambda, y) f(\lambda \vert y) d \lambda 
    \approx \frac{1}{S} \sum\limits_{s=1}^{S} f(p_{\mu} \vert \lambda_s, y).
\end{align}
As a function of $\mu$ and $\lambda$, the p-value for $\mu$ is 
\begin{equation}
    \label{eq:density_p_mu_given_y}
    p_{\mu} = 2 \min\{U_1, 1-U_1 \} = 2 \min\Big\{ \Phi\big( (\mu - \mu_0) \sqrt{\kappa_0 \lambda} \big),\; 1 - \Phi \big( (\mu - \mu_0) \sqrt{\kappa_0 \lambda} \big) \Big\},
\end{equation}
where $\Phi$ is the standard normal CDF.
By Jacobi's formula for transformation of random variables, the conditional density of $p_\mu$ is
\begin{equation}
    \label{eq:density_p_mu_given_lambda_y}
    f(p_\mu \mid \lambda,y) 
    = 
    f_{\mu \mid \lambda,y} \big( g^{-1}_{1,\lambda}(p_{\mu}) \mid \lambda, y \big) \bigg\vert \frac{d}{d p_{\mu}} g^{-1}_{1,\lambda}(p_{\mu}) \bigg\vert
    +
    f_{\mu \mid \lambda,y} \big( g^{-1}_{2,\lambda}(p_{\mu}) \mid \lambda, y\big) \bigg\vert \frac{d}{d p_{\mu}} g^{-1}_{2,\lambda}(p_{\mu}) \bigg\vert,
\end{equation}
where $f_{\mu|\lambda,y}(\cdot\mid \lambda,y)$ is the full conditional of $\mu|\lambda,y$ under the hypothesized model, specifically,
\begin{align}\label{eq:f_mu_full_conditional}
f_{\mu\mid\lambda,y}(\mu\mid \lambda,y) = \mathcal{N}\bigg(\mu \;\bigg\vert\; \frac{\mu_0 \kappa_0 + \sum_{i=1}^n y_i}{\kappa_0 + n},\; \frac{1}{(\kappa_0 + n) \lambda}\bigg),
\end{align}
and

\begin{equation}
\begin{split}
\label{eq:g_f_p_mu_formula}
    g^{-1}_{1,\lambda}(p_{\mu})
        &= \mu_0 + \Phi^{-1}(p_\mu/2)/\sqrt{\kappa_0 \lambda} \\
    \bigg\vert \frac{d}{d p_{\mu}} g^{-1}_{1,\lambda}(p_{\mu}) \bigg\vert
        &= \frac{1}{2\sqrt{\kappa_0 \lambda}} \frac{1}{\varphi(\Phi^{-1}(p_\mu/2))} \\
    g^{-1}_{2,\lambda}(p_{\mu})
        &= \mu_0 + \Phi^{-1}(1- p_\mu/2)/\sqrt{\kappa_0 \lambda} \\
    \bigg\vert \frac{d}{d p_{\mu}} g^{-1}_{2,\lambda}(p_{\mu}) \bigg\vert
        &= \frac{1}{2\sqrt{\kappa_0 \lambda}} \frac{1}{\varphi(\Phi^{-1}(1 - p_\mu/2))}
\end{split}
\end{equation}
where $\varphi$ is the standard normal PDF.

Plugging Equations~\ref{eq:density_p_mu_given_y}, \ref{eq:f_mu_full_conditional}, and \ref{eq:g_f_p_mu_formula} into \cref{eq:density_p_mu_given_lambda_y}, and plugging \cref{eq:density_p_mu_given_lambda_y} into \cref{eq:p_mu_monte_carlo} yields a partially analytic estimate of the density $f(p_{\mu} \vert y)$, which we find to be considerably more accurate than a purely Monte Carlo based estimate. 

\paragraph{Approximating the density of the UPC p-values for $\sigma$.}

To produce smooth plots of the densities of the p-values in \cref{fig:newcomb-upc}, we derive an closed-form expression of the density of $p_\sigma$ given $y$.
The following calculations are only for plotting purposes, and are not needed to implement the UPC method.

Next, we derive a closed-form expression for the posterior density of the p-values for $\sigma$. As in the case of $\mu$, we reparametrize in terms of $\lambda = 1/\sigma^{2}$ to simplify the calculations. This reparametrization does not affect the p-value since there a strictly monotone relationship between $\sigma$ and $\lambda$.
By straightforward calculations, the posterior density of $\lambda$ given $y$ is
\begin{equation}
    \label{eq:density_lambda_given_y}
    f_{\lambda|y}(\lambda) =   \mathrm{Gamma}(\lambda \mid \alpha_n, \beta_n)
\end{equation}
where $\alpha_n = \alpha_0 + n/2$ and 
\[
\beta_n = \beta_0 + \frac{1}{2} \sum\limits_{i=1}^{n} (y_i - \bar{y})^2 + \frac{1}{2} \frac{\kappa_0 n}{\kappa_0 + n} (\bar{y} - \mu_0)^2.
\]
As a function of $\lambda$, the p-value for $\lambda$ is
\begin{align}
\label{eq:p_lambda}
p_{\lambda} = 2 \min\{ U_2, 1 - U_2 \} = 2 \min\Big\{ 
    F_{\mathrm{Gamma}}(\lambda\mid \alpha_0, \beta_0),\,
    1 - F_{\mathrm{Gamma}}(\lambda \mid \alpha_0, \beta_0)
  \Big\}
\end{align}
where $F_{\mathrm{Gamma}}(x \mid \alpha_0, \beta_0) = \gamma(\alpha_0, \beta_0 x)/\Gamma(\alpha_0)$ is the CDF of a gamma distribution with shape $\alpha_0$ and rate $\beta_0$.
Here, $\Gamma$ denotes the gamma function and $\gamma$ is the lower incomplete gamma function.

When $\alpha_0 = \beta_0 = 1$, which are the only values we consider for this example, the posterior density of $p_{\lambda}$ is
\begin{equation}
    \label{eq:density_p_lambda_given_y}
    f(p_{\lambda} \vert y) =    f_{\lambda \vert y} \big(-\log(1 - p_{\lambda}/2)\big) \frac{1}{2 - p_{\lambda}} + f_{\lambda \vert y} \big(-\log(p_{\lambda}/2) \big) \frac{1}{p_{\lambda}}
\end{equation}
by Jacobi's formula for transformation of random variables.
Plugging \cref{eq:density_lambda_given_y,eq:p_lambda} into \cref{eq:density_p_lambda_given_y} yields a closed-form expression for $f(p_\lambda|y)$.

\section{Additional details for the dependent Bernoulli trials example}
\label{appendix:subsection:example_2}

This section contains additional empirical results and implementation details for the example from \cref{sec:bernoulli} on dependent Bernoulli trials.

\subsection{Additional empirical results}
\label{appendix:subsubsection:upc_densities_alternative_priors}

\begin{figure}
\centering

\includegraphics[width=0.3 \textwidth]{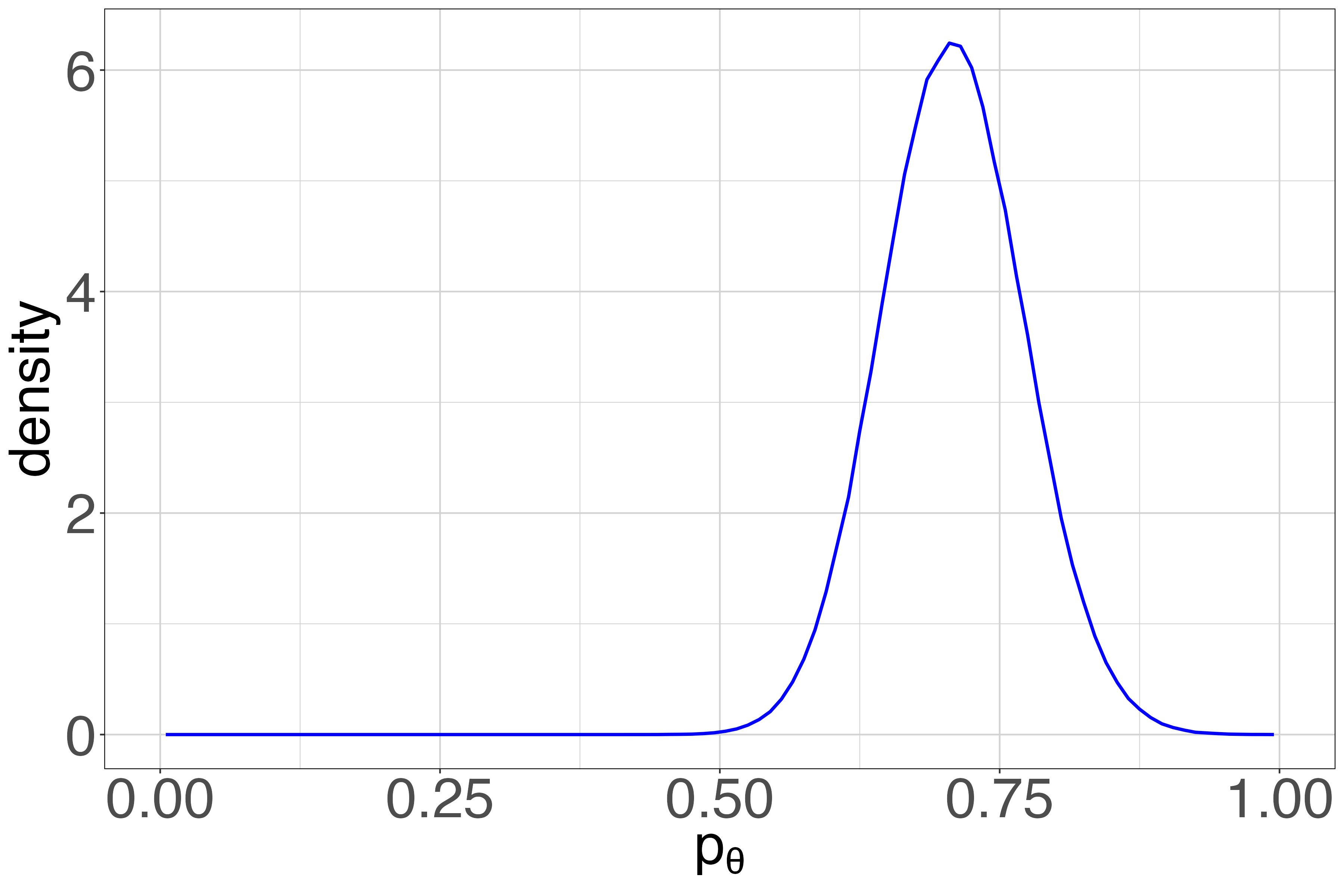}
\hfill 
\includegraphics[width=0.3 \textwidth]{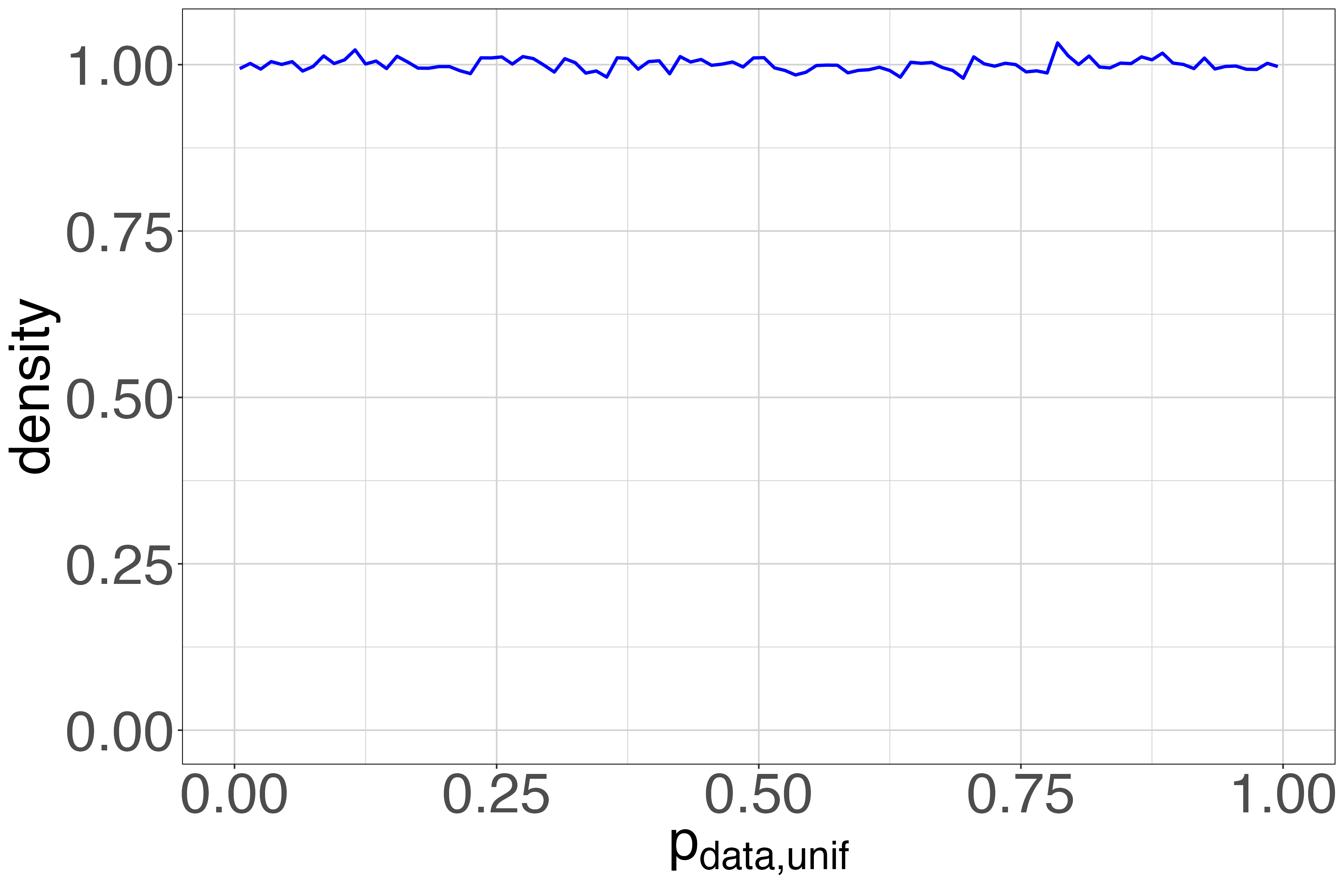}
\hfill
\includegraphics[width=0.3 \textwidth]{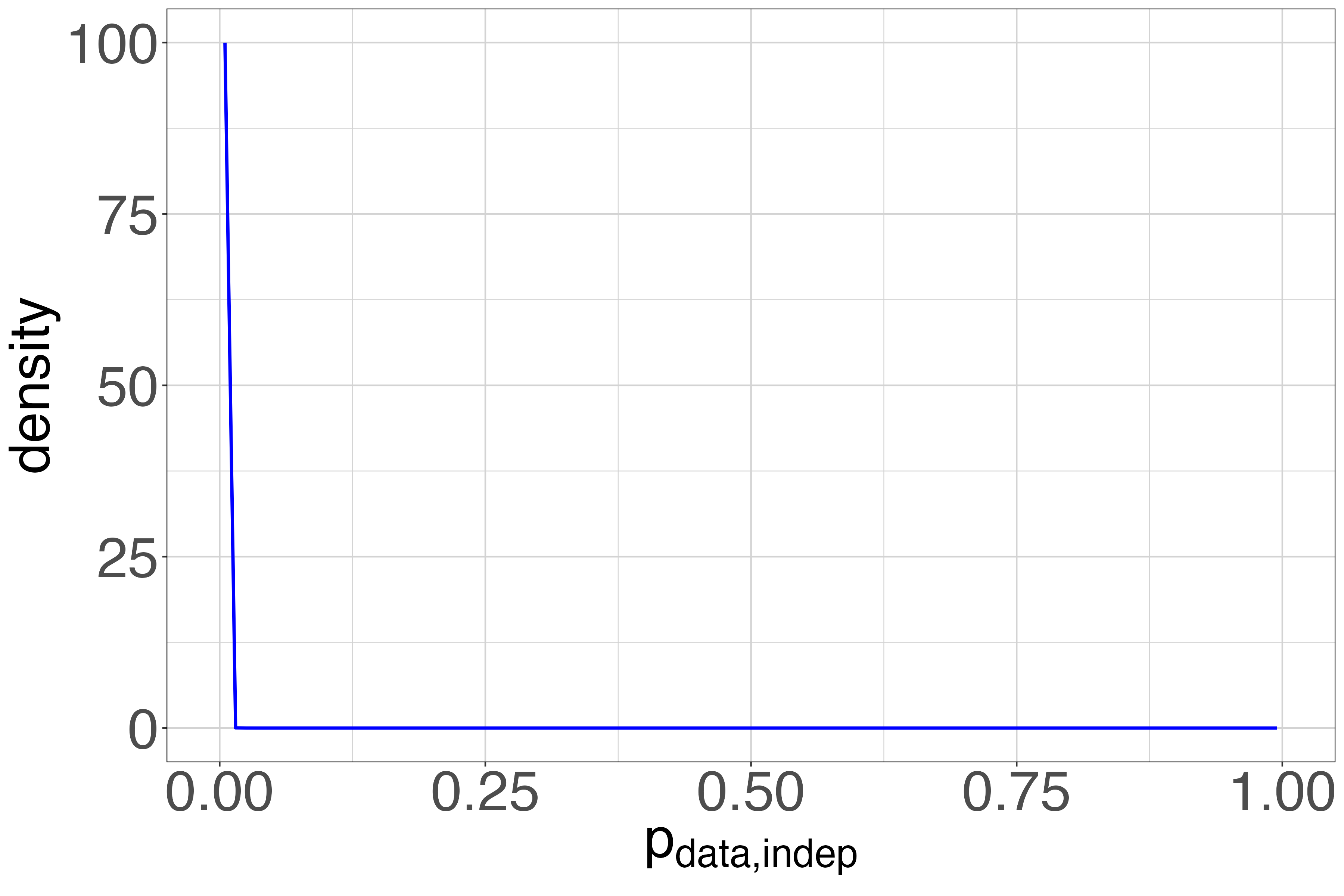}

\includegraphics[width=0.3 \textwidth]{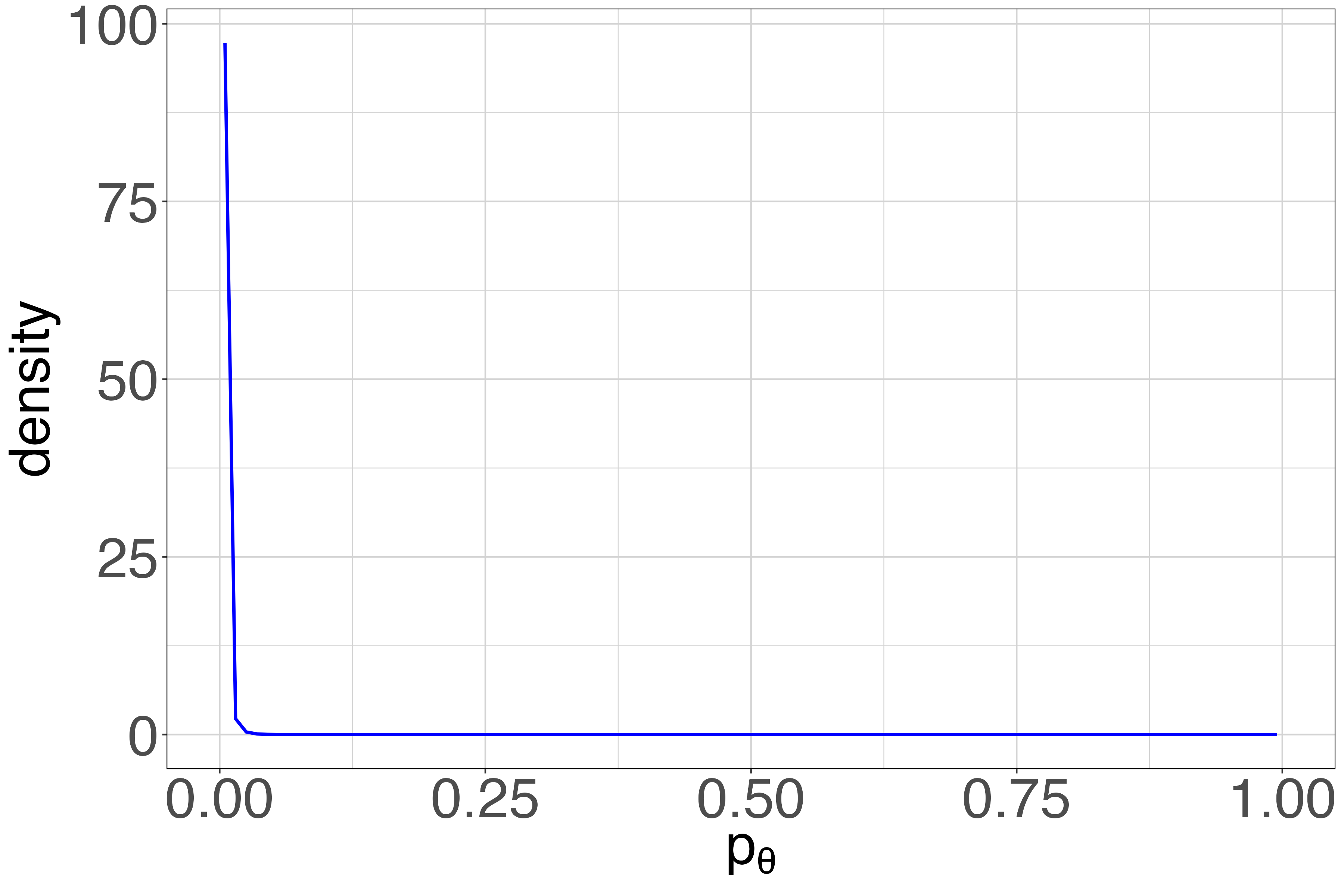}
\hfill 
\includegraphics[width=0.3 \textwidth]{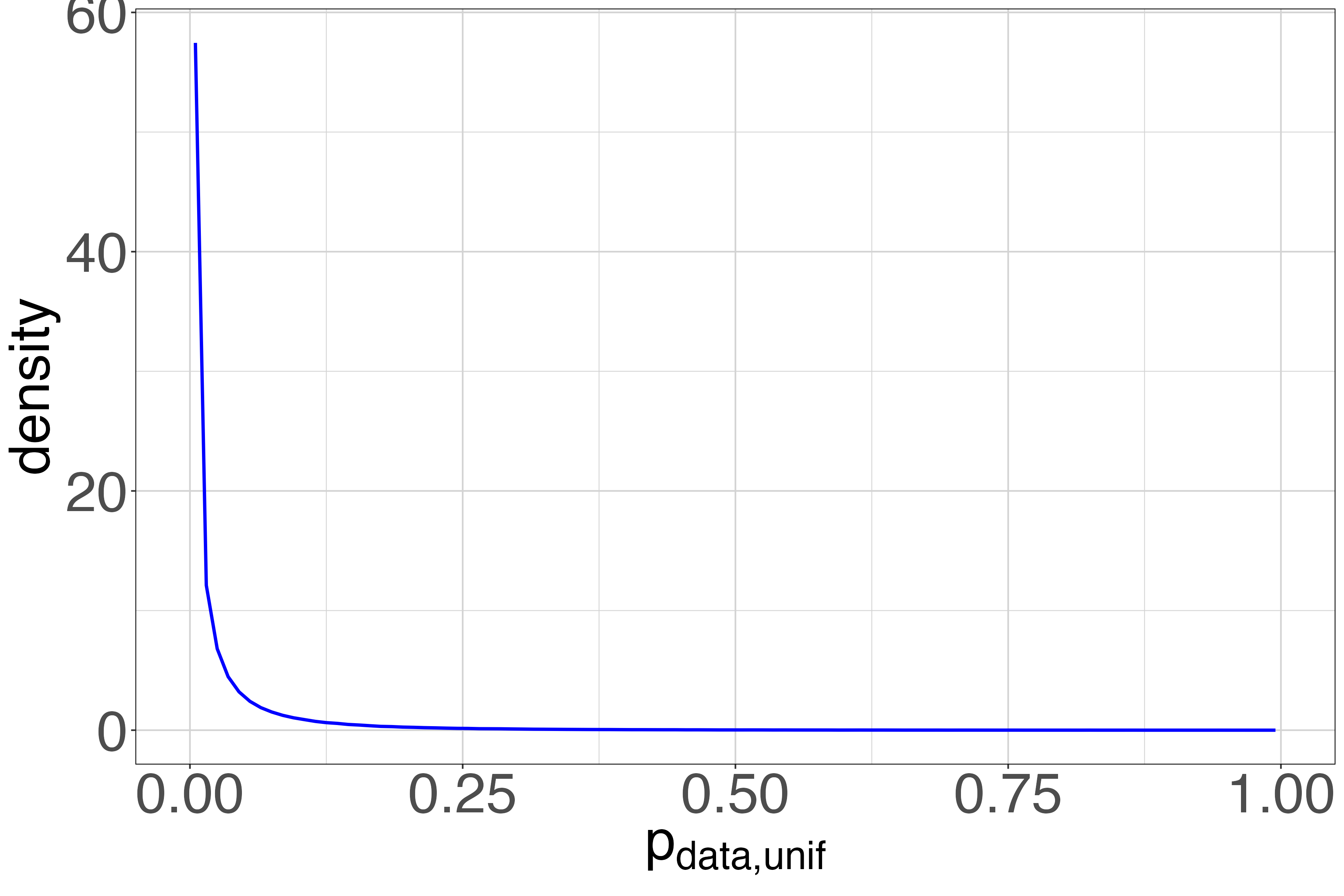}
\hfill
\includegraphics[width=0.3 \textwidth]{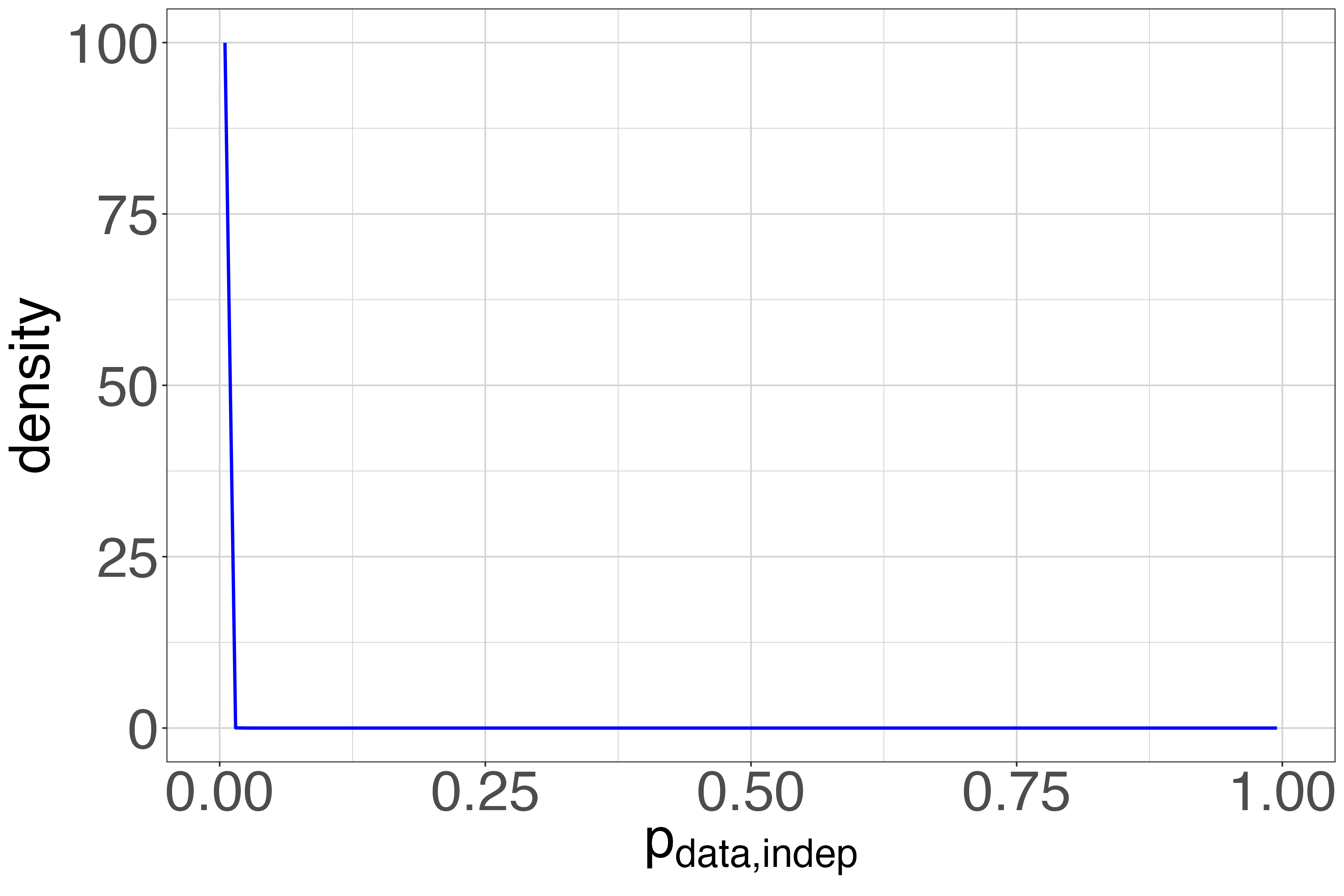}

\caption{Results using UPCs on dependent Bernoulli example. Posterior densities of the p-values $p_\theta$, $p_{\mathrm{data,unif}}$, and $p_{\mathrm{data,indep}}$ given the observed data in \cref{eq:bernoulli_data}, under the Jeffreys prior (top) and poorly chosen prior (bottom). See \cref{fig:bernoulli_upc} (top) for the corresponding plots under the uniform prior.
}
\label{appendix:fig:bernoulli_upc_other_priors}
\end{figure}

Recall that in \cref{table:bernoulli_tests}, we report the aggregated p-values for the dependent Bernoulli trials example, for each of the three choices of prior (uniform, Jeffreys, and poorly chosen).
To visualize how these aggregated p-values arise from the posteriors on UPC p-values, \cref{appendix:fig:bernoulli_upc_other_priors} shows the posterior densities of the p-values $p_\theta$, $p_{\mathrm{data,unif}}$, and $p_{\mathrm{data,indep}}$ given the dataset in \cref{eq:bernoulli_data}, under the Jeffreys and poorly chosen priors; see \cref{fig:bernoulli_upc} for the corresponding plots under the uniform prior.
The posterior of $p_{\mathrm{data,indep}}$ is concentrated near zero for all three priors, correctly indicating that the hypothesized i.i.d.\ Bernoulli model does not adequately capture the dependency in the observed data. This gives rise to the small aggregated p-values $p_{\mathrm{data,indep}}^*$ seen in \cref{table:bernoulli_tests}.  
Under the uniform and Jeffreys priors, the posterior of $p_{\mathrm{data,unif}}$ is nearly uniform, reflecting the fact that the Bernoulli model must be correct marginally, and the parameter $\theta$ has been reasonably well estimated.  
Under the poorly chosen prior, all of the p-values are small; in the case of $p_{\mathrm{data,unif}}$, this is because the prior too strongly biases the inferred value of $\theta$.

\subsection{Hoeffding independence test}

\begin{figure}
    \centering
    \includegraphics[width=0.8 \textwidth]{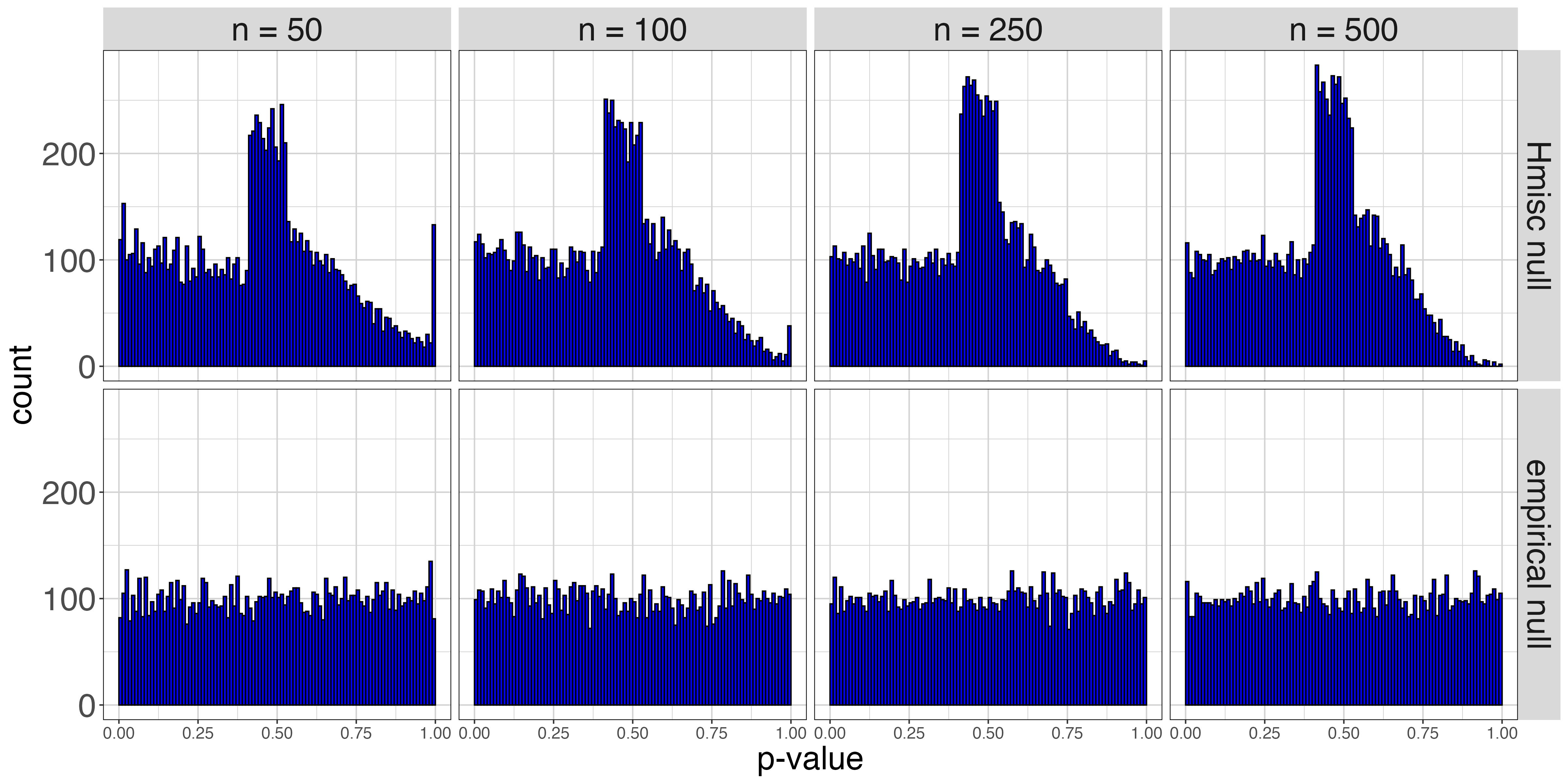}

    \caption{Calibration of the Hoeffding independence test: Distribution of p-values produced by the raw \textsc{Hmisc} procedure and when using our empirical null distribution, for a range of sample sizes.}
    \label{supp:fig:hoeffding_null_comparison}
\end{figure}

To calculate the test statistics for the Hoeffding 
independence test, we use the implementation in the \textsc{Hmisc} R package~\citep{Hmisc}. \textsc{Hmisc} uses a series of rules based on the exact value of the test statistic it calculates to produce a p-value. These rules produce p-values which give valid Type I error rate control under common $\alpha$ values and thus are reasonable to use for most practical use cases. However, they do not yield uniform p-values under the null. Thus, to obtain uniform p-values under the null, we construct an empirical null distribution by Monte Carlo simulation and we base our reported p-values on this instead. 
Specifically, to generate the empirical null, we (1) independently generate $U_1^{(j)},\ldots,U_n^{(j)} \sim \mathrm{Uniform}(0,1)$ i.i.d.\ and $V_1^{(j)},\ldots,V_n^{(j)} \sim \mathrm{Uniform}(0,1)$ i.i.d., for $j = 1,\ldots,J$ where $J = 10^5$, then (2) compute the Hoeffding test statistic value $t_j = T(U_{1:n}^{(j)},V_{1:n}^{(j)})$ for $j = 1,\ldots,J$, and (3) store $t_1,\ldots,t_J$ as defining the empirical null distribution.
Note, in particular, that the empirical null does not depend on the model or data at all -- it only depends on the sample size $n$.
Figure~\ref{supp:fig:hoeffding_null_comparison} shows the distribution of p-values when using the raw \textsc{Hmisc} procedure versus using our empirical null.

\section{Additional details for the logistic regression example}
\label{appendix:section:logistic_regression}

This section contains additional empirical results for the example from \cref{sec:logistic} on the logistic regression model for adolescent smoking data.

\paragraph{Model \#1.}
Recall from the main text that we conduct five different tests on the u-values from Model~\#1. To use the corresponding p-values as part of a decision rule, we choose to control the FWER for this set of tests at the $\alpha = 0.1$ level using the Holm--Bonferroni correction.  
The unadjusted p-values are 
$p_{\mathrm{data},\texttt{wave}}^* = 1.67 \times 10^{-7}$,
$p_{\mathrm{data},\texttt{sex}}^* = 0.72$,
$p_{\mathrm{data},\texttt{parsmk}}^* = 8.47 \times 10^{-3}$,
$p_{\alpha,\texttt{sex}}^* = 0.68$, and
$p_{\alpha,\texttt{parsmk}}^* = 1.81 \times 10^{-11}$ which, after adjustment via the Holm--Bonferroni correction, become 
$p_{\mathrm{data},\texttt{wave}}^{*,\mathrm{adj}} = 6.69 \times 10^{-7}$,
$p_{\mathrm{data},\texttt{sex}}^{*,\mathrm{adj}} \approx 1$,
$p_{\mathrm{data},\texttt{parsmk}}^{*,\mathrm{adj}} = 0.03$,
$p_{\alpha,\texttt{sex}}^{*,\mathrm{adj}} \approx 1$, and
$p_{\alpha,\texttt{parsmk}}^{*,\mathrm{adj}} = 9.07 \times 10^{-11}$.

\paragraph{Model \#2.}

Recall that our two tests for Model \#2 yield unadjusted p-values of $p_{\mathrm{data},\texttt{wave}\times\texttt{parsmk}}^* \approx 1$ and $p_{\alpha,\mathrm{unif}}^* = 3.41 \times 10^{-7}$. After Holm--Bonferroni correction, the adjusted p-values are $p_{\mathrm{data},\texttt{wave}\times\texttt{parsmk}}^{*,\mathrm{adj}} \approx 1$ and $p_{\alpha,\mathrm{unif}}^{*,\mathrm{adj}} = 6.82 \times 10^{-7}$. Note that these p-values are independent under $H_0$, so we could alternatively use Fisher's p-value combination method~\citep{Fisher34}, rather than Holm--Bonferroni. 

\paragraph{Model \#3.}

\begin{figure}
\centering 
\includegraphics[width=0.45\textwidth]{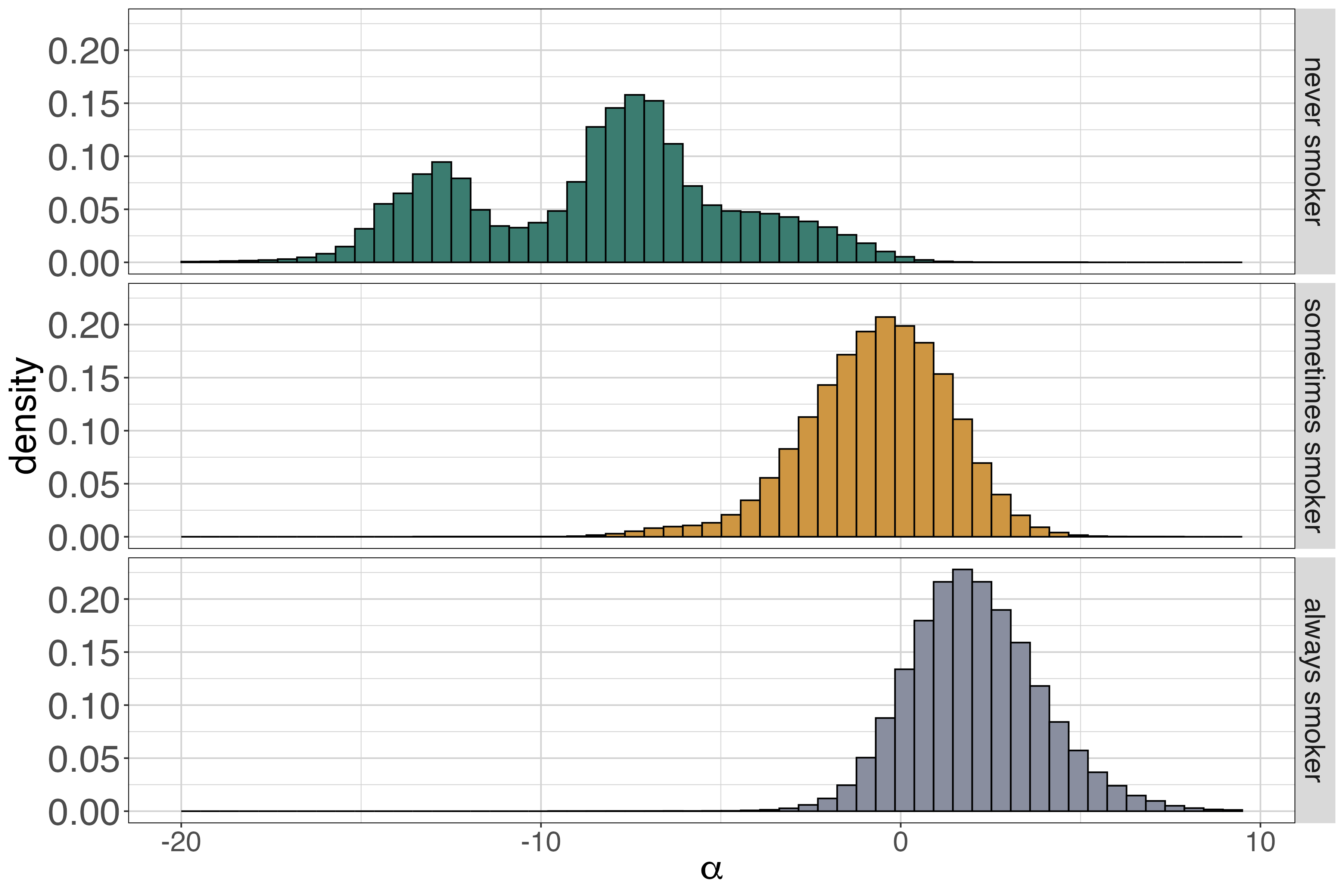}
\hfill
\includegraphics[width=0.45\textwidth]{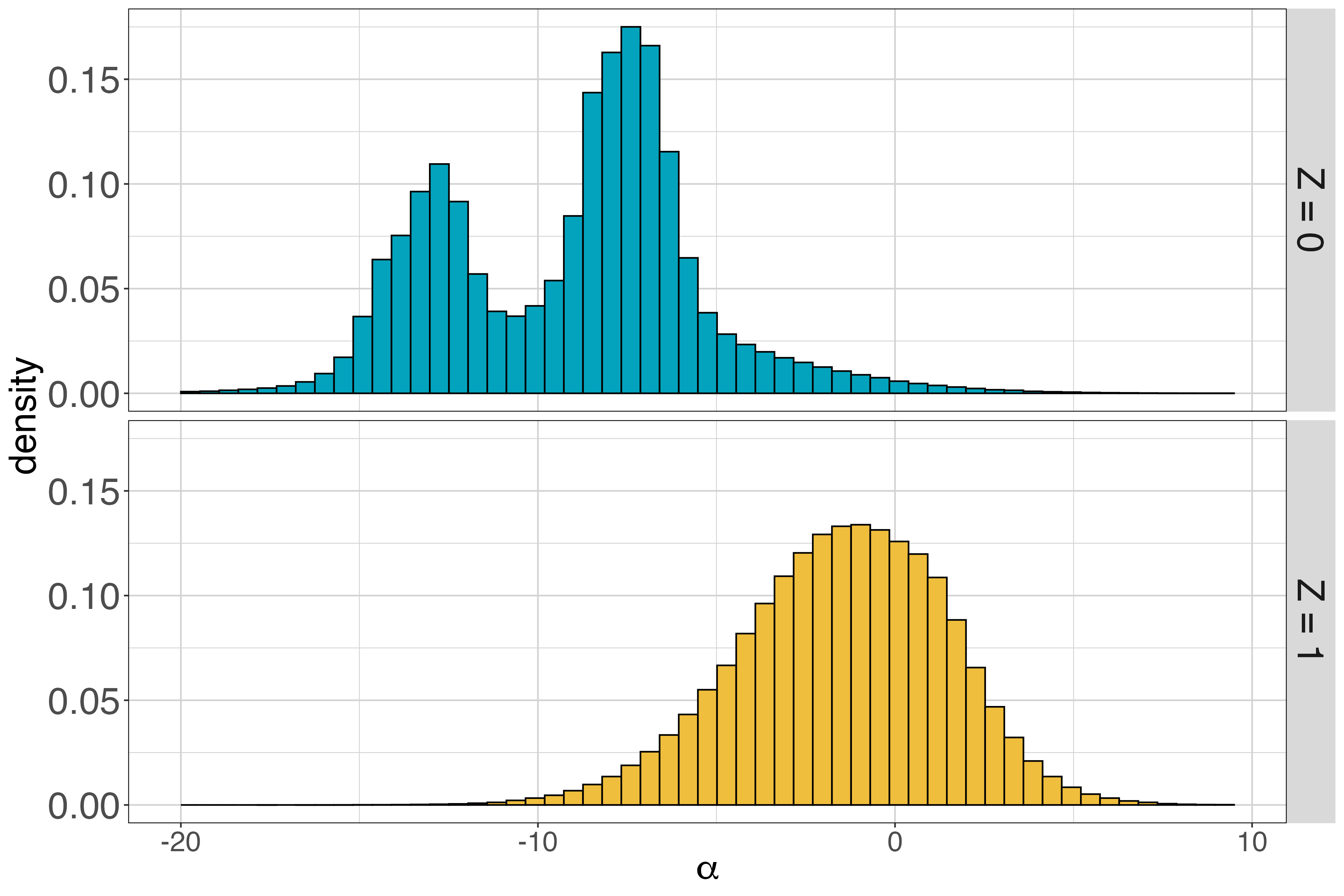}
\caption{Additional results for Model \#3 for logistic regression smoking example: Distribution of $\alpha$ values by smoking status, averaged over $1{,}000$ posterior samples. Distribution of $\alpha$ values conditional on $Z_j$, averaged over $1{,}000$ posterior samples.}
\label{fig:smoking_alpha_by_smoker_status_z}
\end{figure}

To interpret the latent variables $Z_j$ in Model \#3, 
we split the individuals into three groups: 
``always smokers'' (those who smoke regularly at every wave of the study), ``never smokers'' (those who never 
smoke regularly), and ``sometimes smokers'' (those who switch between smoking regularly and not throughout the study); these groups were previously used by \citet[Section 6.3]{gelman2013bayesian} to define PPC test statistics.
\cref{tab:smoking_3} shows the posterior mean of the fraction of individuals in each group, given $Z_j = 0$ and $Z_j = 1$, respectively.
We see that the latent variable $Z_j$ correlates strongly with the ``never smoker'' group, specifically, nearly all of the individuals with $Z_j = 0$ are ``never smokers''.
Thus, the two mixture components in the prior on $\alpha$ represent groups with higher and lower probability of smoking, respectively, beyond the effects due to \texttt{wave}, \texttt{sex}, and \texttt{parsmk} (parent smoking status).
To visualize the relationship between the $\alpha$ values and these groups, \cref{fig:smoking_alpha_by_smoker_status_z} shows histograms of the empirical distribution of $\alpha$ values for individuals in the ``never smoker'', ``sometimes smoker'', and ``always smoker'' groups averaged over 1{,}000 posterior samples.
\cref{fig:smoking_alpha_by_smoker_status_z} also shows histograms of the empirical distribution of $\alpha$ values for individuals with $Z_j = 0$ and $Z_j = 1$, averaged over 1{,}000 posterior samples. 

\begin{figure}
\centering 

\includegraphics[width=0.45\textwidth]{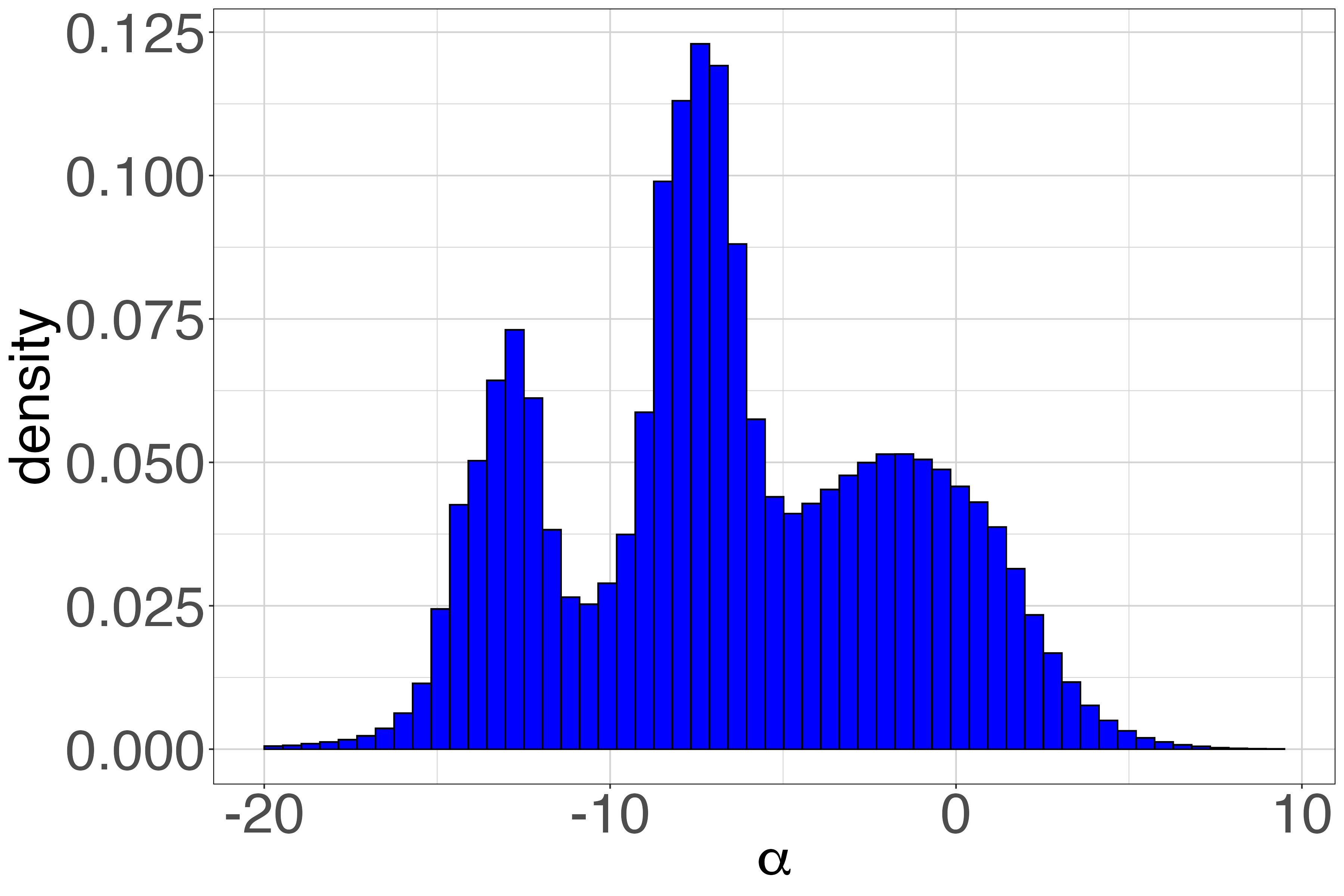}

\hfill

\caption{UPC results for Model \#3 on the logistic regression example. Histogram of the $\alpha$ values over all posterior samples. 
}
\label{fig:smoking_3_alpha}
\end{figure}

\begin{table}
    \centering
    \begin{tabular}{|c|c|c|c|}
        \hline
        \textbf{posterior $Z_j$} & 
        \textbf{never smoker} &
        \textbf{sometimes smoker} &
        \textbf{always smoker} \\
        \hline
        $Z_j = 0$ & $0.96$ & $0.04$ & $6.4 \times 10^{-3}$ \\
        $Z_j = 1$ & $0.41$ & $0.45$ & $0.14$ \\
        \hline
    \end{tabular}

    \caption{ Probabilities of latent smoking status conditional on posterior $Z_j$. }
    \label{tab:smoking_3}
\end{table}

\begin{figure}
    \centering
    \includegraphics[width=0.55\textwidth]{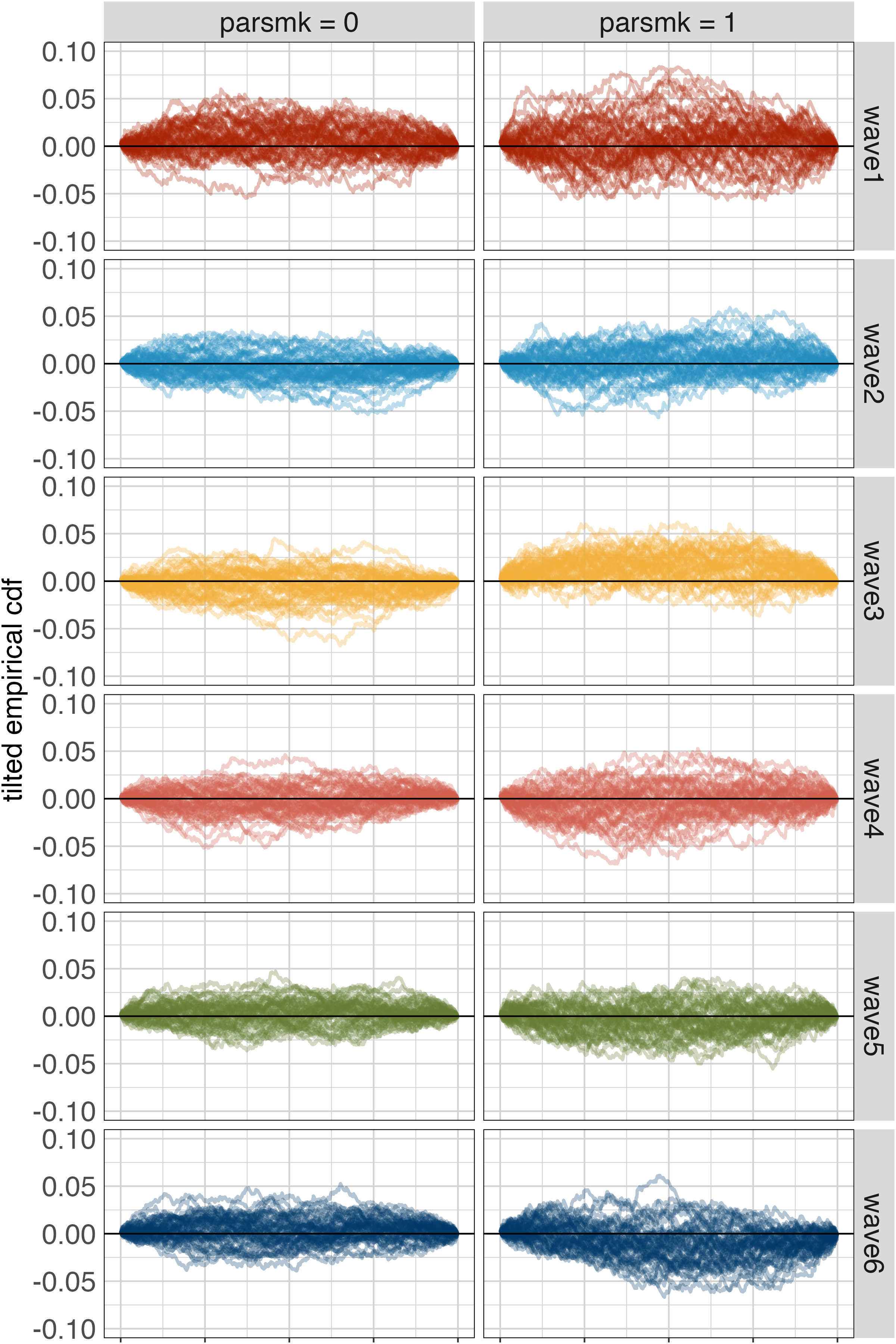}

    \caption{Additional results for Model \#3 for the logistic regression example: Tilted CDFs for data u-values stratified by $\texttt{wave} \times \texttt{parsmk}$. Visually, there is little variation in the data u-value distribution across values of \texttt{wave} $\times$ \texttt{parsmk}, corroborating the fact that $p_{\mathrm{data},\texttt{wave}\times\texttt{parsmk}}^*$ is very large.}
    \label{fig:smoking_3_parsmk_x_wave}
\end{figure}

\end{document}